\documentclass[aps,pra,showpacs,amsmath,amssymb,twocolumn,nofootinbib,superscriptaddress]{revtex4-1}

\usepackage[pdftex]{graphicx}
\usepackage{mathrsfs,amsmath,amssymb,amsthm,graphicx,caption,subcaption}
\usepackage{fancyhdr}
\usepackage[colorlinks=true,
            linkcolor=red,
            urlcolor=blue,
            citecolor=blue]{hyperref}

\begin{document}

\bibliographystyle{apsrev}

\title{Quantum Communication Over Atmospheric Channels: A Framework for Optimizing Wavelength and Filtering}

\author{R. Nicholas Lanning}
\affiliation{Air Force Research Laboratory, Directed Energy Directorate, Kirtland AFB, NM, United States}
\email[AFRL.RDSS.OrgMailbox@us.af.mil\\
Approved for public release; distribution is unlimited. Public Affairs release approval AFRL-2021-0775.]{}


\author{Mark A. Harris}
\affiliation{Leidos, Albuquerque NM, United States}

\author{Denis W. Oesch}
\affiliation{Leidos, Albuquerque NM, United States}

\author{Michael D. Oliker}
\affiliation{Leidos, Albuquerque NM, United States}

\author{Mark T. Gruneisen}
\affiliation{Air Force Research Laboratory, Directed Energy Directorate, Kirtland AFB, NM, United States}

\date{\today}

\begin{abstract}
Despite quantum networking concepts, designs, and hardware becoming increasingly mature, there is no consensus on the optimal wavelength for free-space systems. 
We present an in-depth analysis of a daytime free-space quantum channel as a function of wavelength and atmospheric spatial coherence (Fried coherence length). 
We choose decoy-state quantum key distribution bit yield as a performance metric in order to reveal the ideal wavelength choice for an actual qubit-based protocol under realistic atmospheric conditions. 
Our analysis represents a rigorous framework to analyze requirements for spatial, spectral, and temporal filtering.
These results will help guide the development of free-space quantum communication and networking systems. 
In particular, our results suggest that shorter wavelengths in the optical band should be considered for free-space quantum communication systems.
Our results are also interpreted in the context of atmospheric compensation by higher-order adaptive optics.  
\end{abstract}
		

\maketitle
\pagestyle{fancy}
\cfoot{Approved for public release; distribution is unlimited. Public Affairs release approval AFRL-2021-0775.}
\lhead{}
\chead{}
\rhead{\thepage}

\section{Introduction}
Quantum networking concepts, designs, and hardware are becoming increasingly mature and in many ways transitioning to an engineering phase.
Unlike fiber networks which suffer an exponential attenuation with propagation distance, long-distance free-space networks only suffer a quadratic loss due to geometric aperture-to-aperture coupling when approximated by the Friis equation \cite{friis1971introduction, alexander1997optical}.
In principle, free-space quantum networks can enable global-scale quantum communication via satellite based nodes and quantum ground transceivers. 
This could facilitate distributed quantum computation, blind quantum computation, quantum-assisted imaging, and precise timing, to name just a few proposed applications~\cite{van2014quantum, boone2015entanglement, wehner2018quantum}.
An enduring problem is the ideal wavelength for free-space quantum communication over atmospheric channels, particularly in daytime conditions where filtering sky-noise photons is a formidable challenge~\cite{jacobs1996quantum, buttler2000daylight, hughes2002practical, shan2006free, peloso2009daylight, heim2010atmospheric, garcia2013high, carrasco2014correction, gruneisen2015modeling, gruneisen2016adaptive, gruneisen2017modeling, liao2017long, vasylyev2017free, arteaga2019enabling, gruneisen2020adaptive}. 

\begin{figure}[b]
	\includegraphics[width=1\columnwidth]{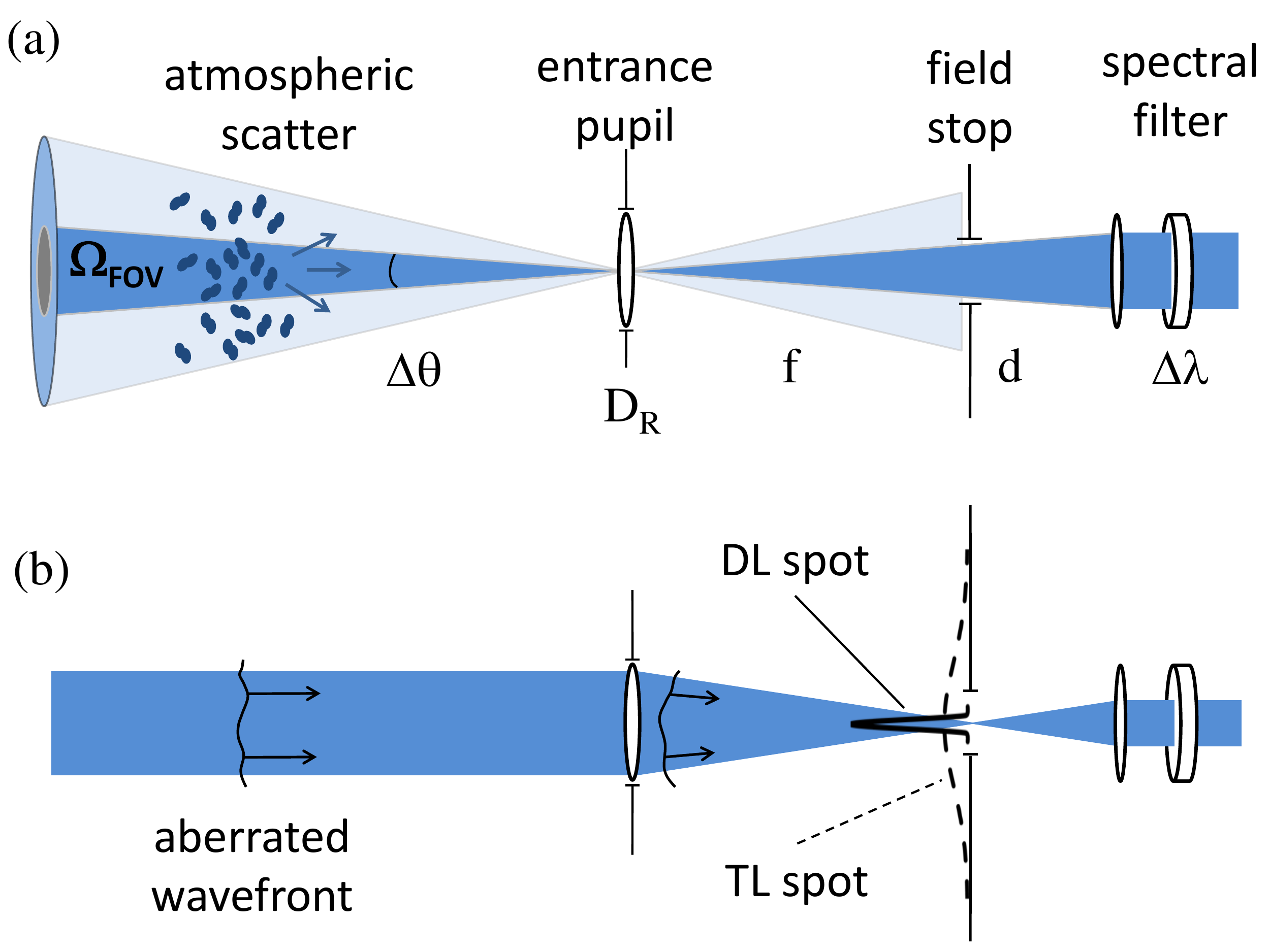}
	\caption{\label{fig:GenSchem}
Schematic illustrating (a) the concept of spatially filtering sky-noise photons created by atmospheric scattering. An entrance pupil ($D_{\mathrm{R}}$) with focal length $f$ and a field stop with diameter $d$ determine the solid-angle FOV ($\Omega_{\mathrm{FOV}}$). Reducing the FOV of the receiver can reduce the number of sky noise-photons transmitted to the spectral filter ($\Delta \lambda$). Schematic (b) illustrates the effect of focusing in the presence of turbulence; the aberrated wavefront results in a broadened turbulence-limited spot (TL spot) as compared to the diffraction-limited spot (DL spot).
	}
\end{figure}

Figure~\ref{fig:GenSchem}(a) illustrates the concept of spatially filtering optical noise at the field stop of an optical receiver.  
A primary optic of diameter $D_{\mathrm{R}}$ defines the entrance pupil and is followed by a field stop situated in the focal plane, a collimating lens, and a spectral filter.
The field stop defines the solid-angle field of view (FOV) and limits the number of sky noise photons $N_{\mathrm{b}}$ transmitted to the quantum detectors.
In a system design, the field stop should be made sufficiently large to minimize losses to the quantum signal but otherwise made small enough to minimize the transmission of sky-noise photons.
The choice of field stop size is closely tied to the choice of quantum-signal wavelength, making wavelength perhaps the most critical design criterion. 
\begin{figure*}[t!]
	\includegraphics[width=1\textwidth]{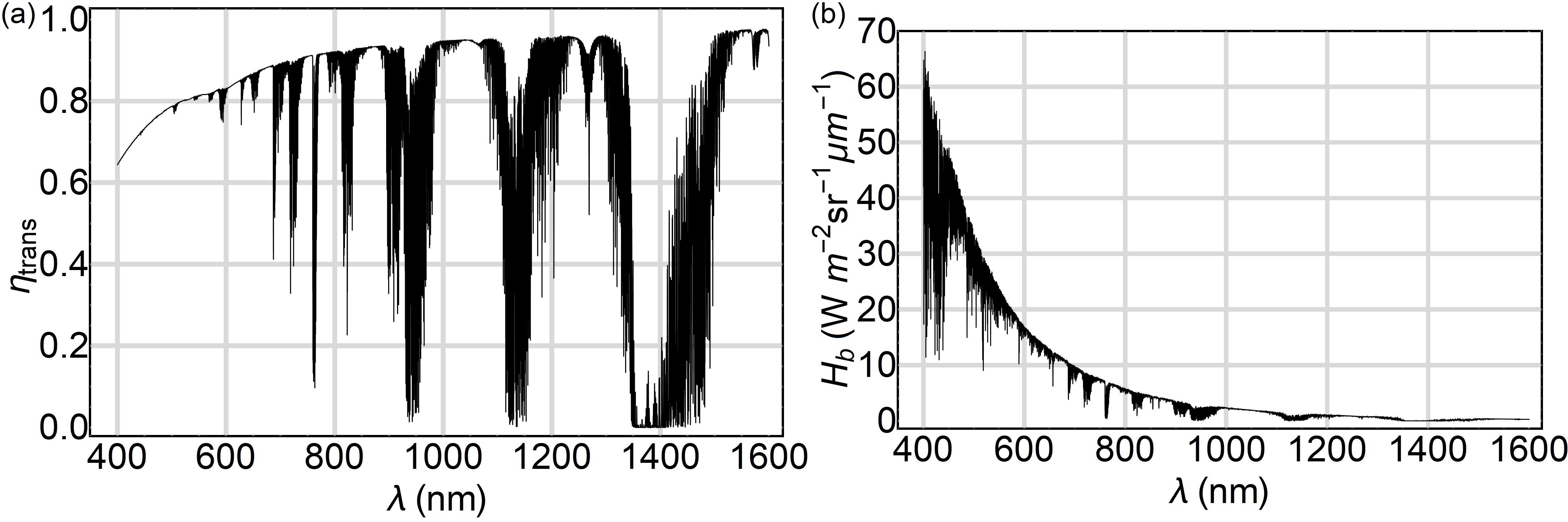}
	\caption{\label{fig:WinSolZenHighVisTransHb}
The site dependent (a) atmospheric transmission $\eta_{\mathrm{trans}}$ and (b) spectral radiance $H_{\mathrm{b}}$ for a receiver pointed at zenith on the winter solstice at 1:00 PM with 50-km visibility. 
	}
\end{figure*}
Still, the quantum-networking community has not settled on an ideal free-space quantum communication wavelength.
For example, wavelengths of interest for satellite-Earth quantum communication have included both the 1550-nm telecom wavelength and shorter wavelengths near 775 nm~\cite{hughes2002practical, nordholt2002present, bourgoin2013comprehensive, gruneisen2015modeling, gruneisen2016adaptive, gruneisen2017modeling}.
To date, most daytime free-space quantum-communication demonstrations have included rudimentary analysis on the wavelength dependence of quantum communication.
For example, some have concluded that 1550 nm is ideal due to lower solar radiance, the wavelength-dependent nature of scattering, and compatibility with current telecom technology~\cite{liao2017long}.
However, in this case they did not consider the effects of geometric coupling, spatial filtering strategies, or the wavelength dependence of the focused spot size as we do in this article.
The effort herein is to provide a careful analysis that is linked to actual space-Earth channel atmospheric parameters and can help guide system development as the quantum-networking community progresses toward the global-scale quantum internet.

To analyze the wavelength-dependence of free-space quantum communication, we choose the decoy-state BB84 quantum key distribution (QKD) \cite{bennett1984proc, bennett2020quantum, ma2005practical} bit yield as our performance metric.
We simulate a satellite down-link scenario in the presence of daytime spectral radiance and atmospheric turbulence characterized by the Fried coherence length $r_0$~\cite{fried1966optical}.
The Fried coherence length characterizes the dominating phenomenon determining the signal throughput at the receiver field stop. 
Scintillation is another consequence of atmospheric propagation and leads to spatial and temporal intensity variations in the focal-plane where spatial filtering occurs.  
In classical optical communication, scintillation leads to signal fading that can be problematic.  
However, in quantum communication protocols, scintillation itself is not necessarily problematic \cite{shapiro2011scintillation}.  
Over sufficiently long integration times, the statistics of quantum measurements are not affected by scintillation but instead governed by the time-averaged channel efficiency which, as we will show, can be modeled using the spatial coherence $r_0$.
Thus, scintillation will be neglected and we will derive the bit yield as a function of wavelength and spatial coherence. 

We show that shorter wavelengths generally outperform longer wavelengths.
We also investigate how site-specific atmospheric conditions can affect spectral filtering requirements.
We show that more aggressive spectral filtering can be used to mitigate the effects of a more challenging atmosphere while still taking advantage of the optimal spatial filtering and aperture coupling at shorter wavelengths.
We also cast the optimization problem in terms of higher-order adaptive optics (AO).
Although realistic daytime atmospheres will be quite challenging, higher-order AO allows one to correct for atmospheric turbulence and in effect operate their optical receiver closer to the diffraction limit (see Fig.~\ref{fig:GenSchem}(b)).
This allows one to use very tight spatial filtering and relax other filtering requirements if required~\cite{gruneisen2015modeling, gruneisen2016adaptive,gruneisen2017modeling}. 
AO will likely be necessary for high-performance entanglement-based protocols where narrow-spectral filtering would significantly block the quantum signal. 
Altogether, our results represent an application-specific, yet modifiable framework for designing free-space quantum channels including specifying the optimal wavelength and the necessary filtering to achieve high performance.

\section{Theory}
In this section we develop the theory necessary for establishing QKD bit yield as a performance metric.
We progress in a pedagogic way, examining the wavelength dependence of each component contributing to the key-bit yield. 
We define a low-Earth-orbit (LEO) satellite down-link architecture similar to our earlier numerical and experimental simulations demonstrating the benefits of AO \cite{gruneisen2016adaptive, gruneisen2017modeling, gruneisen2020adaptive}.
Using MODTRAN, we generate wavelength dependent transmission and spectral radiance profiles for different site-dependent downlink scenarios. 
In the foothills of the Manzano mountains outside Albuquerque NM, we experience an arid high-desert climate with desert albedos and urban aerosols.  
For example, in Fig.~\ref{fig:WinSolZenHighVisTransHb} we plot the atmospheric transmission $\eta_{\mathrm{trans}}$ and spectral radiance $H_{\mathrm{b}}$ for a receiver pointed at zenith on the winter solstice at 1:00 PM with 50-km visibility.
We will use this site condition for the rest of the plots in this section. 
Considering the high-resolution structure of the atmospheric transmission and spectral radiance is a critical step in the optimization problem.
In a system design analysis, the detailed spectral structure, that is, the Fraunhofer lines, should be included around the target wavelength. 
For example, whereas the spectral radiance is relatively constant near 1550 nm, we judiciously chose dips in radiance near 780  and 430 nm to analyze in this section.
Specifically, we use the central wavelengths 1549.91, 780.945, and 430.886 nm, but hereafter we will round to the nearest nanometer when discussing these wavelengths.

We assume a $D_{\mathrm{T}}=10$-cm transmitter in a 600-km LEO and a $D_{\mathrm{R}}=1$-m ground receiver with efficiency $\eta_{\mathrm{rec}}=0.5$ and spectral filter efficiency $\eta_{\mathrm{spec}}=0.9$.
Due to advances in superconducting nanowire single-photon detector technology, we show no partiality and assume detector efficiency $\eta_{\mathrm{det}}=0.8$ and detector dark count rate $f_{\mathrm{dark}}$=10 Hz for all wavelengths. 
The signal pulse rate $R_{\mathrm{p}}$ is assumed to be 10 MHz, and the spectral/temporal filters are chosen to be $\Delta \lambda=1$ nm and $\Delta t =1$ ns, respectively.  
Additional constants must be set in order to discuss the QKD protocol, specifically, we assume that the system noise error rate is $e_0=0.5$, the polarization cross-talk error is $e_\mathrm{d}=0.01$, and the error correction efficiency is $f_\mathrm{ec}=1.22$. 
Under a majority of the channel conditions we consider, we find that the optimal signal and decoy-state mean photon numbers (MPNs) are $\mu=0.7$ and $\nu=0.1$, respectively.

\subsection{Spatial Filtering}

The performance of a free-space QKD system ultimately depends on the amount of signal and noise that passes through the receiver field stop.
Therefore, we will first introduce the physics related to the focused spot size and how this relates to field stop spatial-filtering strategies.
As one might expect, these choices permeate throughout the calculation and we will show how they effect the channel efficiency, noise probability, signal-to-noise, and error rate.

\subsubsection{Field of View}
For a receiver with no central obscuration operating at the diffraction limit, one can set the field stop to transmit the central peak of the diffraction-limited Airy pattern, thus transmitting 84$\%$ of the signal light.
Accordingly, the spatial filter diameter is 
\begin{equation}\label{eq:DLSPOT}
d_{\mathrm{spot}}^{(\mathrm{DL})} = 2.44 \dfrac{\lambda f}{D_{\mathrm{R}}},
\end{equation}
where $f$ is the receiver focal length and $D_\mathrm{R}$ is the receiver aperture diameter. 
This choice of spatial-filter diameter gives the diffraction limited (DL) solid-angle FOV
\begin{equation}\label{eq:DLFOV}
\Omega_{\mathrm{FOV}}^{(\mathrm{DL})}=\pi \big( \dfrac{1.22  \lambda }{D_\mathrm{R}} \big)^2 .
\end{equation}
The solid lines in Fig.~\ref{fig:WinSolZenHighVisLAFOV1550} give the solid-angle DL FOV $\Omega_{\mathrm{FOV}}^{(\mathrm{DL})}$ for 1550, 781, and 431 nm (black, red, and blue respectively).
One can see, for example, that 431 nm has $\sim$13$\times$ smaller FOV than 1550 nm.
Therefore, although the sky tends to be much brighter at shorter wavelengths (see Fig.~\ref{fig:WinSolZenHighVisTransHb}(b)), a receiver operating near the diffraction limit would be much more restrictive of that noise.
This is the first of many competing phenomenon that are inherent to the optimal-wavelength problem, and this is further complicated when considering the effects of turbulence.

In the presence of turbulence, reduced spatial coherence at the entrance pupil broadens the spot size in the focal plane (see Fig.~\ref{fig:GenSchem}(b)).
The so-called turbulence limited (TL) spot size can be approximated by \cite{tyson2015principles}
\begin{equation}\label{eq:TLSPOT}
d_{\mathrm{spot}}^{(\mathrm{TL})} = d_{\mathrm{spot}}^{(\mathrm{DL})} / \sqrt S,
\end{equation}
where 
\begin{equation}\label{eq:SR}
S = \Big[
1+ \Big(\dfrac{D_{\mathrm{R}}}{r(\lambda)} \Big) ^{5/3}
\Big]^{- 6/5}
\end{equation}
is the on-axis Strehl ratio resulting from atmospheric turbulence with no wavefront correction~\cite{sasiela2012electromagnetic}, 
\begin{equation}\label{eq:FRIED}
r(\lambda)=r_0 (\lambda / \lambda_0)^{6/5}
\end{equation}
is the Fried coherence length, $\lambda$ is the wavelength, and $r_0$ is the value measured at $\lambda_0$=500 nm~\cite{sasiela2012electromagnetic}.
Throughout this article, the symbol $r_0$ specifies the value of the Fried coherence length at 500 nm and can be considered a wavelength independent measure of the strength of turbulence.  
Choosing a spatial filter corresponding to the broadened spot size in Eq.~\ref{eq:TLSPOT} leads to the so-called TL solid-angle FOV 
\begin{equation}\label{eq:TLFOV}
\Omega_{\mathrm{FOV}}^{(\mathrm{TL})}=\pi \bigg( 1.22 \dfrac{ \lambda }{D_\mathrm{R}} \Big[
1+ \Big(\dfrac{D_{\mathrm{R}}}{r(\lambda)} \Big) ^{5/3}
\Big]^{ 3/5} \bigg)^2 .
\end{equation}
The dashed lines in Fig.~\ref{fig:WinSolZenHighVisLAFOV1550} give the solid-angle TL FOV $\Omega_{\mathrm{FOV}}^{(\mathrm{TL})}(r_0)$ for 1550, 781, and 431 nm (black, red, and blue respectively).
The wavelength dependence of the Fried coherence length $r(\lambda)$ introduces another critical phenomenon, that is, the perceived turbulence is more intense at shorter wavelengths.
This can be observed by examining the wavelength dependence of the FOV. 
Although the DL FOV is much smaller for the shorter wavelength, we see that the TL FOV increases more rapidly as $r_0$ gets small with respect to the receiver aperture size, that is, as the spatial coherence deteriorates.
For example, when $r_0=30$ cm, the TL FOV is $\sim$18$\times$ larger than the DL FOV at 431 nm, but only $\sim$2$\times$ larger at 1550 nm.

\subsubsection{Effective $r_0$ After AO Correction}
In a higher-order AO system, the aberrations of the incoming wavefront are corrected via a fast steering mirror (FSM) and a deformable mirror (DM) imprinted with the conjugate of the wavefront error.
The performance of an AO system ultimately depends on the ability to spatially resolve the wavefront characterized by $r_0$ and keep pace with the temporal fluctuations characterized by the Greenwood frequencies. 

\begin{figure}[t]
	\includegraphics[width=1\columnwidth]{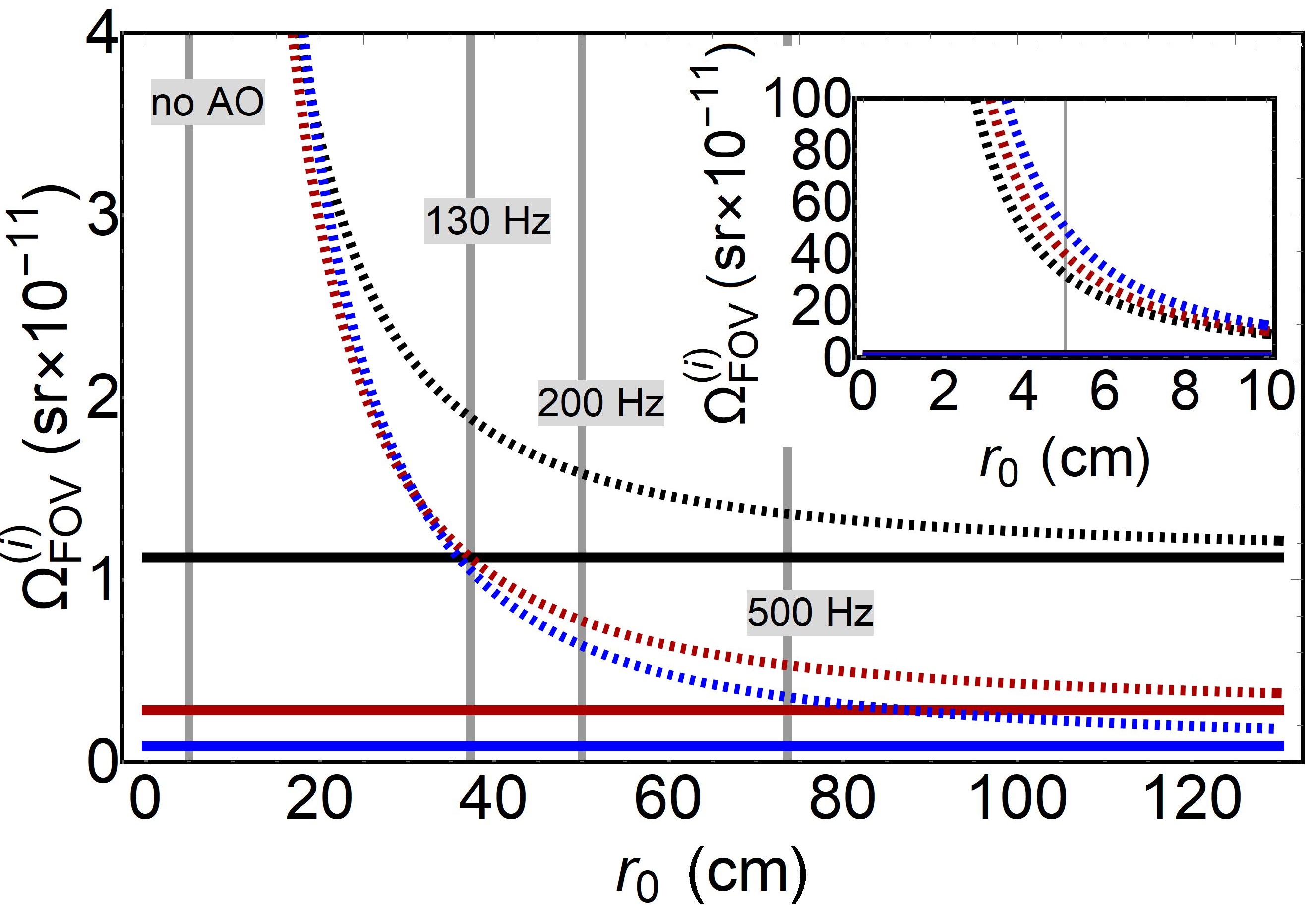}
	\caption{\label{fig:WinSolZenHighVisLAFOV1550}
Solid-angle field of view $\Omega_{\mathrm{FOV}}^{(i)}$ for 1550, 781, and 431 nm (black, red, and blue respectively), plotted as a function of $r_0$. 
The solid and dashed curves indicate the DL and TL FOVs, respectively. The vertical lines indicate higher order AO system performance. The line at 5 cm corresponds to no AO correction whereas the lines at $r_{0}^{(\mathrm{CL})}=37$, 50, and 74 cm  correspond to full AO with 130-, 200-, and 500-Hz closed-loop bandwidths, respectively.       
	}
\end{figure}

In Fig.~\ref{fig:WinSolZenHighVisLAFOV1550} we plot a wide range of $r_0$'s to show the trend in $\Omega_{\mathrm{FOV}}^{(\mathrm{TL})}$ as the spatial coherence approaches the 1-m receiver diameter.
However, realistic conditions will likely range from 5 cm $<r_0<15$ cm.
For example, assuming a Hufnagel-Valley ($\mathrm{HV}_{5/7}$) \cite{andrews2004field} turbulence profile and slew dependent wind dynamics we find a spatial coherence of $r_0 \equiv r(500\,\mathrm{nm}) \approx 5$ cm and a higher-order temporal coherence characterized by the Greenwood frequency $f_{\mathrm{G}}(500\,\mathrm{nm}) \approx 301$ Hz (see App.~\ref{sec:appendixA} and Ref.~\cite{gruneisen2020adaptive} for more details). 
For reference, in this plot and throughout the article, we include a vertical line at $r_0 = 5$ that indicates the uncompensated atmospheric condition.
We also include an \textit{effective $r_0$ that corresponds to the residual wavefront error after AO compensation}.
The latter depends on the relationship between the Greenwood frequencies and the AO bandwidths, and we will discuss this in the following.

To achieve a high degree of wavefront compensation one should design an AO system that is on the order of or several times faster than the Greenwood frequencies.
However, for this simulation we first asses the closed-loop bandwidth of the system we built for our field experiment reported in Ref.~\cite{gruneisen2020adaptive}.
Despite this system only being designed to compensate for turbulence observed in a 1.6-km horizontal channel with stationary transmit/receive stations, we show that it could provide a relevant QKD system performance increase even if used in a space-Earth down-link architecture where slewing substantially increases the temporal atmospheric fluctuations.
Hence, assuming the closed-loop bandwidth $f_c=130$ Hz, we calculate the effective closed-loop spatial coherence $r_0^{(\mathrm{CL})}$$\approx \;$37 cm (see App.~\ref{sec:appendixAEffective}).
We also consider two other design reference points from previous numerical simulations \cite{gruneisen2016adaptive, gruneisen2017modeling}.
Namely, a 200-Hz closed-loop-bandwidth system yielding  $r_0^{(\mathrm{CL})}$$\approx \;$50 cm and a 500-Hz system yielding  $r_0^{(\mathrm{CL})}$$\approx \;$74 cm as seen in Fig.~\ref{fig:WinSolZenHighVisLAFOV1550}. 
Throughout the rest of this article we will only include vertical lines at $r_0^{(\mathrm{OL})} = 5$ cm and $r_0^{(\mathrm{CL})} = 50$ cm, but one use can the equations in Appendix~\ref{sec:appendixAEffective} to assess the performance of different closed-loop-bandwidth systems.

\subsubsection{Spatial Filtering Strategies}
Whereas the DL FOV is constant, the TL FOV is a function of $r_0$ and grows with increasing turbulence strength, in effect, maintaining signal throughput at the expense of permitting more noise photons through the spatial filter.
The wavefront correction introduced by AO creates a tighter focused spot, and thus allows a narrower FOV.
For example, the intersections with the vertical line at $r_0=50$ cm indicates the TL FOV one could operate at with a 200-Hz AO system.
In fact, in this case AO would allow one to make their FOV $\sim$20$\times$ smaller at 1550 nm, $\sim$52$\times$ smaller at 781 nm, and $\sim$78$\times$ smaller at 431 nm.
This is significant because it would provide a considerable reduction in noise while maintaining signal throughput.
The focused-spot size, the FOV, and the resulting spatial-filtering are perhaps the most crucial phenomenon ultimately affecting the QKD system performance.
Therefore, in each of the subsequent subsections, one must pay careful attention to the wavelength and spatial filtering dependence of each contribution to the key-bit yield.
For more details about the competing phenomenon that effect the focused spot size and the FOV, see App.~\ref{sec:appendixASpot}.

In this article we will consider two spatial filtering strategies.
The first strategy is to choose the FOV according to ones focused spot size, for example, free-space coupling to detectors using a field stop diameter corresponding to the average spot size for ones site condition. 
The second strategy is to operate the receiver at the DL FOV regardless of $r_0$; this corresponds to, for example, coupling into single-mode fiber prior to detection.
These two strategies are defined more rigorously in the following section where we introduce the channel efficiency and background probability.

\begin{figure*}[]
	\includegraphics[width=1\textwidth]{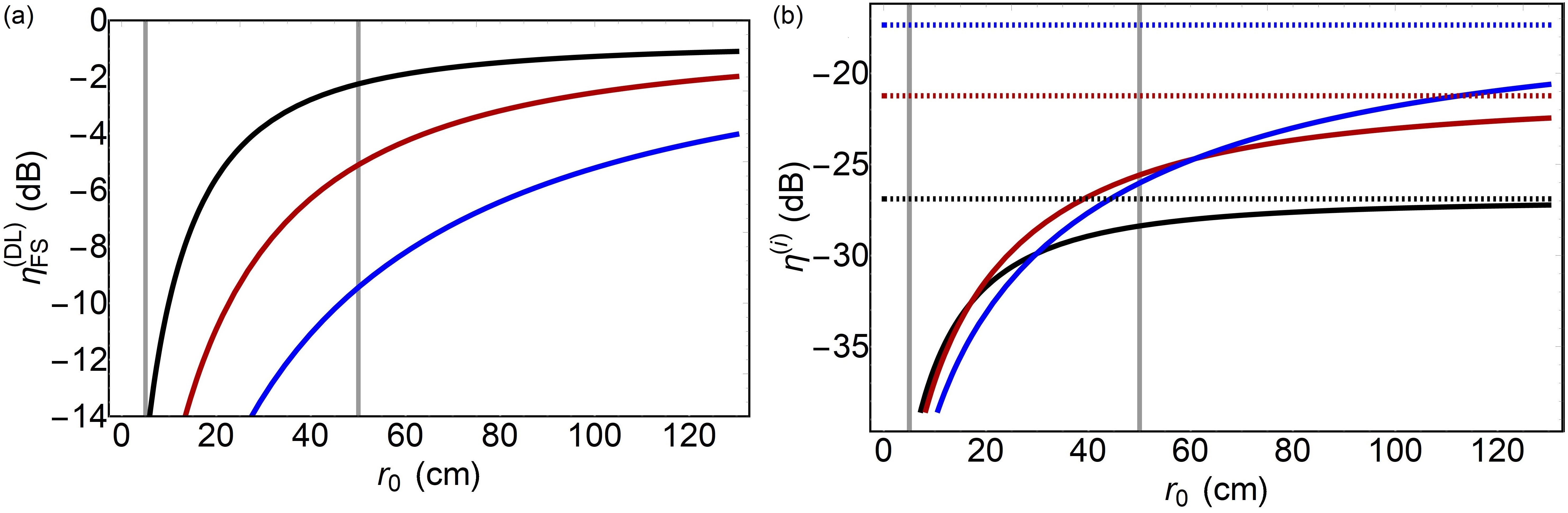}\\
	\caption{\label{fig:WinSolZenHighVisEta}
The (a) DL strategy field stop transmission efficiency $\eta_{\mathrm{FS}}^{(\mathrm{DL})}$ and (b) total system efficiency $\eta^{(\mathrm{DL})}$ for 1550, 781, and 431 nm (black, red, and blue respectively), plotted as a function of $r_0$. The solid and dashed curves indicate the DL and TL strategies, respectively. The vertical line at 5 cm corresponds to no AO correction and the line at 50 cm corresponds to the effective $r_0$ of a $f_c =200$-Hz AO system.   
	}
\end{figure*}
\subsection{Channel Efficiency}
The geometric aperture-to-aperture coupling is approximated by the Gaussian beam equation~\cite{tomaello2011intersatellite} 
\begin{equation}
\eta_\mathrm{geo}= 1-\exp \big( -\dfrac{1}{2} \dfrac{D_\mathrm{R}^2}{w^2(\lambda, z)} \big),
\end{equation}
where $w^2(\lambda)=w_0^2  ( 1 + z^2 / z_{\mathrm{R}}^2(\lambda))$ is the waist function, $z_{\mathrm{R}}=\pi w_0^2 / \lambda$ is the Rayleigh range, $w_0 \equiv 0.7 D_{\mathrm{T}}/2$ is the waist of the signal beam, and $D_{\mathrm{T}}$ is the transmitter aperture diameter.
Aperture-to-aperture coupling loss due to turbulence-induced beam spreading is negligible in the down-link architecture \cite{bonato2009feasibility}. 

Loss at the spatial filter is defined in terms of the two spatial-filtering strategies.
The first strategy is to account for the broadened spot and increase the size of the field stop to always pass $84 \%$ of the signal.
The second strategy is to keep the size of the field stop at the diffraction limit regardless of the broadened spot size.
In the later case, the turbulence broadened spot may be partially blocked by the field stop spatial filter.
To model this we take the ratio of the spot-size areas and write the TL and DL field-stop transmissions as
\begin{equation}
\begin{split}
\eta_{ \mathrm{FS} }^{(\mathrm{TL})}(r_0, \lambda) &= 0.84 \\
\eta_{ \mathrm{FS} }^{(\mathrm{DL})} (r_0, \lambda) &=0.84 \times S,
\end{split}
\end{equation}
where the Strehl $S$ is defined in Eq.~\ref{eq:SR}.
These allow us to define the total channel efficiency
\begin{equation}
\begin{split}
\eta^{(i)} (r_0, \lambda) &= \eta_{\mathrm{geo}}(\lambda) \, \eta_{\mathrm{trans}}(\lambda) \, \eta_{ \mathrm{FS} }^{(\mathrm{i})}(r_0, \lambda) \\ 
& \times \eta_{\mathrm{spec}} \, \eta_{\mathrm{rec}} \, \eta_{\mathrm{det}},
\end{split}
\end{equation}
where $i$ indicates either the DL or TL strategy, and $\eta_{\mathrm{trans}}$ is the atmospheric transmission efficiency predicted by MODTRAN and plotted in Fig.~\ref{fig:WinSolZenHighVisTransHb}(a). 

In Fig.~\ref{fig:WinSolZenHighVisEta} we plot the $r_0$ dependence of $\eta_{\mathrm{FS}}^{(\mathrm{DL})}$ and $\eta^{(i)}$ for 1550, 781, and 431 nm (black, red, and blue respectively).
Figure~\ref{fig:WinSolZenHighVisEta}(a) shows the $r_0$ dependence of the field stop transmission and reveals the wavelength dependent nature of focusing in the presence of turbulence, that is, the effects of turbulence are weaker at longer wavelengths and 1550 nm appears to have the advantage.
Again, the intersections with the vertical lines at $r_0^{(\mathrm{OL})}$$=\;$5 cm and $r_0^{(\mathrm{CL})}$$=\;$50 cm indicate the achievable field-stop transmission efficiency for a $f_c = 200$-Hz AO system under open- and closed-loop operation, respectively.  

Figure~\ref{fig:WinSolZenHighVisEta}(b) shows the $r_0$ dependence of the total system transmission which reveals the limitation of longer wavelengths, that is, the geometric coupling prevails and lends an advantage to the shorter wavelengths for the TL strategy $\eta^{\mathrm{(TL)}}$.
We also see that despite the relatively low efficiency at the field stop, with AO correction the total channel efficiencies for the DL strategy $\eta^{\mathrm{(DL)}}$ are actually higher for 431 and 781 nm. 
Figure~\ref{fig:WinSolZenHighVisEta}(b) also serves as an illustration of the two field stop strategies.
For example, $\eta^{(\mathrm{TL})}$ is constant over the entire range of $r_0$ while $\eta^{(DL)}$ decreases once the spatial coherence degrades and begins to broaden the spot size. 
One might realize the benefit of the TL strategy if their site conditions are relatively constant in $r_0$, or more ambitiously, by developing higher-order AO systems with adaptive spatial-filters.
For example, utilizing a spatial filter that adjusts according to the effective $r_0$ of the AO system, thereby dynamically maximizing signal throughput while maintaining noise filtering at the field stop.   

\subsection{Background Probability}
\begin{figure}[t!]
	\includegraphics[width=1\columnwidth]{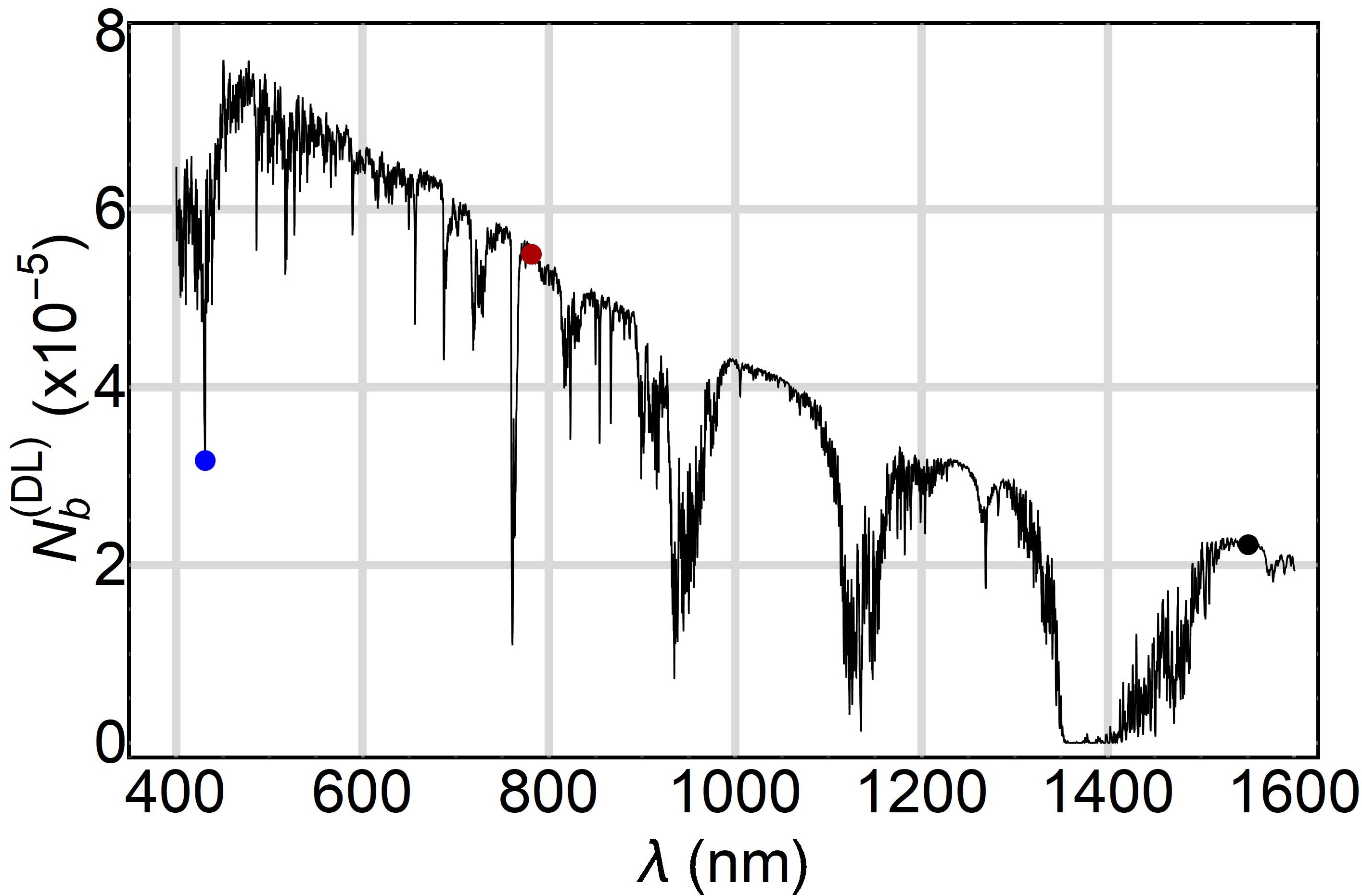}\\
	\vspace{0.25cm} 
 	\includegraphics[width=1\columnwidth]{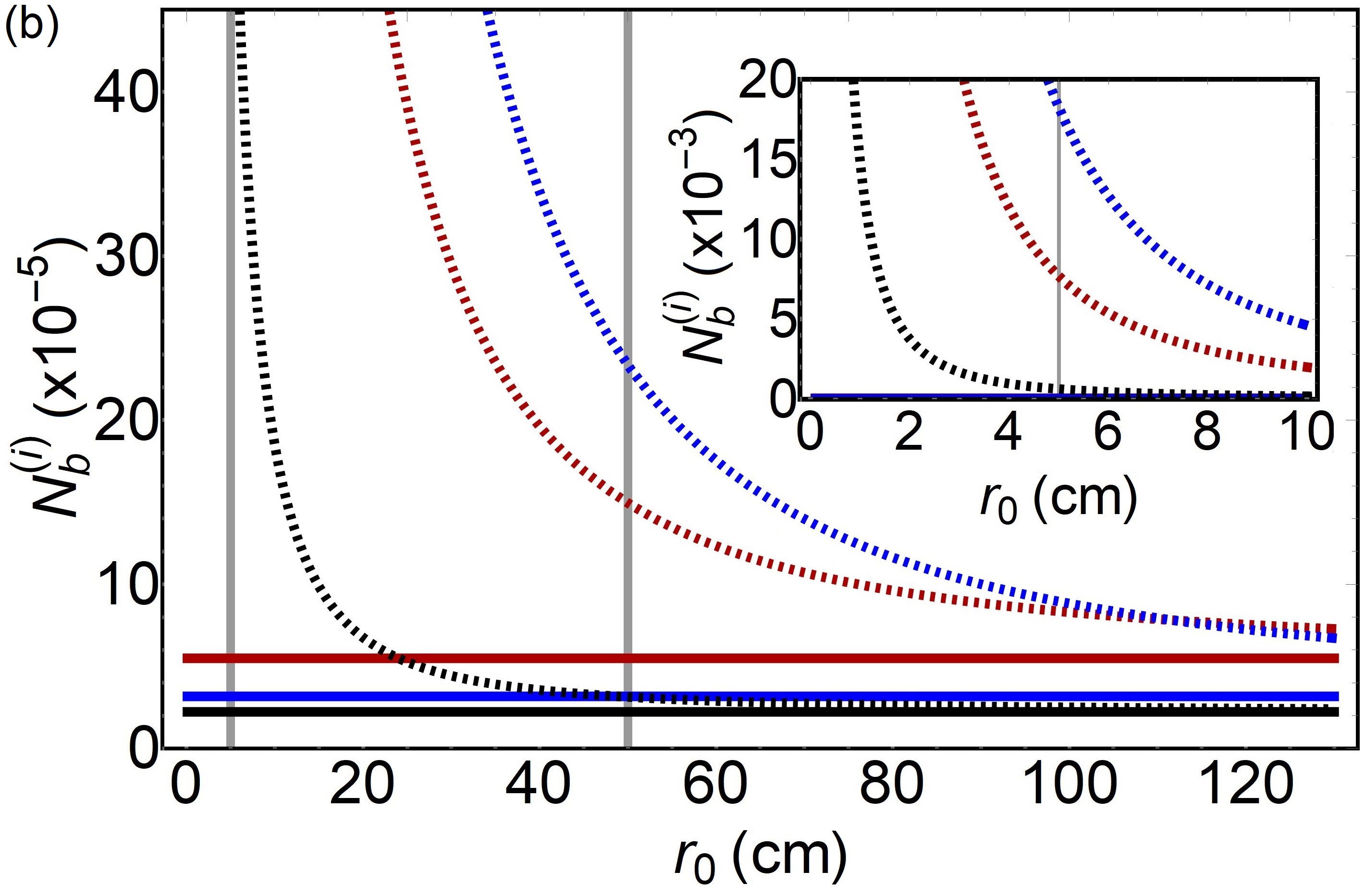}
	\caption{\label{fig:WinSolZenHighVisNb}
The (a) wavelength dependence of the number of background photons for the DL strategy $N_{\mathrm{b}}^{(\mathrm{DL})}$ and (b) the $r_0$ dependence of $N_{\mathrm{b}}^{(i)}$ for 1550, 781, and 431 nm (black, red, and blue respectively), plotted as a function of $r_0$. In (b), the solid and dashed curves indicate the DL and TL strategies, respectively. The vertical line at 5 cm corresponds to no AO correction and the line at 50 cm corresponds to the effective $r_0$ of a $f_c =200$-Hz AO system.   
	}
\end{figure}

For either strategy, the number of noise photons transmitted by the field stop within a time, wavelength, and FOV window is given by the radiometric equation
\begin{equation}\label{eq:Nb}
N_{\mathrm{b}}^{(i)} = \int d\lambda \,  \dfrac{\lambda}{4h c} H_{\mathrm{b}} (\lambda) \, \Omega_{\mathrm{FOV}}^{(i)}(r_0, \lambda_0) \, \pi \, D_{\mathrm{R}}^2 \, \Delta t,
\end{equation} 
where the integral is performed over a notch filter with width $\Delta \lambda=1$ nm and central wavelength $\lambda_0$, the spectral radiance $H_{\mathrm{b}} (\lambda)$ is predicted by MODTRAN and given in Fig.~\ref{fig:WinSolZenHighVisTransHb}(b), $\Delta t=1$ ns is the temporal detection window, $h$ is Planck's constant, $c$ is the speed of light, and $i$ indicates either the DL or TL strategy. 
In Fig.~\ref{fig:WinSolZenHighVisNb}(a) we plot the wavelength dependence of $N_{\mathrm{b}}^{(\mathrm{DL})}$ and in Fig.~\ref{fig:WinSolZenHighVisNb}(b) we plot the $r_0$ dependence of $N_{\mathrm{b}}^{(i)}$ for 1550, 781, 431 nm.
Comparing Fig.~\ref{fig:WinSolZenHighVisNb}(a) and Fig.~\ref{fig:WinSolZenHighVisTransHb}(b) reveals a reduction of the \textit{number} of photons at shorter wavelengths.
For example, under the atmospheric conditions assumed in this section the sky is approximately 20 times brighter at 781 nm as compared to 1550 nm, but there are only 2.5 times the number photons for the DL strategy.
This reduction in number of photons is a result of the higher photon energy $h\,c/\lambda$ and the more effective spatial filtering for the shorter wavelengths, that is, the smaller FOV as a consequence of the $\lambda^2$ dependence of Eq.~\ref{eq:DLFOV}. 
One should note that the relative brightness is highly dependent on site conditions.
For example, in Sec.~\ref{sec:VisStudy} we discuss how lower visibility conditions close the gap in relative brightness.

Figure \ref{fig:WinSolZenHighVisNb}(b) reveals the negative effects of widening the FOV with the TL strategy by giving the $r_0$ dependence of $N_{\mathrm{b}}^{(i)}$ for 1550, 781, and 431 nm.
For example, when $r_0 \approx$ 50 cm and $\lambda=431$ nm, $N_{\mathrm{b}}^{(\mathrm{TL})}/N_{\mathrm{b}}^{(\mathrm{DL})} \approx 7.4$.
This might suggest that the TL strategy is fundamentally flawed due to the excess noise, but it also accommodates a boost in signal due to the high channel efficiency as revealed in Fig.~\ref{fig:WinSolZenHighVisEta}(b).
Therefore, the question which remains to be answered is whether the excess noise actually translates to increased errors and lower QKD system performance. 
In the following, we reveal the performance of the TL strategy by investigating the error rate, signal-to-noise probability, and ultimately the key-bit yield. 
To do so we must define the background probability~\cite{ma2005practical}
\begin{equation}
Y_0^{(i)} (r_0, \lambda) = N_{\mathrm{b}}^{(i)} (r_0, \lambda) \, \eta_{\mathrm{spec}} \, \eta_{\mathrm{rec}} \, \eta_{\mathrm{det}} + 4 f_{\mathrm{dark}} \, \Delta t,
\end{equation}
which has a contribution from the detector dark counts, but is dominated by the number of photons $N_{\mathrm{b}}^{(i)}$ and the choice of strategy.

\subsection{Quantum Bit Error Rate}
\begin{figure}[t]
	\includegraphics[width=1\columnwidth]{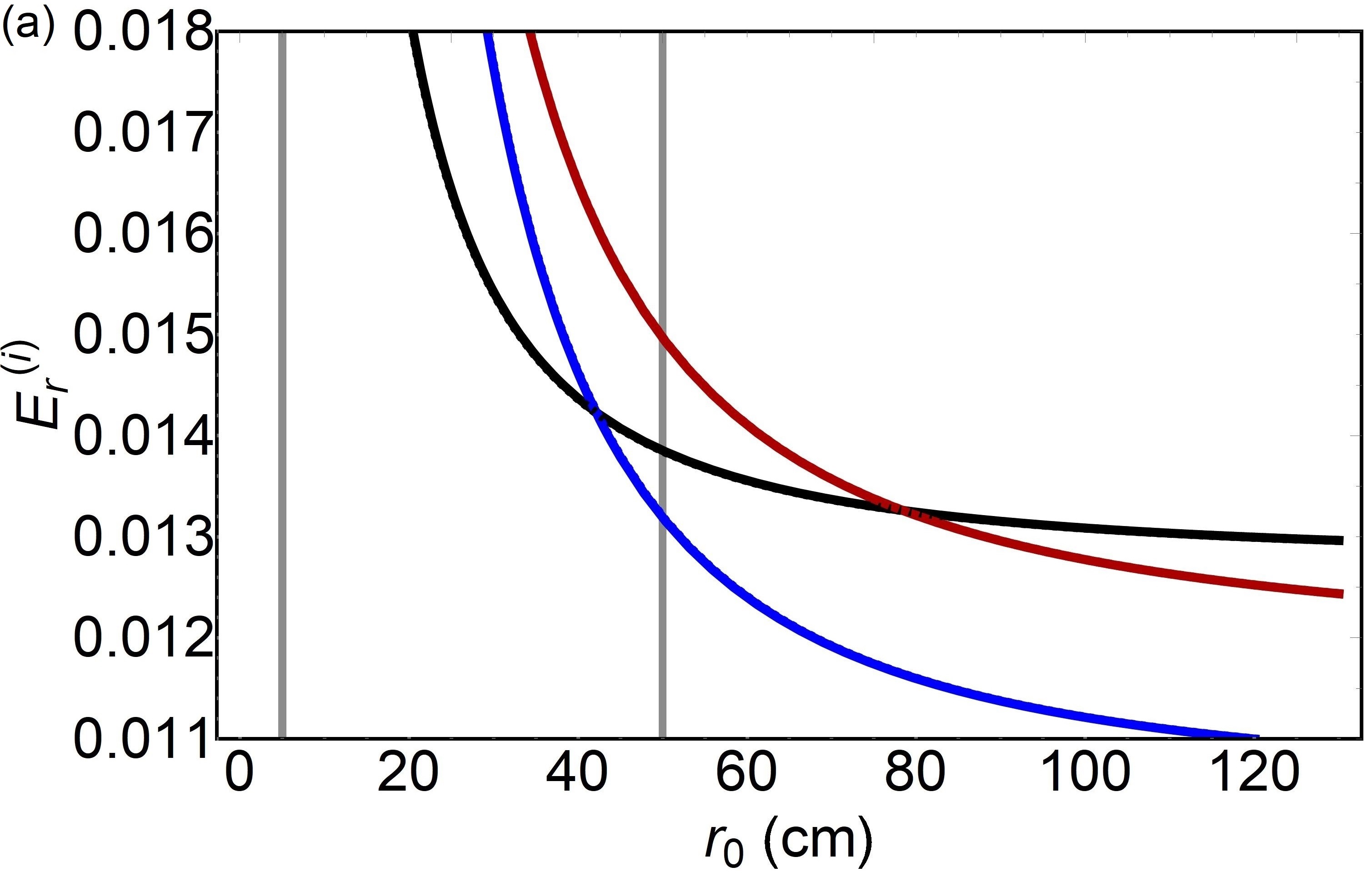}\\
	\vspace{0.25cm} 
	\includegraphics[width=1\columnwidth]{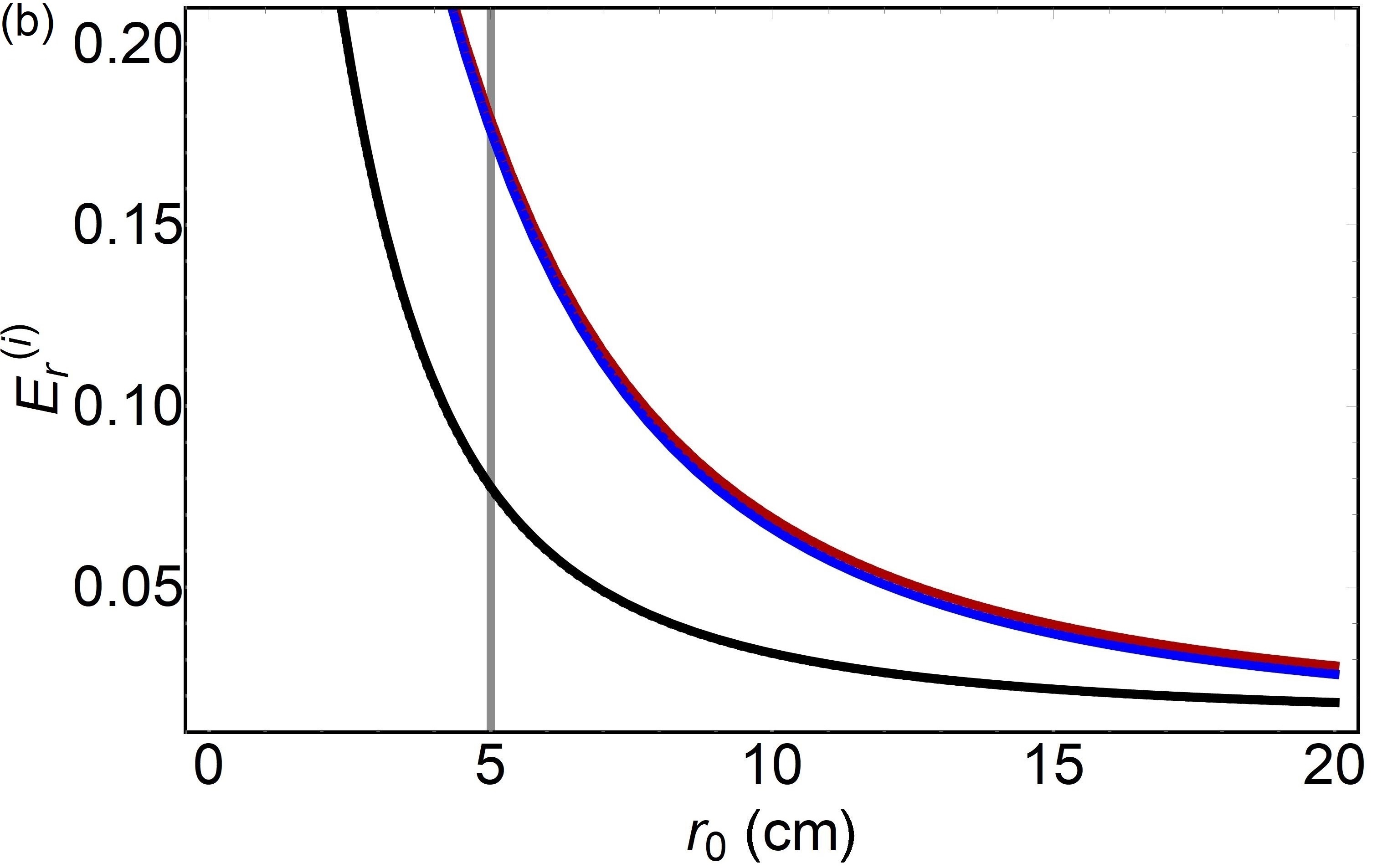}
	\caption{\label{fig:WinSolZenHighVisEr}
The $r_0$ dependence of $E_{\mathrm{r}}^{(i)}$ for 1550, 781, and 431 nm (black, red, and blue respectively). The solid and dashed curves indicate the DL and TL strategies, respectively. In (a) the vertical line at 5 cm corresponds to no AO correction and the line at 50 cm corresponds to the effective $r_0$ of a $f_c =200$-Hz AO system. In (b) we plot the range of $r_0$ representative of no AO correction. The dashed lines are nearly totally obscured by the solid lines. 
	}
\end{figure}

Next, we define the decoy/signal quantum bit error rate (QBER)~\cite{ma2005practical}
\begin{equation} \label{eq:QBER}
E_{\mathrm{r}, n}^{(i)} (r_0, \lambda ) = \dfrac{e_0 Y_0^{(i)} (r_0, \lambda) +  e_\mathrm{d} ( 1 -  e^{ - \eta^{(i)} (r_0, \lambda) \, n} )}
{ Y_0^{(i)} (r_0, \lambda) +  1 -  e^{ - \eta^{(i)} (r_0, \lambda) \, n} },
\end{equation}
where $n$ is the MPN of the signal or decoy state.
In Fig.~\ref{fig:WinSolZenHighVisEr}(a) and \ref{fig:WinSolZenHighVisEr}(b) we plot the $r_0$ dependence of $E_{\mathrm{r},\mu}^{(i)}$ for 1550, 781, and 431 nm and two different ranges of $r_0$.
Interestingly, despite $N_{\mathrm{b}}^{(i)}(\lambda)$ (and correspondingly $Y_{0}^{(i)}(\lambda)$) being larger for the TL strategy, the dashed lines are nearly totally obscured by the solid lines.
This is because the probability of detecting a signal photon $1 -  e^{ - \eta^{(i)} (r_0, \lambda) \, \mu} $ also increases with the TL strategy.
Apparently, the increase in noise is directly compensated by the increase in signal, and in effect, the ratio in Eq.~\ref{eq:QBER} remains nearly constant. 
This can be investigated by keeping the first term in the expansion of $1 -  e^{ - \eta^{(i)} (r_0, \lambda) \, \mu} $.
Making this substitution in Eq.~\ref{eq:QBER} and rearranging terms one can find that the QBER for DL strategy is 
\begin{figure}[t!]
	\includegraphics[width=1\columnwidth]{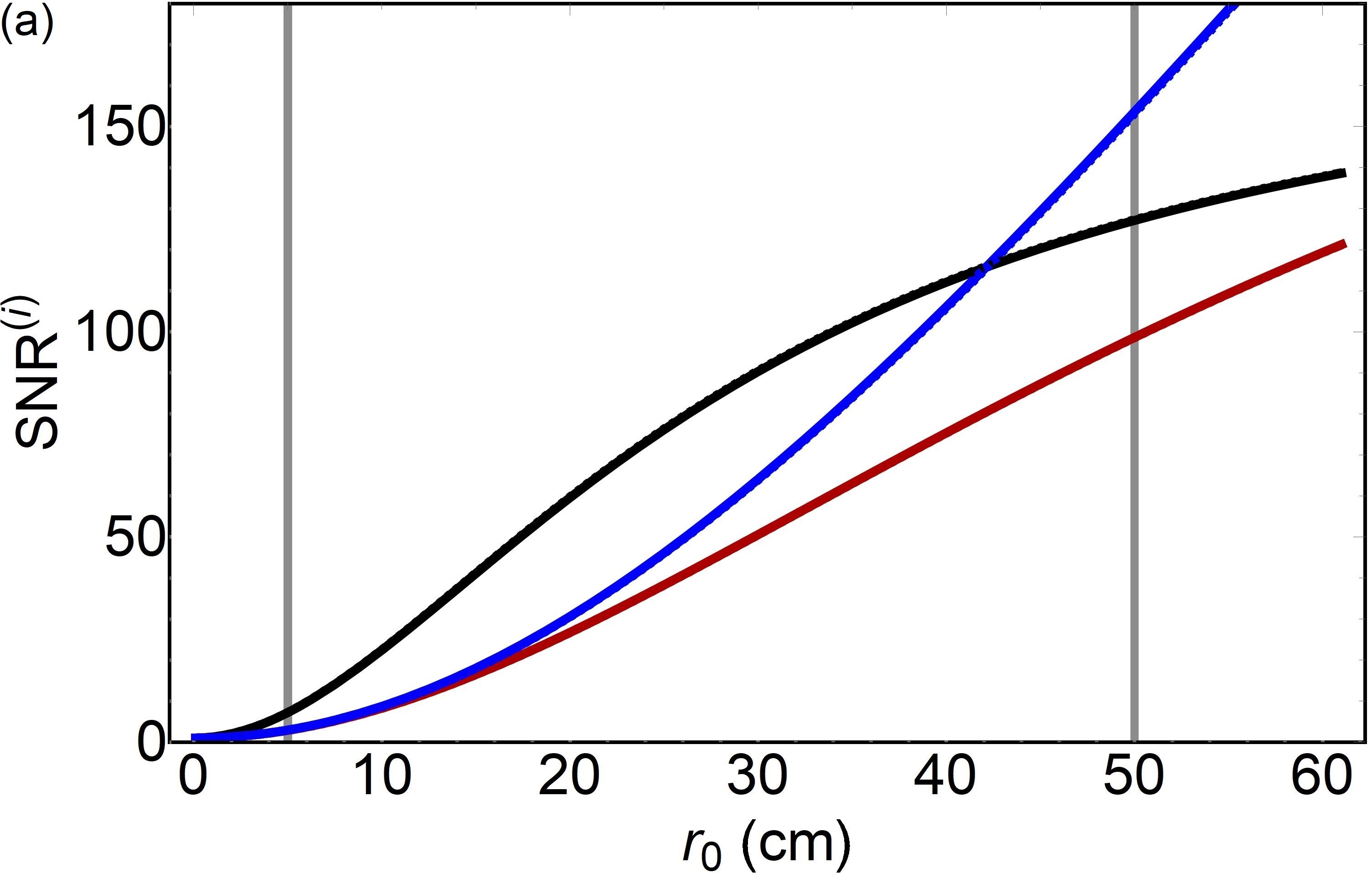}\\
	\vspace{0.25cm} 
	\includegraphics[width=1\columnwidth]{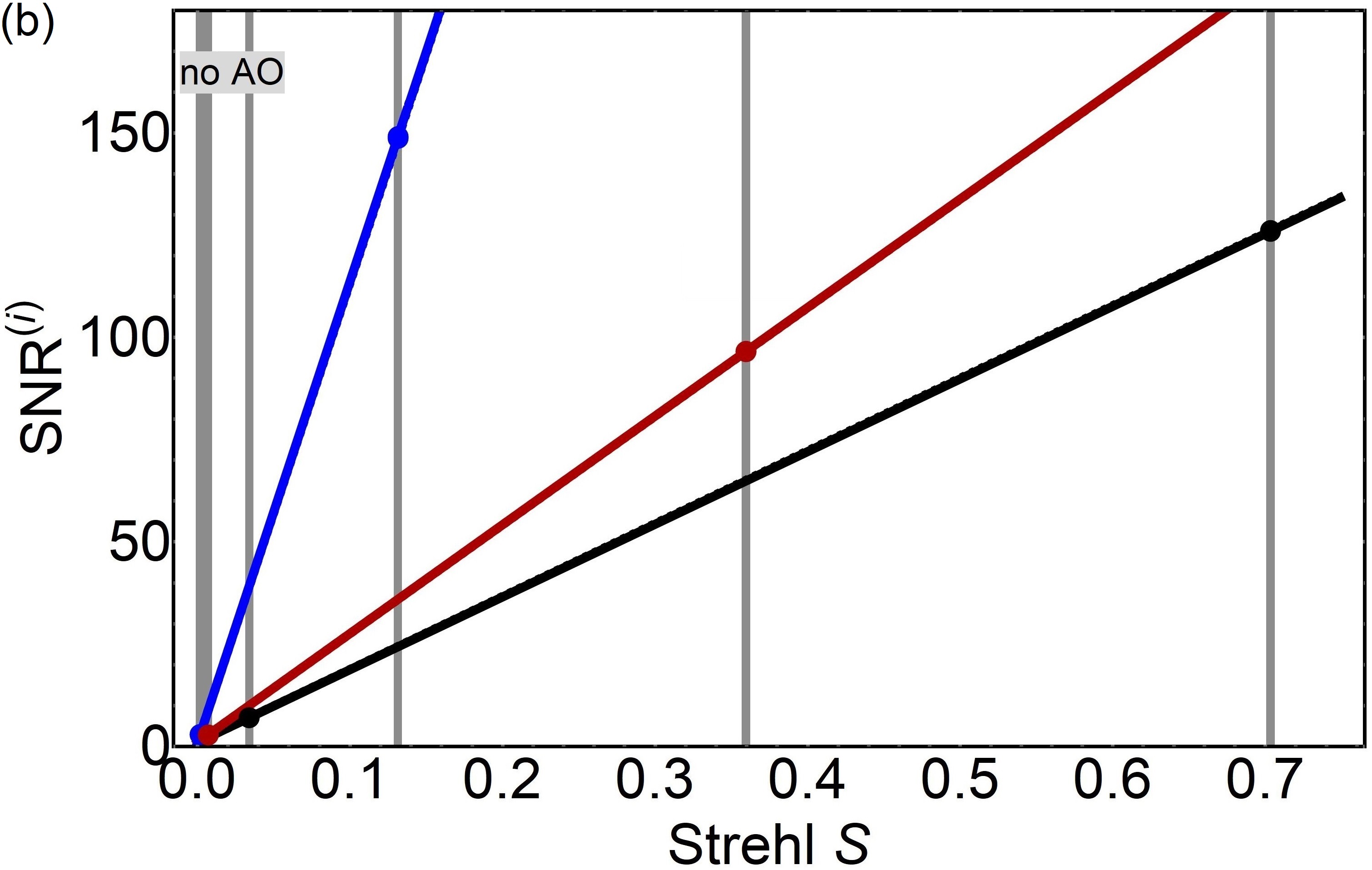}
		\caption{\label{fig:WinSolZenHighVisSNRBoth}
The (a) $r_0$ and (b) $S$ dependence of $SNR_{\mu}^{(i)}(\lambda)$ for 1550, 781, and 431 nm (black, red, and blue respectively). The solid and dashed curves indicate the DL and TL strategies, respectively. In (a) the vertical line at 5 cm corresponds to no AO correction and the line at 50 cm corresponds to the effective $r_0$ of a $f_c =200$-Hz AO system. In (b) the $\lambda$ dependence of $S$ produces six vertical lines. The grouping near $S=0$ correspond to no AO and the vertical lines at $S=0.71$, 0.37, and 0.14 correspond to a $f_c =200$-Hz AO system operating at 1550, 781, and 431 nm, respectively. The dashed lines are nearly totally obscured by the solid lines.  
	}
\end{figure}
\begin{equation} \label{eq:QBER2}
E_{\mathrm{r}, n}^{(\mathrm{DL})}  \approx \dfrac{e_0 ( Y_0^{(\mathrm{TL})} +  \epsilon )  +  e_\mathrm{d} \, \eta^{(\mathrm{TL})}  \, n}
{ Y_0^{(\mathrm{TL})} + \epsilon + \eta^{(TL)}  \, n },
\end{equation}
where $\epsilon \equiv 4 f_{\mathrm{dark}} \, \Delta t (1/S - 1)$.
Therefore, one can see that the functional dependence of the QBER for the TL and DL strategies differ by an additive noise term $\epsilon$ in the numerator and denominator. 
When $r(\lambda)$ increases relative to the receiver diameter $D_{\mathrm{R}}$, the Strehl approaches unity, the noise term $\epsilon \rightarrow 0$, and $E_{\mathrm{r}, n}^{(\mathrm{DL})} = E_{\mathrm{r}, n}^{(\mathrm{TL})}$.
Fortunately, this noise term is always very small due to the dependence on the narrow temporal filtering $\Delta t$ and low dark-count rate $f_{\mathrm{dark}}$, and this explains the nearly identical $r_0$ dependence of the QBER for the two strategies.

\subsection{Signal-To-Noise}
Next, we will investigate the signal-to-noise ratio (SNR) which we define as the ratio of signal gain to background probability
\begin{equation}
SNR^{(i)}_n (r_0 , \lambda) = Q_{n}^{(i)} (r_0, \lambda)/Y_0^{(i)} (r_0, \lambda),
\end{equation}
where 
\begin{equation}
Q_{n}^{(i)} (r_0, \lambda) = Y_0^{(i)} (r_0, \lambda) + 1 - e^{ - \eta^{(i)} (r_0, \lambda) n },
\end{equation}
is the signal or decoy state gain and $n$ is the MPN of the signal or decoy state~\cite{ma2005practical}.
In Fig.~\ref{fig:WinSolZenHighVisSNRBoth}(a) we plot the $r_0$ dependence of $SNR_{\mu}^{(i)}$ for 1550, 781, and 431 nm.
Similar to the QBER, we do not see a significant difference between the two strategies, thus confirming the relationship between the signal and noise for the TL strategy.
To reiterate, the TL strategy accommodates a proportional boost in signal and noise in such a way as to not increase the error rate.

It is also useful to investigate the SNR as a function of Strehl via the relation in Eq.~\ref{eq:SR}.
In Fig.~\ref{fig:WinSolZenHighVisSNRBoth}(b) we see that a 431 nm system can in principle outperform longer wavelength sytems for all values of Strehl.
Since Strehl is a function of $r_0$ \textit{and} $\lambda$, we now have three separate vertical lines indicating QKD system performance with AO correction (see App.~\ref{sec:appendixAStrehl}).
In effect, a given AO correction yields a lower Strehl for the shorter wavelengths.
For this reason it might be tempting to assume that the longer wavelengths will yield better QKD system performance.
Strictly in terms of the SNR of a system with AO correction, we see that the longer wavelengths have much better Strehl, but the SNR can be higher at 431 nm due to the low relative noise as is evident by the low QBER (see Fig.~\ref{fig:WinSolZenHighVisEr}(a) at $r_0^{(\mathrm{CL})}=50$ cm).  
The SNR helps build intuition regarding the performance of different wavelength systems, but ultimately one must adopt a quantum metric to make any legitimate claim. 
Therefore, we will now investigate the key-bit probability (KBP) and reveal how the wavelength dependent trends in channel efficiency, QBER, and SNR translate to actual QKD system performance. 

\begin{figure*}[t!]
	\includegraphics[width=1\textwidth]{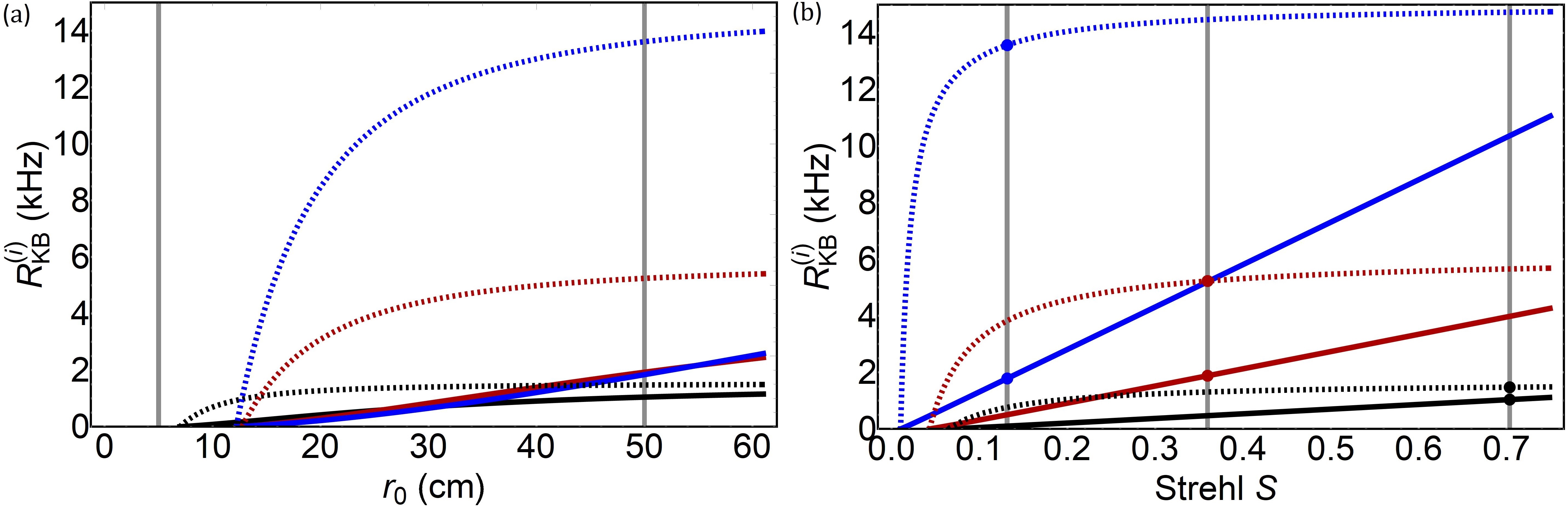}
\caption{\label{fig:WinSolZenHighVisSKBP}
The (a) $r_0$ and (b) $S$ dependence of the key-bit rate $R_{\mathrm{KB}}^{(i)}$ for 1550, 781, and 431 nm (black, red, and blue respectively). The solid and dashed curves indicate the DL and TL strategies, respectively. In (a) the vertical line at 5 cm corresponds to no AO correction and the line at 50 cm corresponds to the effective $r_0$ of a $f_c =200$-Hz AO system. In (b) the $\lambda$ dependence of $S$ produces three vertical lines indicating AO correction.   
	}
\end{figure*}
\subsection{Key Bit Probability}
To establish the KBP we must define a few more important quantities~\cite{ma2005practical}.
Although these are well known, we include them here for completeness and convenience.
The single photon gain is
\begin{equation}
\begin{split}
Q_{1}^{(i)} &(r_0, \lambda) =  \dfrac{\mu^2 e^{-\mu}}{\mu \nu - \nu^2} \bigg( Q_{\nu}^{(i)} (r_0, \lambda) \, e^{\nu} \\ 
 &- Q_{\mu}^{(i)} (r_0, \lambda) \, e^{\mu} \, \dfrac{\nu^{2}}{\mu^{2}} - \dfrac{\mu^2 - \nu^2 }{\mu^2} \, Y_0^{(i)} (r_0, \lambda) \bigg),
\end{split}
\end{equation}
and similarly the single photon state yield is
\begin{equation}
\begin{split}
Y_{1}^{(i)} &(r_0, \lambda) =  \dfrac{\mu}{\mu \nu - \nu^2} \bigg( Q_{\nu}^{(i)} (r_0, \lambda) \, e^{\nu} \\ 
 &- Q_{\mu}^{(i)} (r_0, \lambda) \, e^{\mu} \, \dfrac{\nu^{2}}{\mu^{2}} - \dfrac{\mu^2 - \nu^2 }{\mu^2} \, Y_0^{(i)} (r_0, \lambda) \bigg).
\end{split}
\end{equation}
Lastly, the single photon state error rate is 
\begin{equation}
\begin{split}
e_{1}^{(i)} (r_0, \lambda) &= \dfrac{E_{\mathrm{r}, \nu}^{(i)} (r_0, \lambda) Q_{\nu}^{(i)} (r_0, \lambda) e^{\nu} }{Y_1^{(i)} (r_0, \lambda) \, \nu} \\
&-\dfrac{ e_0 \, Y_0^{(i)} (r_0, \lambda)}{Y_1^{(i)} (r_0, \lambda) \, \nu}.
\end{split}
\end{equation}
Finally, we define the KBP~\cite{ma2005practical}
\begin{equation}
\begin{split}\label{eq:SKBP}
P_{\mathrm{KB}}^{(i)} (r_0, \lambda) &= \dfrac{1}{2} \bigg( -Q_{\mu}^{(i)} (r_0, \lambda) f_{ec} \, H_2 \big[ E_{\mathrm{r}, \mu}^{(i)}(r_0, \lambda) \big] \\
&+ Q_{1}^{(i)} (r_0, \lambda) \big( 1- H_2 \big[ e_{1}^{(i)}(r_0, \lambda) \big] \big)
\bigg),
\end{split}
\end{equation}
where $H_2 (x) = -x \log_2 (x) - (1-x) \log_2 (1-x) $ is the Shanon binary entropy formula.
The KBP is used to define the key-bit rate
\begin{equation}
R_{\mathrm{KB}}^{(i)}(r_0, \lambda) = R_{\mathrm{p}}(1-N_{\nu \mu})P_{\mathrm{SKB}}^{(i)} (r_0, \lambda),
\end{equation}
where $N_{\nu \mu}=0.3$ is the ratio of decoy plus vacuum pulses. 
 
In Fig.~\ref{fig:WinSolZenHighVisSKBP}(a) we plot the key-bit rate as a function of $r_0$.
Interestingly, despite the two FOV strategies yielding nearly identical results for QBER and SNR, we see a drastic difference when examining the key-bit rate. 
As compared to the DL strategy, the TL strategy experiences a dramatic increase in key-bit rate. 
This is because, quite fortuitously, the TL strategy permits a large boost in signal gain $Q_{\mu}$ without any significant boost in QBER.
Another striking feature is the relatively high performance of the shorter wavelengths. 
Despite turbulence effects being weaker and the sky being dimmer at 1550 nm, one can see that performance at 781 and 431 nm is better for $r_0>13$ cm.
The only scenario where 1550 nm is a reasonable option is when $r_0<13$ cm \textit{and} narrower spectral filtering, narrower temporal filtering, or tighter spatial filtering in conjunction with AO are not viable options. 

\begin{figure}[t]
	\includegraphics[width=0.95\columnwidth]{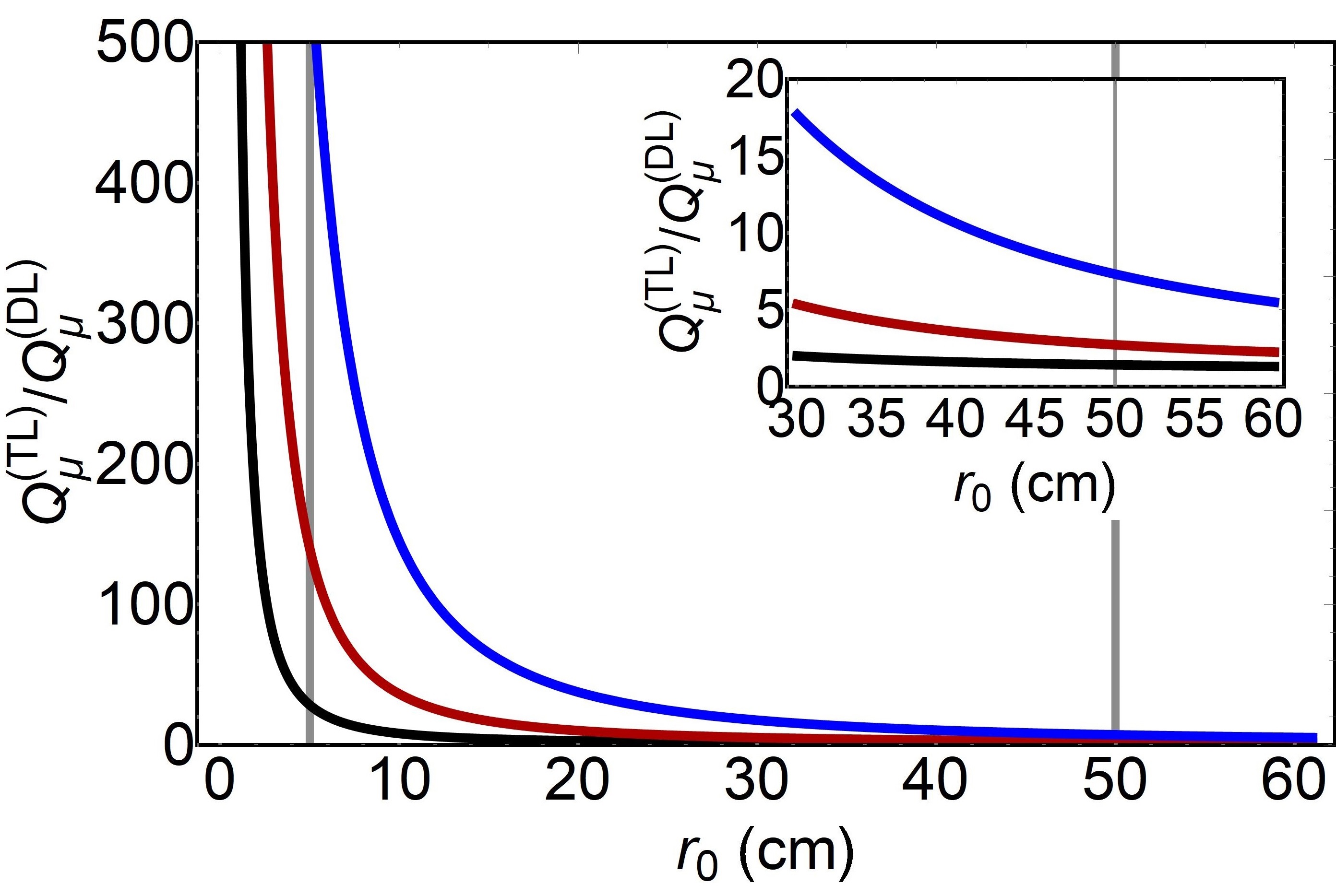}
	\caption{\label{fig:WinSolZenHighVisBoost}
The ratio $Q_{\mu}^{(\mathrm{TL})}/Q_{\mu}^{(\mathrm{DL})}$ for 1550, 781, and 431 nm (black, red, and blue respectively), plotted as a function of $r_0$. The vertical line at 5 cm corresponds to no AO correction and the line at 50 cm corresponds to the effective $r_0$ of a $f_c =200$-Hz AO system.   
	}
\end{figure}

To investigate the relatively high performance of the TL FOV strategy one can rearrange Eq.~\ref{eq:SKBP} and find
\begin{equation}
\begin{split}\label{eq:SKBP2}
P_{\mathrm{KB}}^{(i)} (r_0, \lambda) &= \dfrac{1}{2} \, Q_{\mu}^{(i)} (r_0, \lambda) \bigg( - c_1^{(i)} (r_0, \lambda) \\
&+ q^{(i)} (r_0, \lambda) \, c_2^{(i)} (r_0, \lambda) \bigg),
\end{split}
\end{equation}
where
\begin{equation}
\begin{split}
q^{(i)} (r_0 , \lambda) &= Q_{1}^{(i)} (r_0, \lambda) / Q_{\mu}^{(i)} (r_0, \lambda) \\
c_1^{(i)} (r_0 , \lambda) &= f_{ec} \, H_2 \big[ E_{\mathrm{r}, \mu}^{(i)}(r_0, \lambda) \big]\\
c_2^{(i)} (r_0 , \lambda) &= \big( 1- H_2 \big[ e_{1}^{(i)} (r_0, \lambda) \big] \big).
\end{split}
\end{equation}
The ratio of gains $q^{(i)}$ remains relatively constant upon switching between the filtering strategies because the fractional loss induced by the field stop does not effect single- and multi-photon pulses differently at such low MPNs and channel efficiencies.
Furthermore, $c_1^{(i)}$ and $c_2^{(i)}$ remain relatively constant by virtue of the behavior of the error rates $E^{(i)}_{r, \mu}$ and $e^{(i)}_1$, respectively.
Therefore, the dominating trend in the Eq.~\ref{eq:SKBP2} comes from the leading factor $Q_{\mu}^{(i)}$. 
In Fig.~\ref{fig:WinSolZenHighVisEta}(b) we showed how loss at the spatial filter significantly reduces the channel efficiency for the DL strategy with respect to the TL strategy.
In Fig.~\ref{fig:WinSolZenHighVisBoost} we show how this affects the signal gain by plotting the ratio $Q_{\mu}^{(\mathrm{TL})}/Q_{\mu}^{(\mathrm{DL})}$ as a function of $r_0$ for 1550, 781, and 431 nm.
This clearly shows the drastic improvement in signal gain for small $r_0$ and the TL strategy. 
This effect, in conjunction with Eq.~\ref{eq:SKBP2}, reveals how the TL strategy permits such relatively high performance.

The vertical line at 5 cm in Fig.~\ref{fig:WinSolZenHighVisSKBP}(a) reveals that the assumed atmospheric and system conditions would suppress the key-bit yield at the selected wavelengths. 
To increase performance, one has no choice but to implement some form of more aggressive noise filtering.
For example, the vertical line $r_0=50$ cm reveals the key-bit rates attainable with AO and correspondingly tighter spatial filtering.
One can see that 431 nm and the TL strategy is the optimal choice giving $\sim$13$\times$ improvement over 1550 nm.
In fact, we judiciously chose 431 nm because it gives the largest key-bit rate in the 400 - to 1600-nm range when using a 1-nm filter (see App.~\ref{sec:appendixB}).
For reference, this level of performance is comparable to a system without AO compensation but utilizing a $50\times$ narrower filter, that is, $\Delta \lambda$$=$0.02 nm.

In the case with AO, our simulation shows that high QKD performance can be achieved with relatively low system Strehls. 
This is somewhat counterintuitive since it is customary to relate system Strehl to system performance. 
However, in Fig.~\ref{fig:WinSolZenHighVisSKBP}(b) we see that with the TL strategy and a corrected Strehl near 0.1, high performance can be achieved with 431 nm, in effect beating a 1550-nm system operating at the diffraction limit by a factor of $\sim$9.
Moreover, for a 431-nm system, a majority of the TL strategy performance is achievable with a low Strehl, that is, with a Strehl of 0.1, one can achieve $\sim$89$\%$ of the maximum key-bit rate at that wavelength. 
This is in contrast to a 1550-nm system that only achieves $\sim$34$\%$ of the maximum key-bit rate at a Strehl of 0.1. 
Next, we will discuss how these effects relate to the speed of the AO system.

As we outlined in App.~\ref{sec:appendixA}, the degree of AO correction ultimately depends on the ability of the AO system to keep pace with the temporal component of atmospheric turbulence, that is, the tracking and higher-order Greenwood frequencies $f_{\mathrm{TG}}$ and $f_{\mathrm{G}}$, respectively. 
Similar to the Fried coherence length, the observed Greenwood frequencies are more challenging at shorter wavelengths (see Eqs.~\ref{eq:fTG} and \ref{eq:fG}).
For example, when $f_{\mathrm{G}}$(500 nm)$=$301 Hz, we find that $f_{\mathrm{G}}$(1550 nm)$\approx$77 Hz and $f_{\mathrm{G}}$(431 nm)$\approx$360 Hz.
Therefore, one might have assumed that an AO system integrated with a free-space QKD system will need to be faster to operate at 431 nm and provide adequate performance.
However, the preceding study showed that a useful degree of AO wavefront compensation can be achieved with AO bandwidths below the rate of change of turbulence.  
In this example, even the $f_c=130$-Hz system offered a substantial performance benefit and this is further emphasized in the following.

\begin{figure}[t!]
	\includegraphics[width=1\columnwidth]{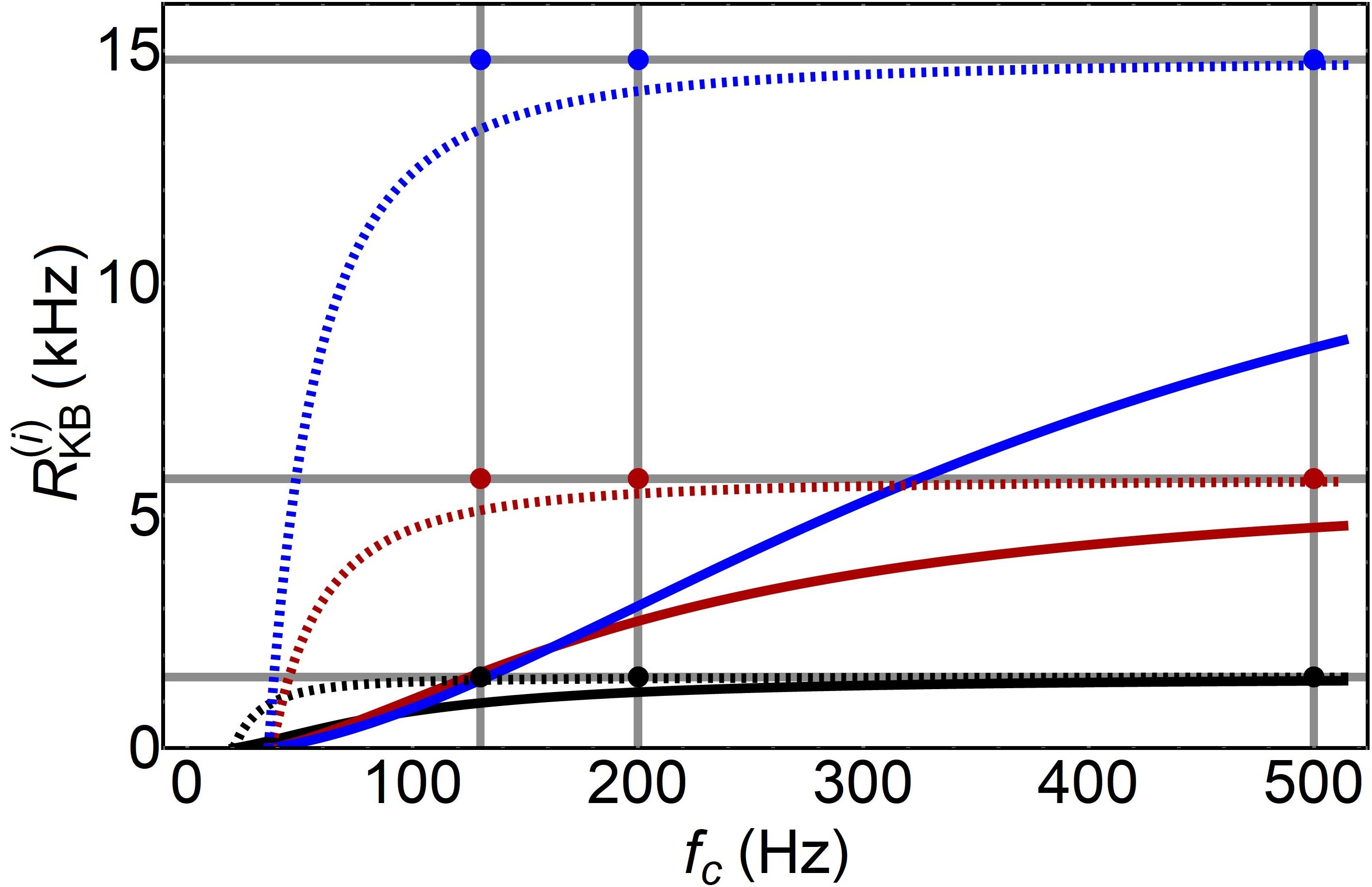}
\caption{\label{fig:WinSolZenHighVisSKBP2}
The closed-loop bandwidth $f_c$ dependence of the key-bit rate $R_{\mathrm{KB}}^{(i)}$ for 1550, 781, and 431 nm (black, red, and blue respectively). The solid and dashed curves indicate the DL and TL strategies, respectively. The vertical line at $f_c=130$-Hz indicates the closed-loop bandwidth of the system built for our field experiment \cite{gruneisen2020adaptive} whereas $f_c=200$ Hz and 500 Hz indicate bandwidths assumed in previous numerical simulations \cite{gruneisen2016adaptive, gruneisen2017modeling}. The horizontal lines and color coded markers indicate the maximum key-bit rate for each wavelength, that is, the key-bit rate for the receiver operating at the diffraction limit.}
\end{figure}

In Fig.~\ref{fig:WinSolZenHighVisSKBP2} we plot the key-bit rate as a function of the closed-loop bandwidth of the AO system.
First, this plot illustrates a prevailing concept well understood in the AO community, that is, given a certain closed-loop bandwidth, the corrected wavefront quality relative to the diffraction limit will be better at longer wavelengths.
This is evident from examining the solid DL-FOV-strategy curves in relation to that maximum key-bit rate for $f_c=500$ Hz.
The 1550-nm system would be practically operating at the diffraction limit where as the 781- and 431-nm systems fall noticeably short of diffraction limited performance, indicated by the horizontal lines and color coded markers. 
This is evident from the gap between the solid curves and the respective dots indicating the maximum key-bit rates for each of the wavelengths.
On the other hand, Fig.~\ref{fig:WinSolZenHighVisSKBP2} illustrates a less common concept, that is, perfect wavefront correction is not necessarily needed for high performance.
The dashed TL-FOV-strategy curves show that with relatively slow AO, high performance can still be achieved despite the larger FOV and increased number of noise photons.
For example, at $f_c=130$ Hz, the TL FOV key-bit rate $R_{\mathrm{KB}}^{(\mathrm{TL})}(\lambda)$ is already quite close to the maximum key-bit rate for each wavelength.
Therefore, one should weigh the increased technological challenges faced when building a faster AO system with the diminishing returns evident when already using the TL FOV strategy.

Lastly, these results strongly contradict conventional wisdom in regards to wavelength selection.  
Naively, one might presuppose that 1550 nm is the optimal choice because the sky is dimmer and atmospheric effects are more benign. 
Especially in the case of AO where, as we have shown, given a certain measure of correction, the 1550-nm receiver will be operating much closer to the diffraction limit as compared to shorter wavelengths.
Nonetheless, we have shown that a small amount of AO correction can permit the advantages of the shorter wavelengths to dominate.
Namely, the larger photon energy and smaller spot size reduce the number of background photons, and the increased aperture-to-aperture coupling boosts the signal. 
Our results in Fig.~\ref{fig:WinSolZenHighVisSKBP2} reveal that a $f_c \sim 40$-Hz system operating at 431-nm with the TL FOV could in principle outperform a 1550-nm system operating at the diffraction limit.  
In the following section we will investigate how performance changes with more challenging atmospheric conditions, namely diminishing visibility on both the winter and summer solstices, and explore how narrower spectral filtering can restore performance.

\section{Visibility Study}\label{sec:VisStudy}
The parameter space of atmospheric conditions is seemingly infinite for the down-link scenario.
However, many of the effects can be studied with relatively simple analyses.
For our purposes here, it suffices to investigate the effects due to changes in visibility while at zenith. 
This is because as zenith angle increases, the longer propagation path reduces transmission and increases scattering into the channel, but these are the same effects observed with decreased visibility.
An angle-dependent study would be redundant, except in the case of large zenith angles where $r_0$ and $f_{\mathrm{G}}$ will be appreciably different than at zenith.  
In terms of Greenwood frequency, the more challenging scenario is \textit{near} zenith since the slew rate is highest.
Conversely, the spatial coherence degrades with increasing zenith angle.
However, $r_0$ only ranges from 5 to 4 cm for pointing angles ranging from zenith to 45 degrees. 
Therefore, a visibility study at zenith is quite representative of a challenging down-link scenario.
We will consider two sun positions, namely the winter and summer solstices at 1:00 PM.

\subsection{Winter Solstice}
\begin{figure}[t!]
  \centering
  \begin{tabular}{@{}p{0.475\linewidth}@{\quad}p{0.475\linewidth}@{}}
    \includegraphics[width=\linewidth]{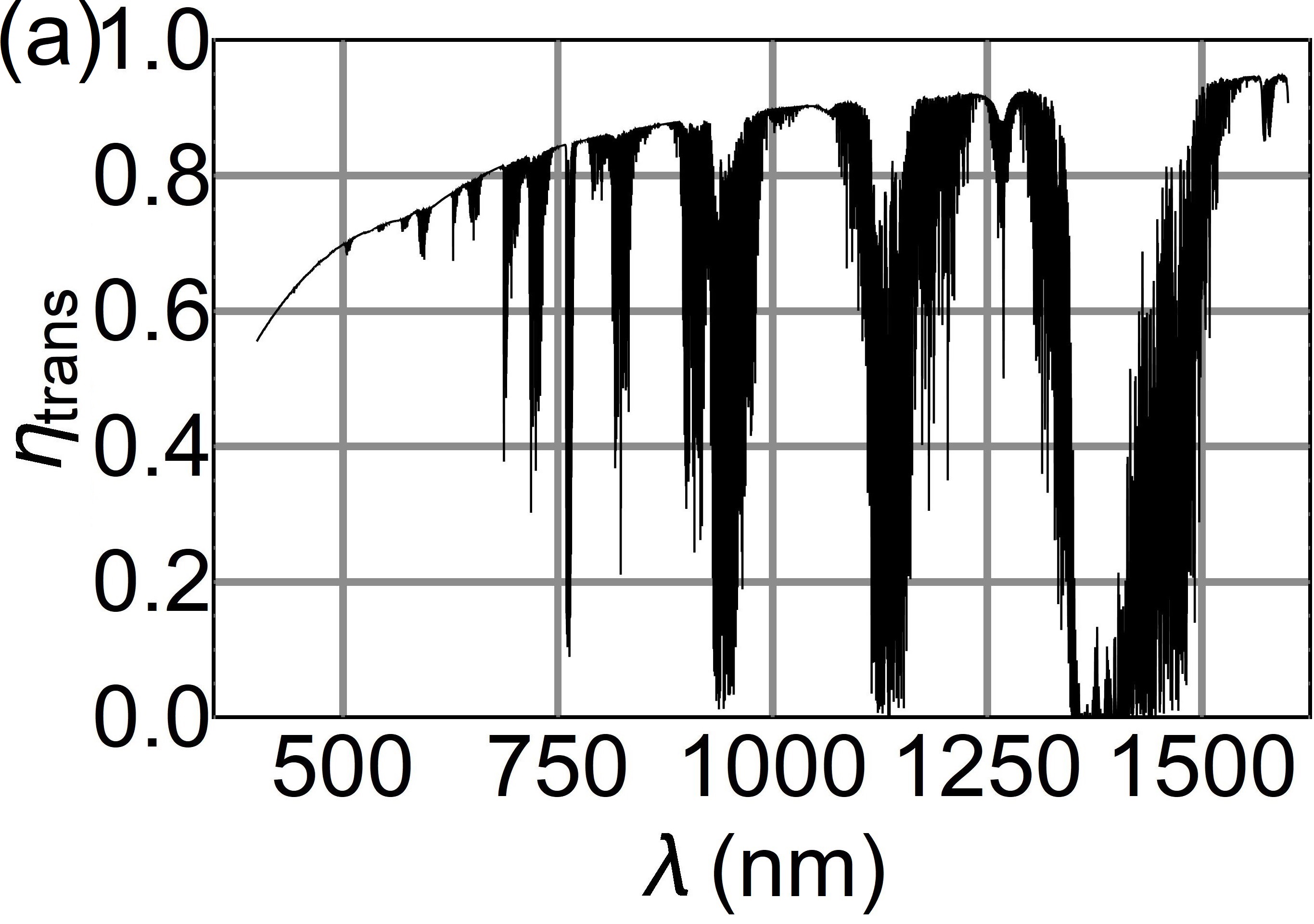} &
    \includegraphics[width=\linewidth]{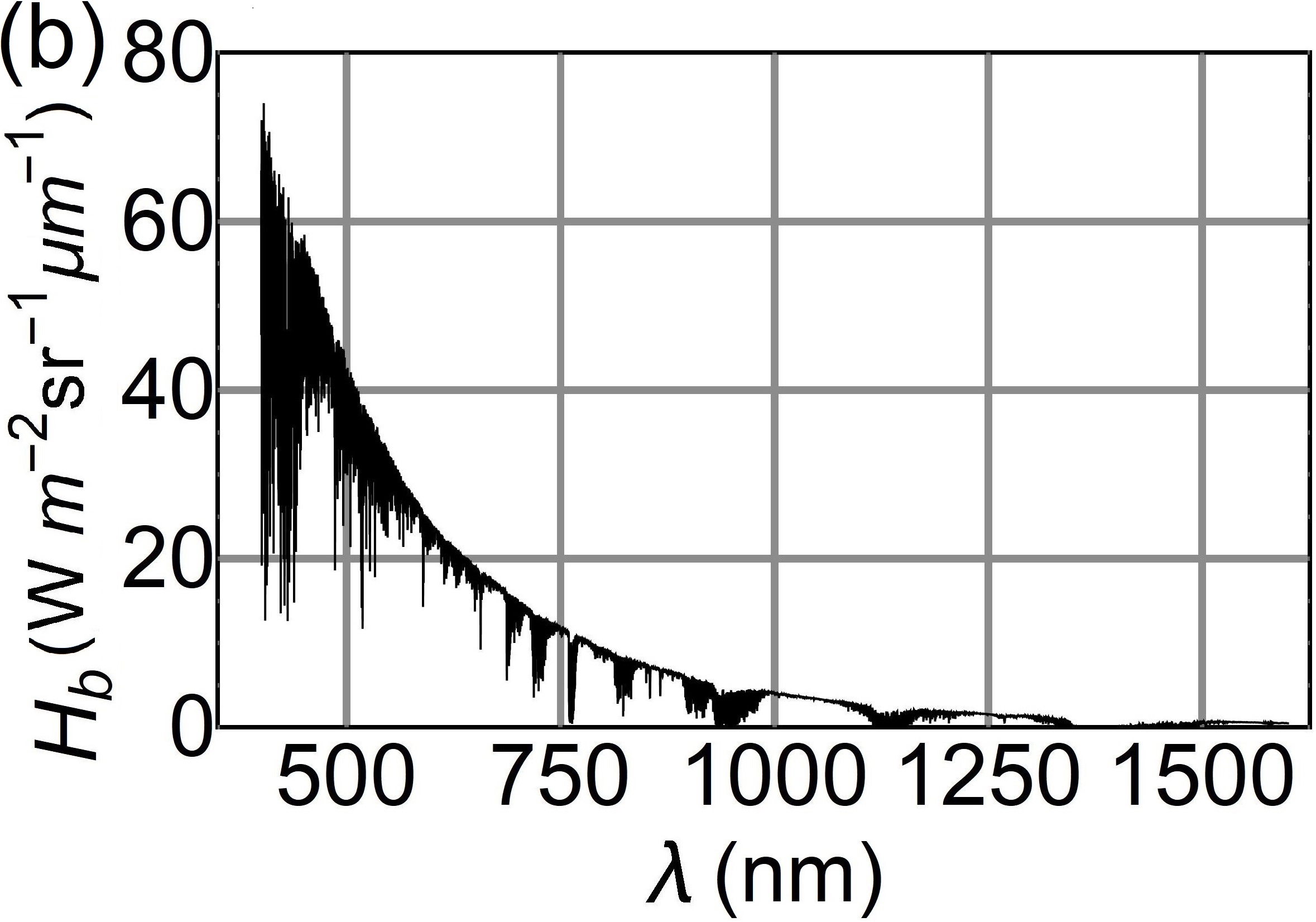} \\
    \includegraphics[width=\linewidth]{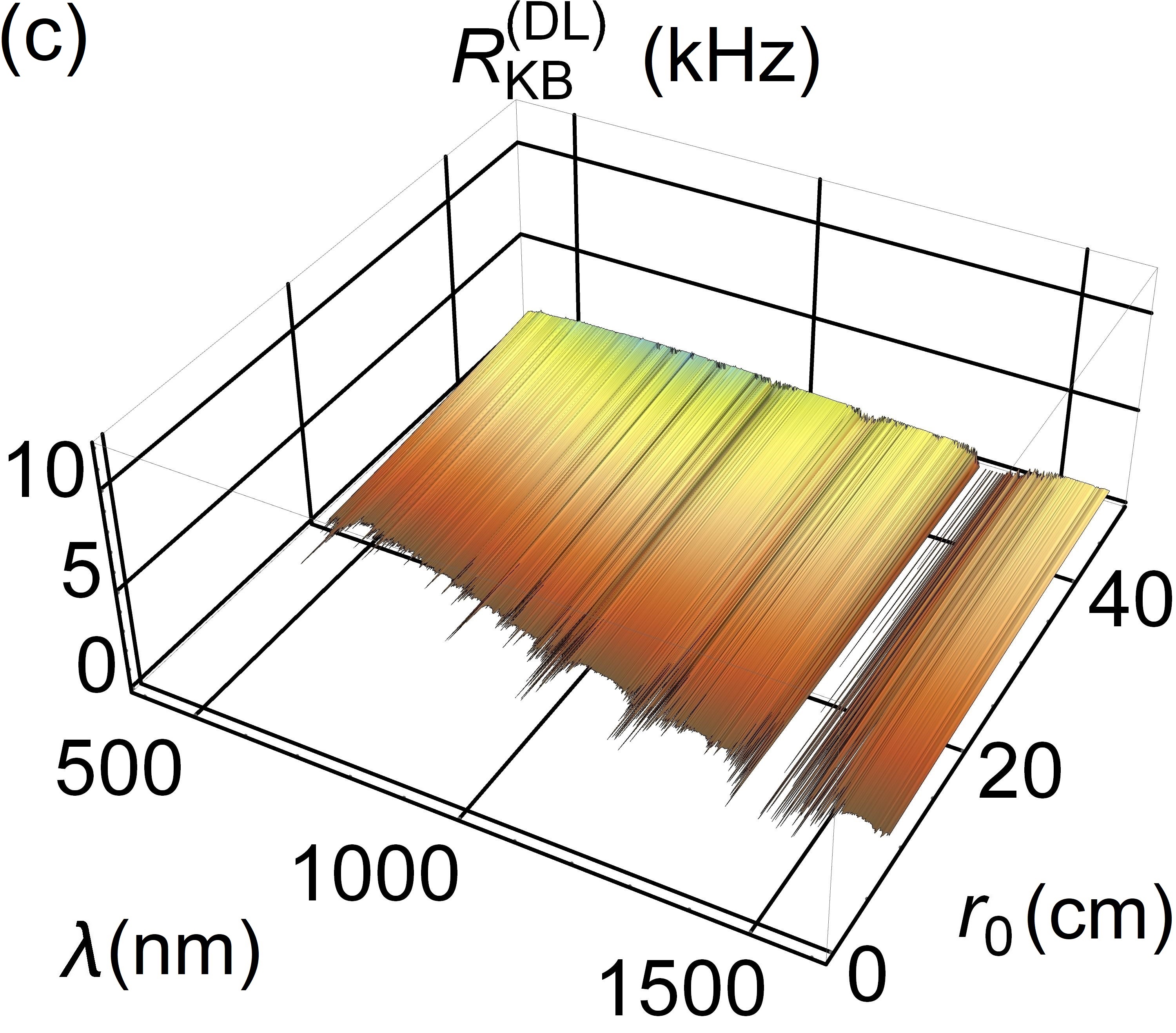} &
    \includegraphics[width=\linewidth]{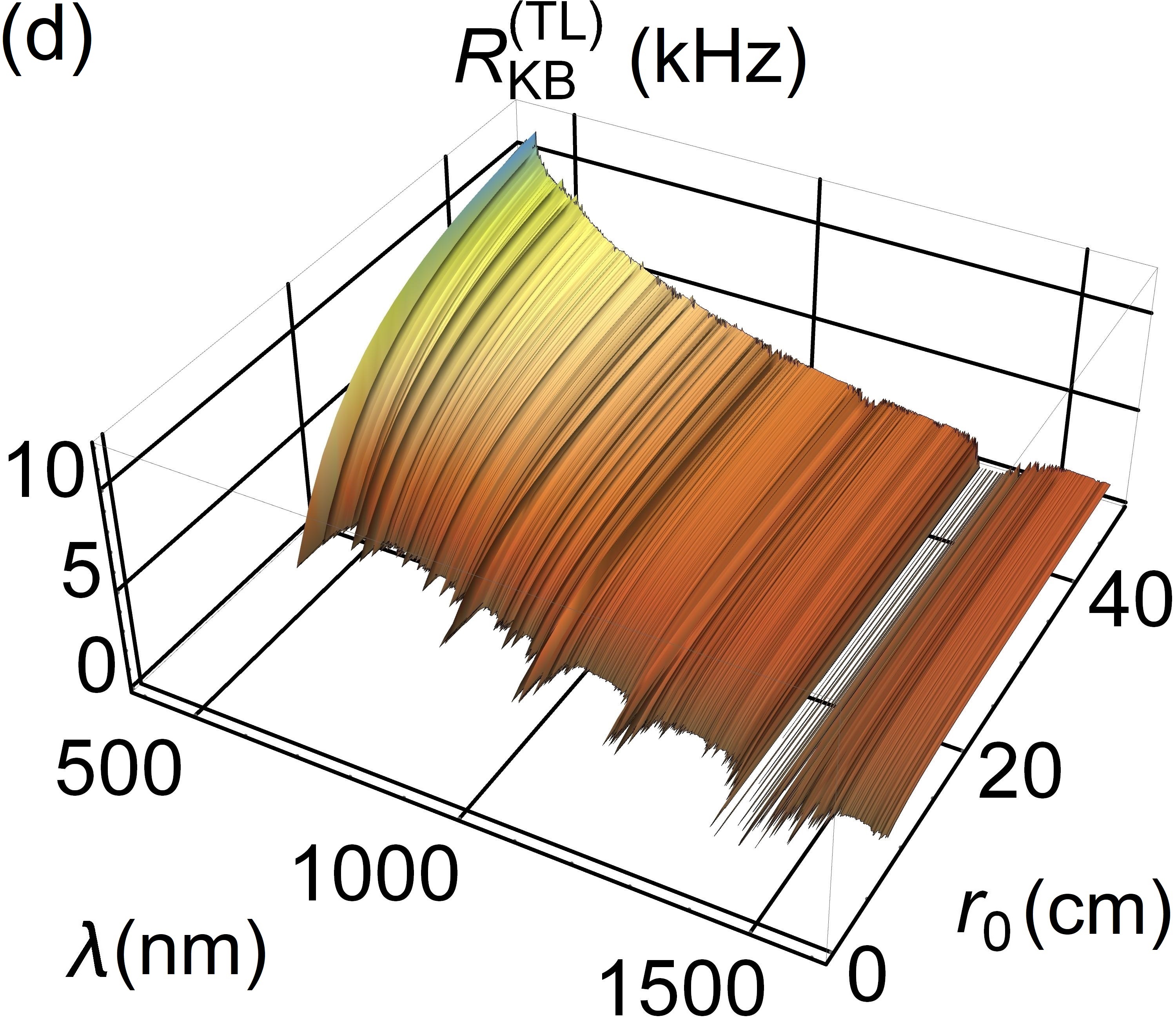} \\
  \end{tabular}
 \begin{tabular}{@{}p{\linewidth}}
    \includegraphics[width=0.95\linewidth]{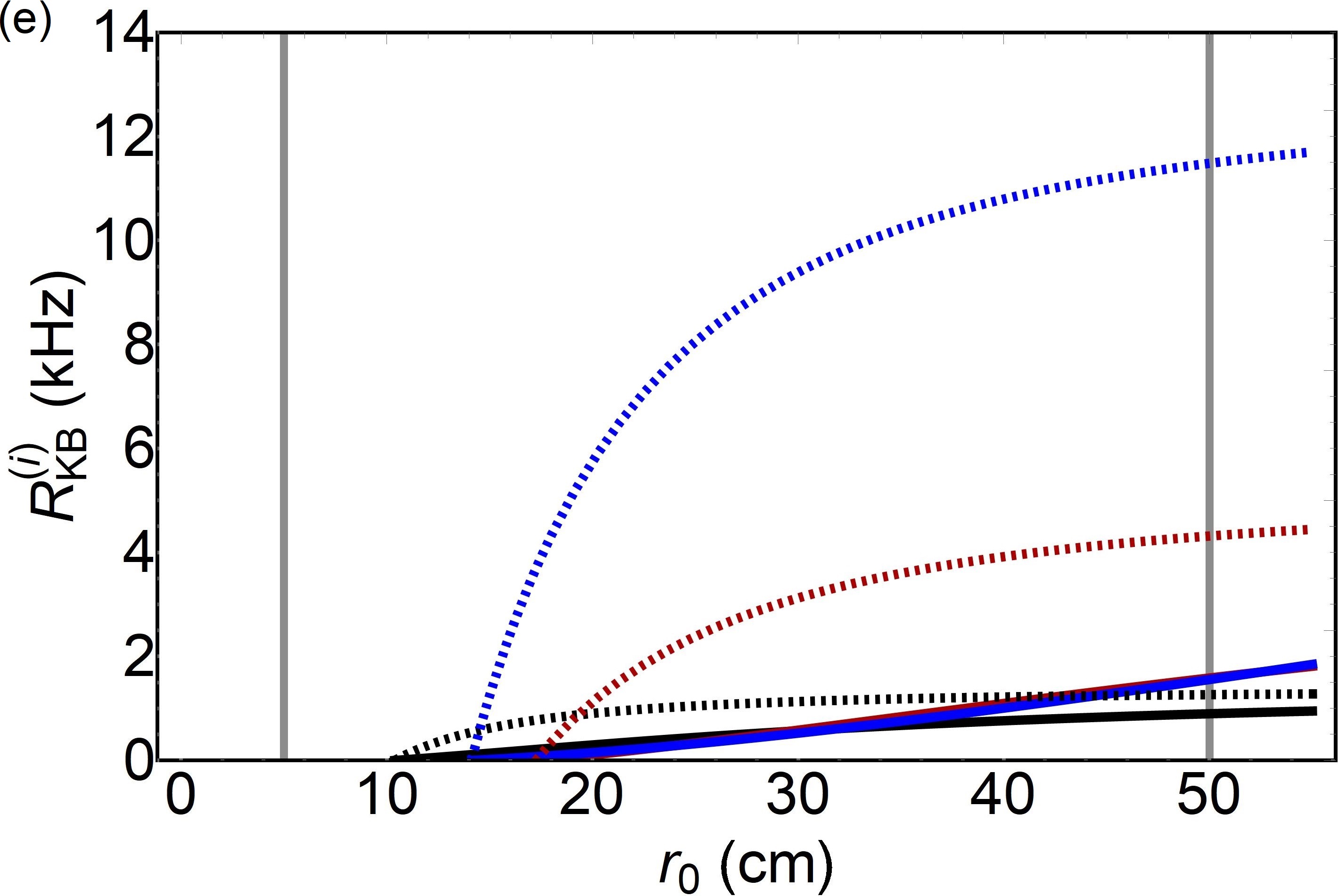} \\
    \includegraphics[width=0.95\linewidth]{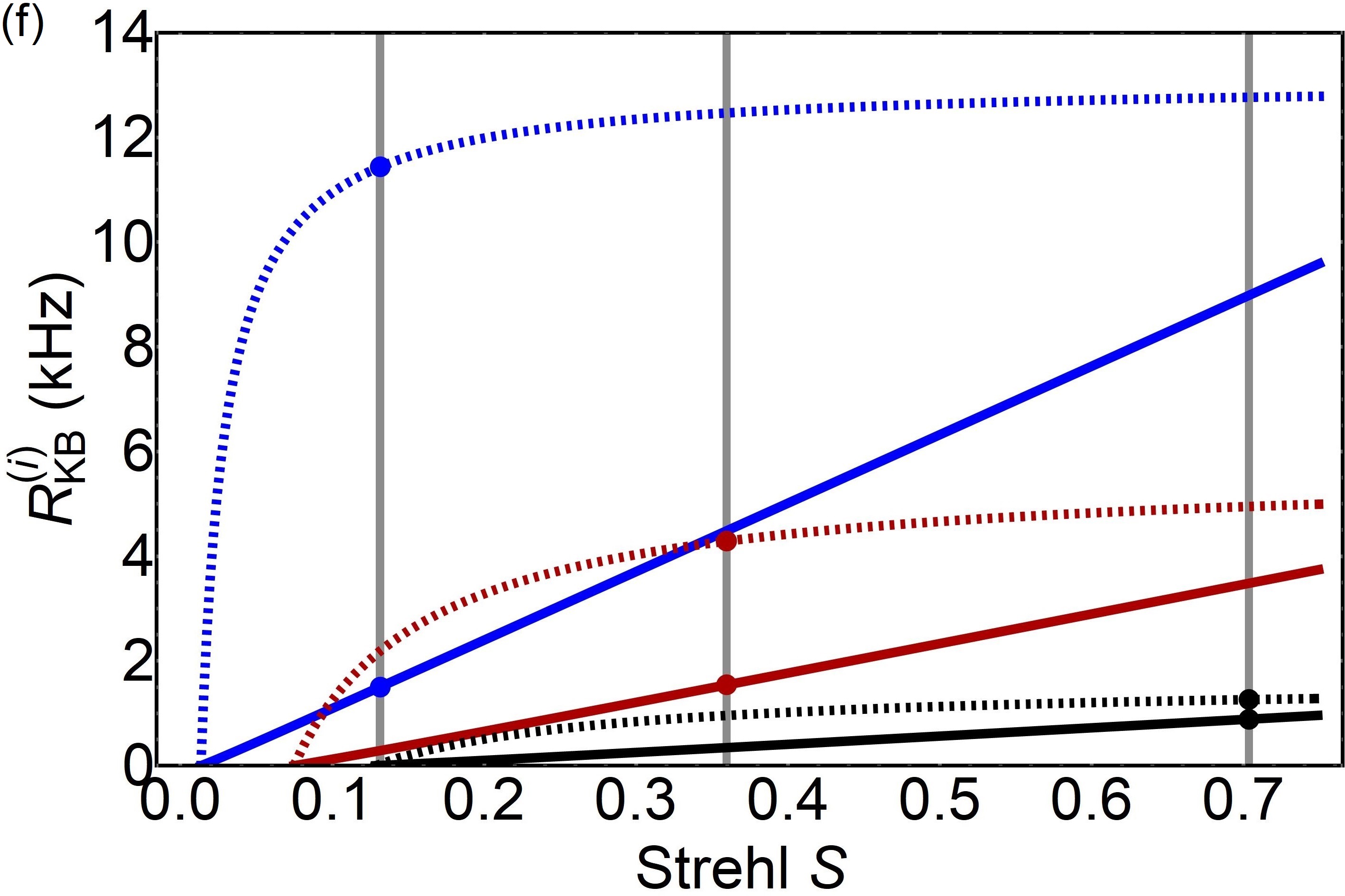}
  \end{tabular}
  \caption{
The (a-b) atmospheric transmission $\eta_{\mathrm{trans}}$ and spectral radiance $H_{\mathrm{b}}$, and (c-d) the key-bit rate $R_{\mathrm{KB}}^{(i)}$ for winter solstice with medium visibility (23-km), plotted as a function of $r_0$. In (e-f) we have chosen 1550, 781, and 431 nm (black, red, and blue respectively), and the solid and dashed curves indicate the DL and TL strategies, respectively. In (e) the vertical line at 5 cm corresponds to no AO correction and the line at 50 cm corresponds to the effective $r_0$ of a $f_c =200$-Hz AO system. In (f) the $\lambda$ dependence of $S$ produces three vertical lines indicating AO correction. 
} \label{fig:WinSolZenMed}
\end{figure}

\begin{figure}[t!]
  \centering
  \begin{tabular}{@{}p{0.475\linewidth}@{\quad}p{0.475\linewidth}@{}}
    \includegraphics[width=\linewidth]{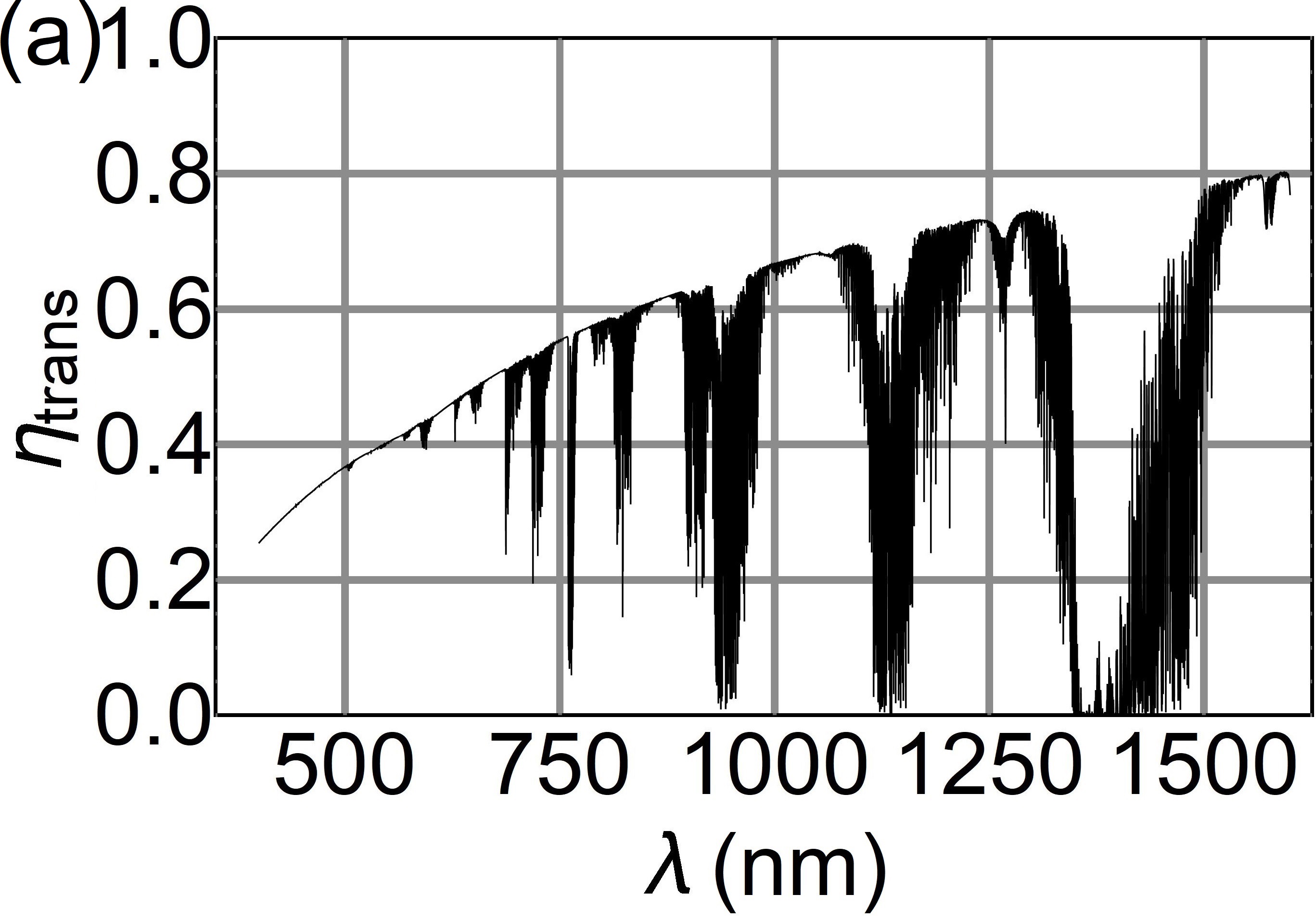} &
    \includegraphics[width=\linewidth]{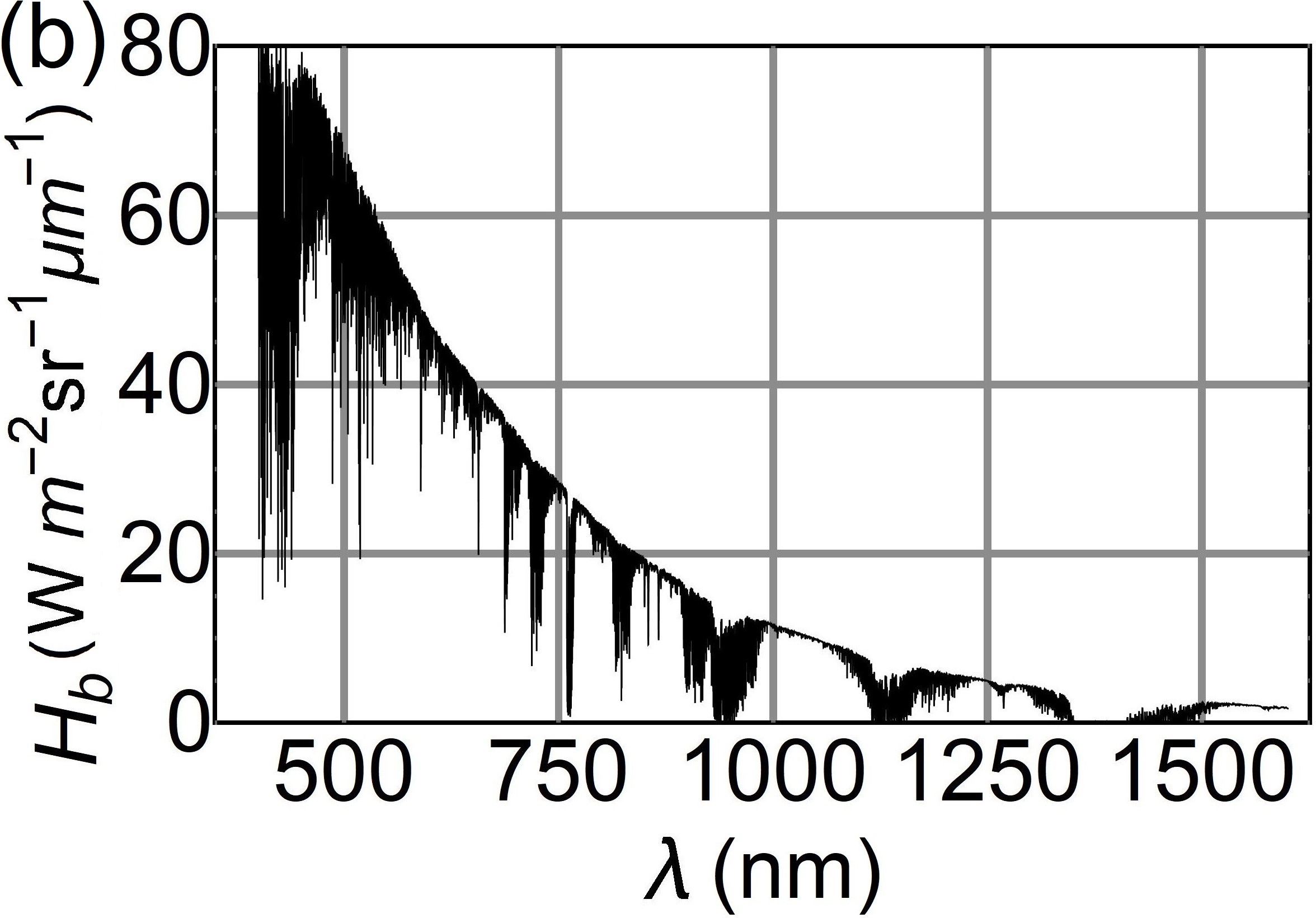} \\
    \includegraphics[width=\linewidth]{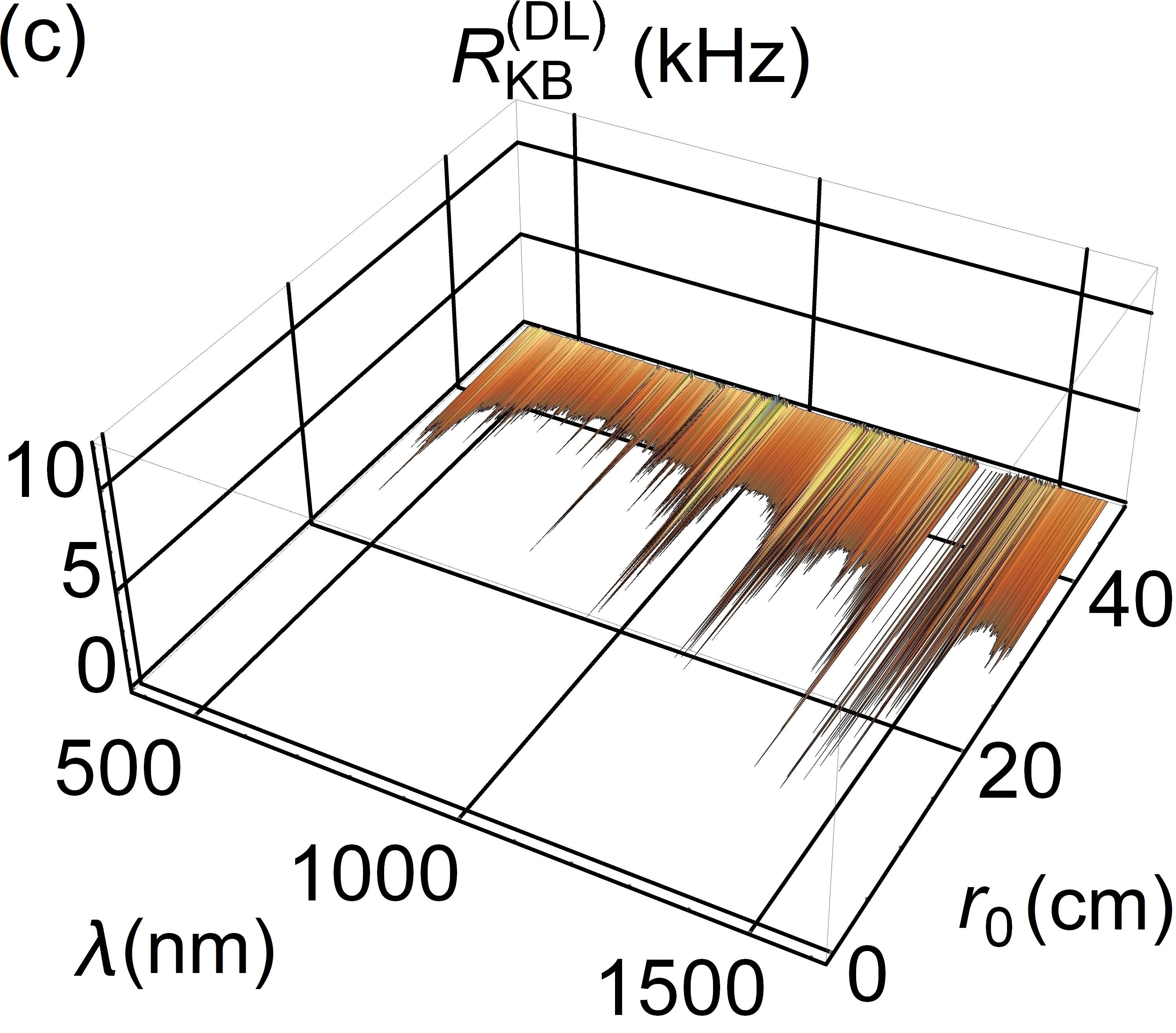} &
    \includegraphics[width=\linewidth]{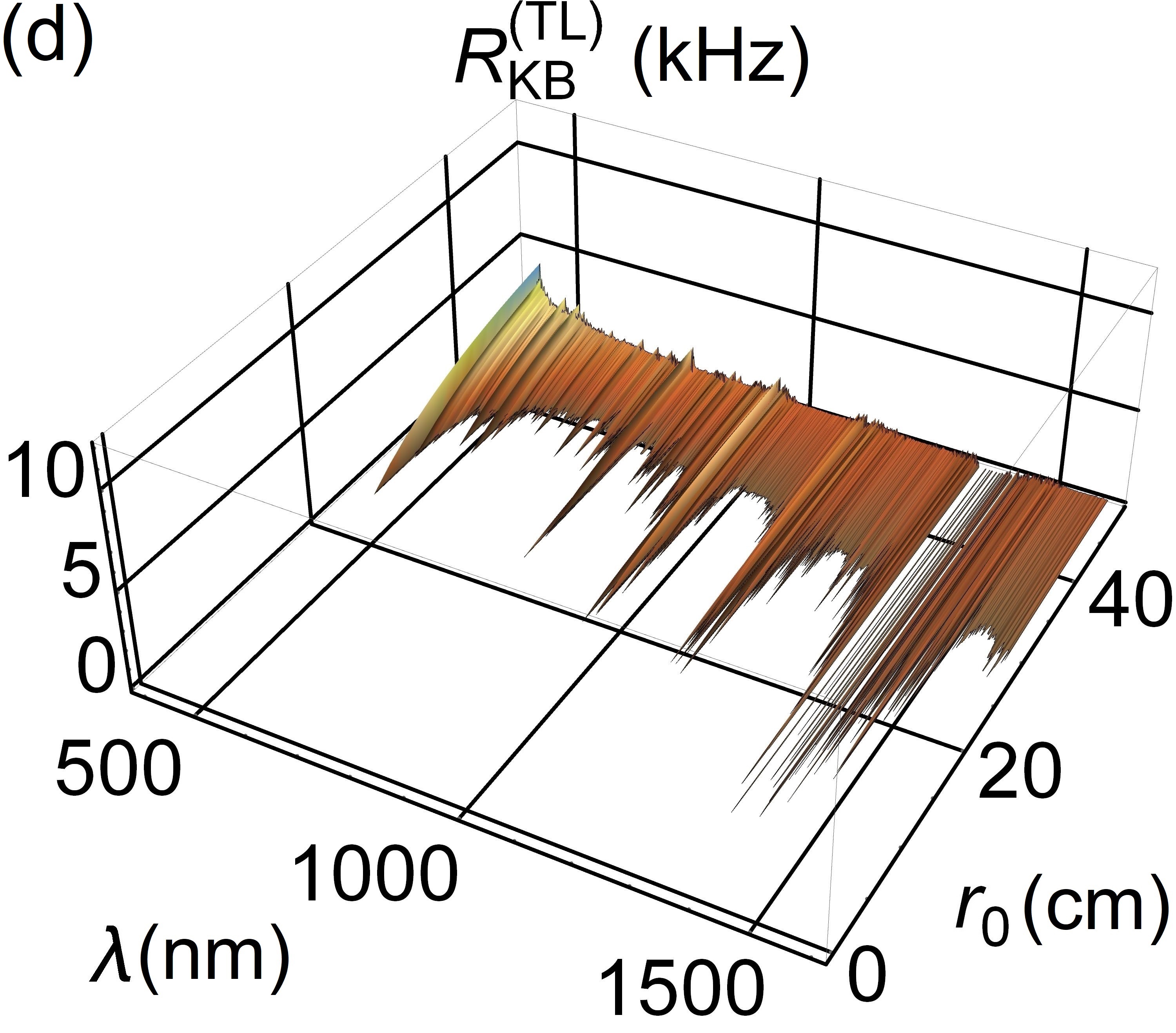} \\
  \end{tabular}
 \begin{tabular}{@{}p{\linewidth}}
    \includegraphics[width=0.95\linewidth]{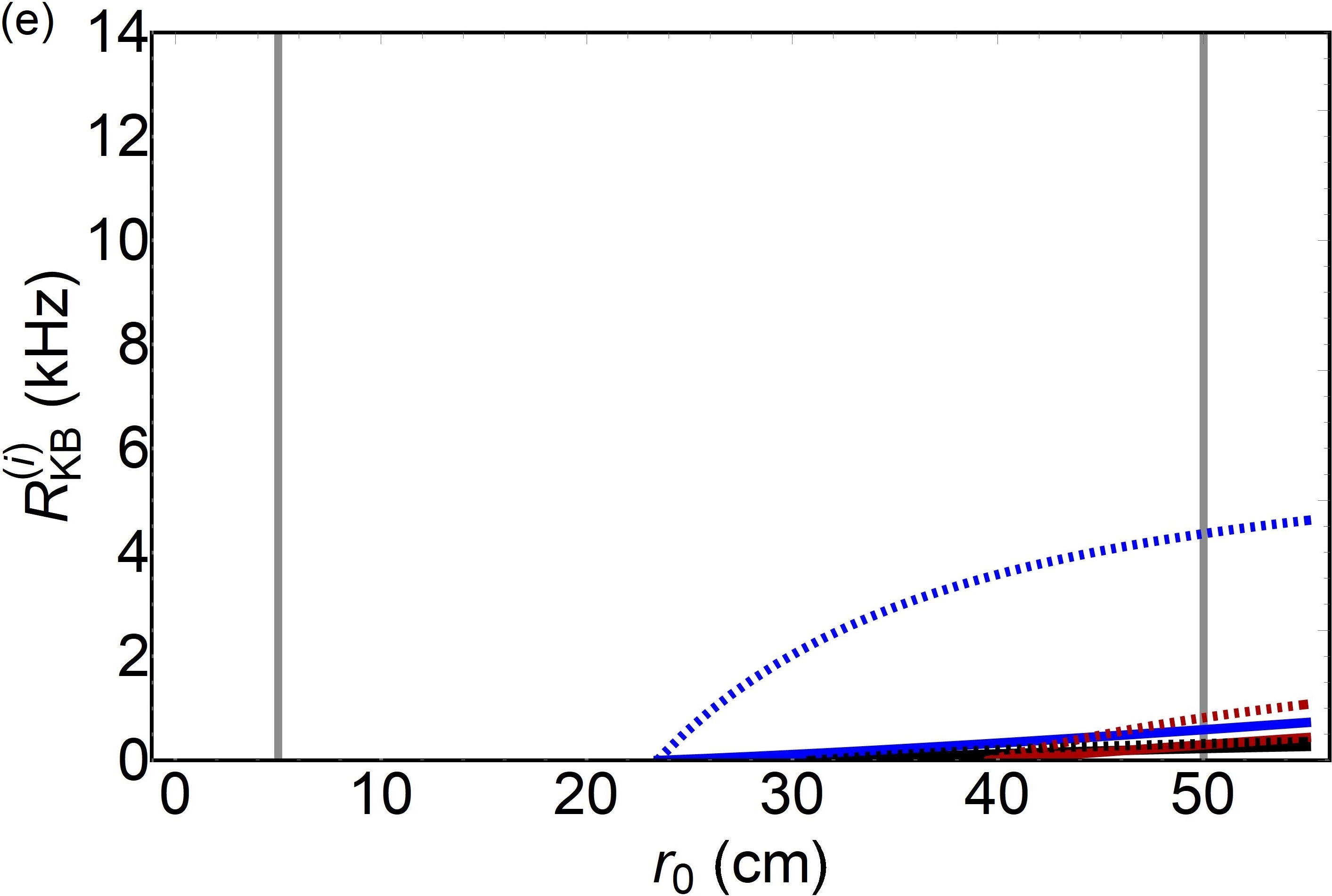}\\
    \includegraphics[width=0.95\linewidth]{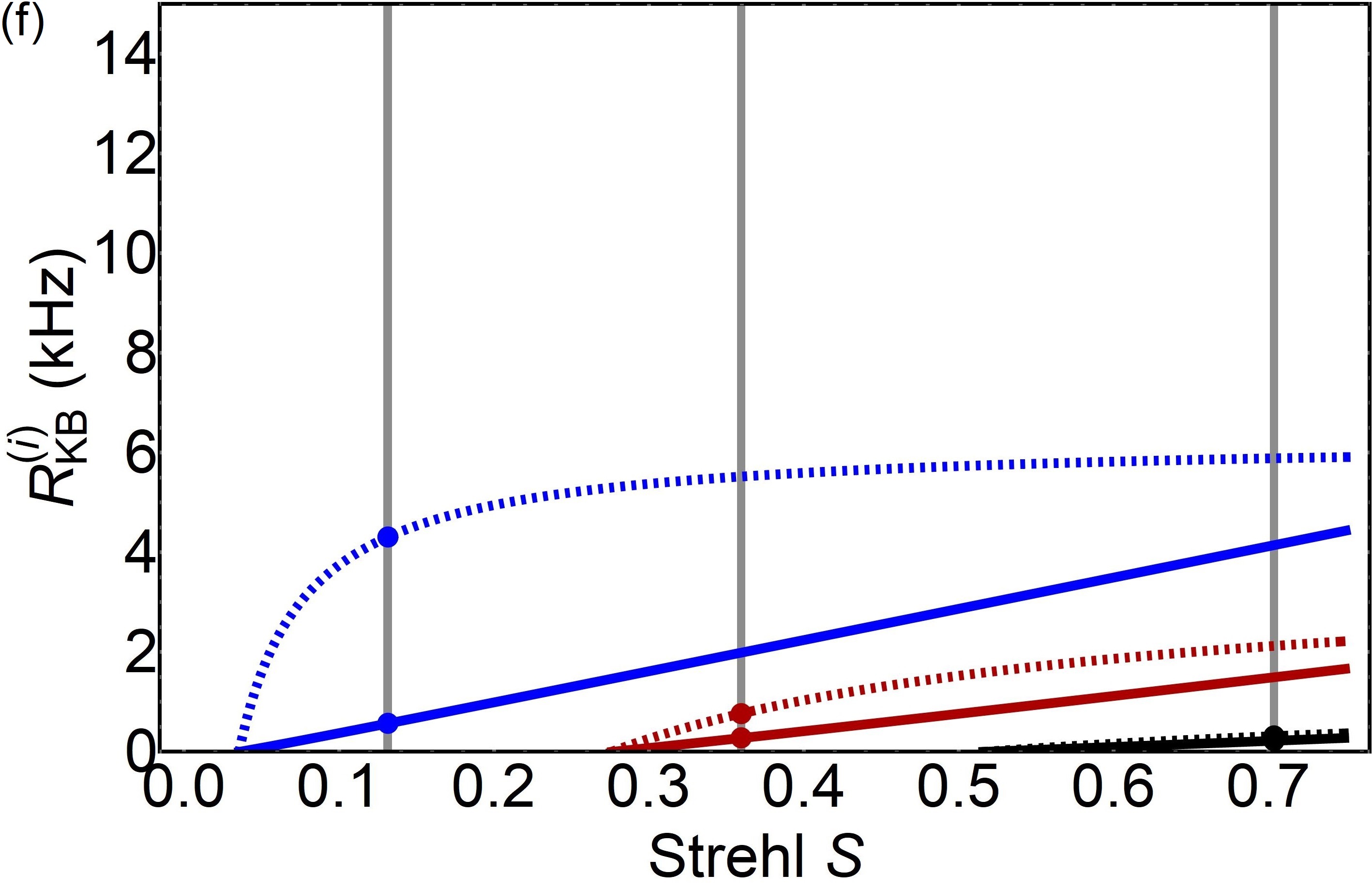} 
  \end{tabular}
  \caption{
The (a-b) atmospheric transmission $\eta_{\mathrm{trans}}$ and spectral radiance $H_{\mathrm{b}}$, and (c-d) the key-bit rate $R_{\mathrm{KB}}^{(i)}$ for winter solstice with low visibility (5-km), plotted as a function of $r_0$. In (e-f) we have chosen 1550, 781, and 431 nm (black, red, and blue respectively), and the solid and dashed curves indicate the DL and TL strategies, respectively. In (e)the vertical line at 5 cm corresponds to no AO correction and the line at 50 cm corresponds to the effective $r_0$ of a $f_c =200$-Hz AO system. In (f) the $\lambda$ dependence of $S$ produces three vertical lines indicating AO correction.
} \label{fig:WinSolZenLow}
\end{figure}

In this subsection we consider medium (23-km) and low (5-km) visibilities at 1:00PM on the winter solstice (see Figs.~\ref{fig:WinSolZenMed}(a-b) and \ref{fig:WinSolZenLow}(a-b), respectively).
One will see that a lower visibility condition decreases the atmospheric transmission and correspondingly increases the spectral radiance due to increased scattering of sunlight. 
In Fig.~\ref{fig:WinSolZenMed}(c-e) and \ref{fig:WinSolZenLow}(c-e) we plot the $r_0$ dependence of $R_{\mathrm{KB}}^{(i)}$ for the two winter solstice visibility conditions.
One will notice that the absence of intersections with the vertical line at $r_0$=5 cm indicates that it is not possible to generate key with a large spectral filter and without AO compensation. 
With AO enabled, one will see that the relative high performance of shorter wavelengths and the TL strategy persists.

In Fig.~\ref{fig:WinSolZenMed}(f) and \ref{fig:WinSolZenLow}(f) we plot $R_{\mathrm{KB}}^{(i)}$ as a  function of system Strehl.
Analyzing the performance as a function of Strehl again reveals that a relatively slow AO system
\pagebreak
\clearpage


\clearpage

\pagebreak
\clearpage

\begin{figure}[t!]
  \centering
  \begin{tabular}{@{}p{0.475\linewidth}@{\quad}p{0.475\linewidth}@{}}
    \includegraphics[width=\linewidth]{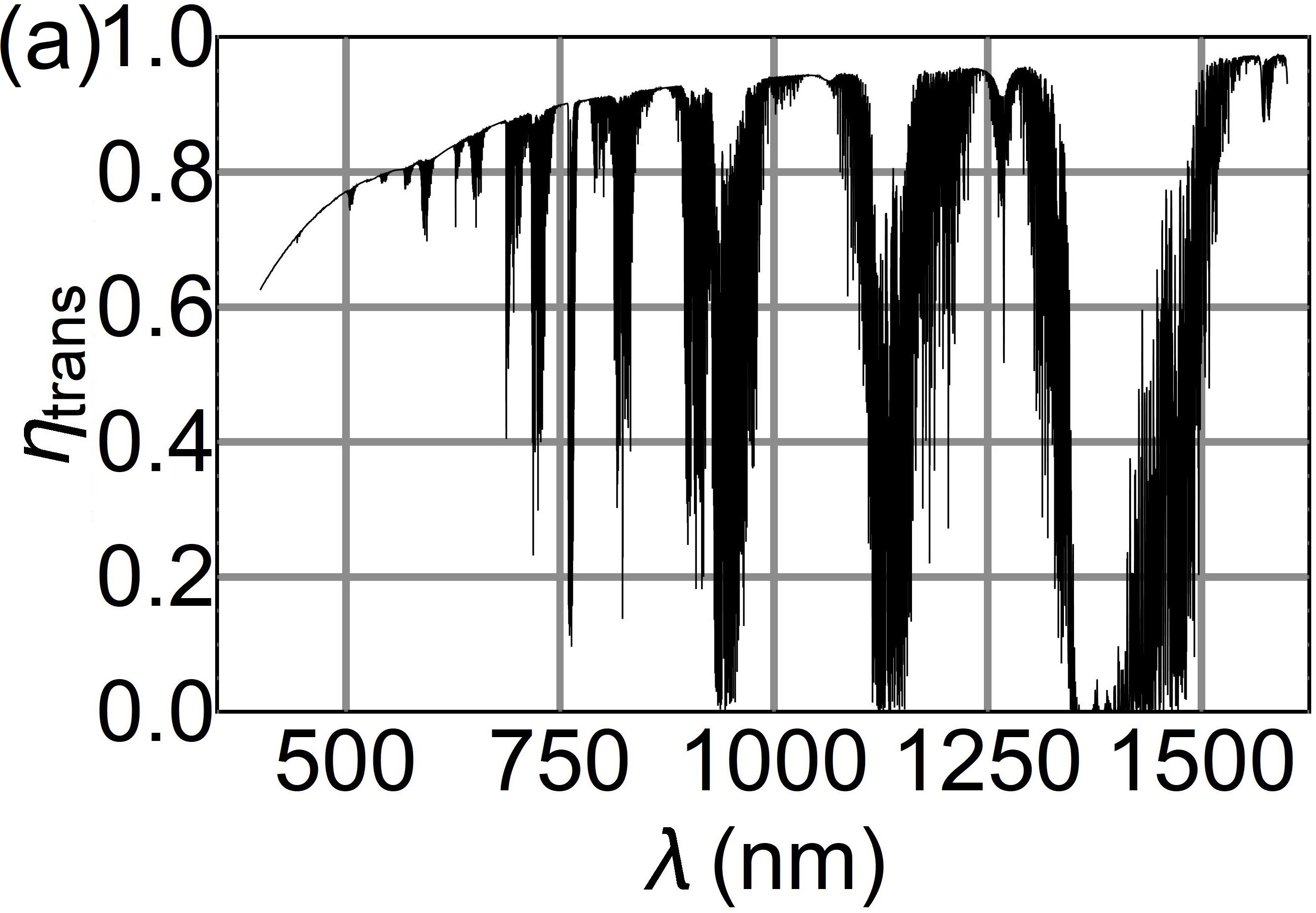} &
    \includegraphics[width=\linewidth]{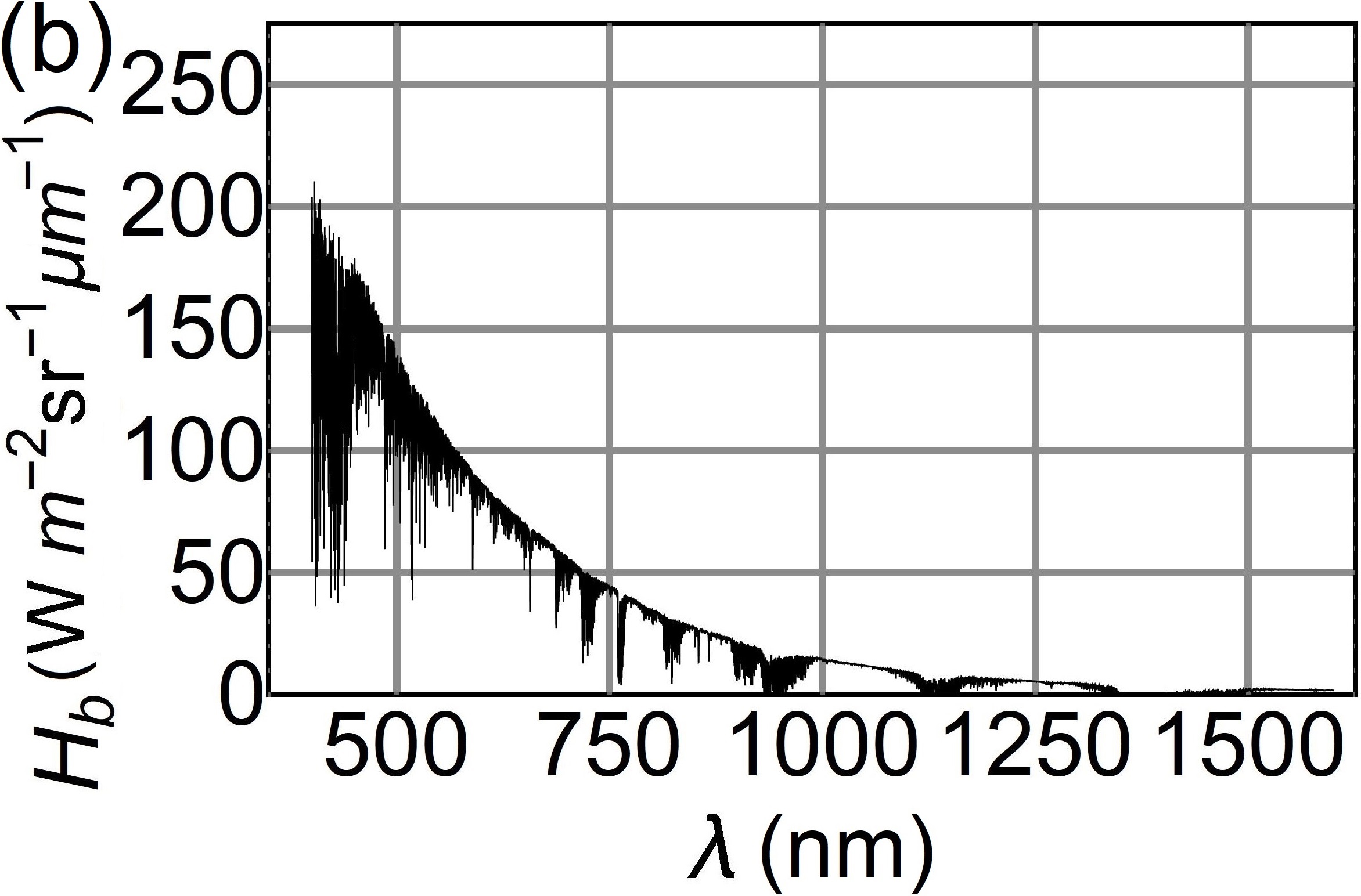} \\
    \includegraphics[width=\linewidth]{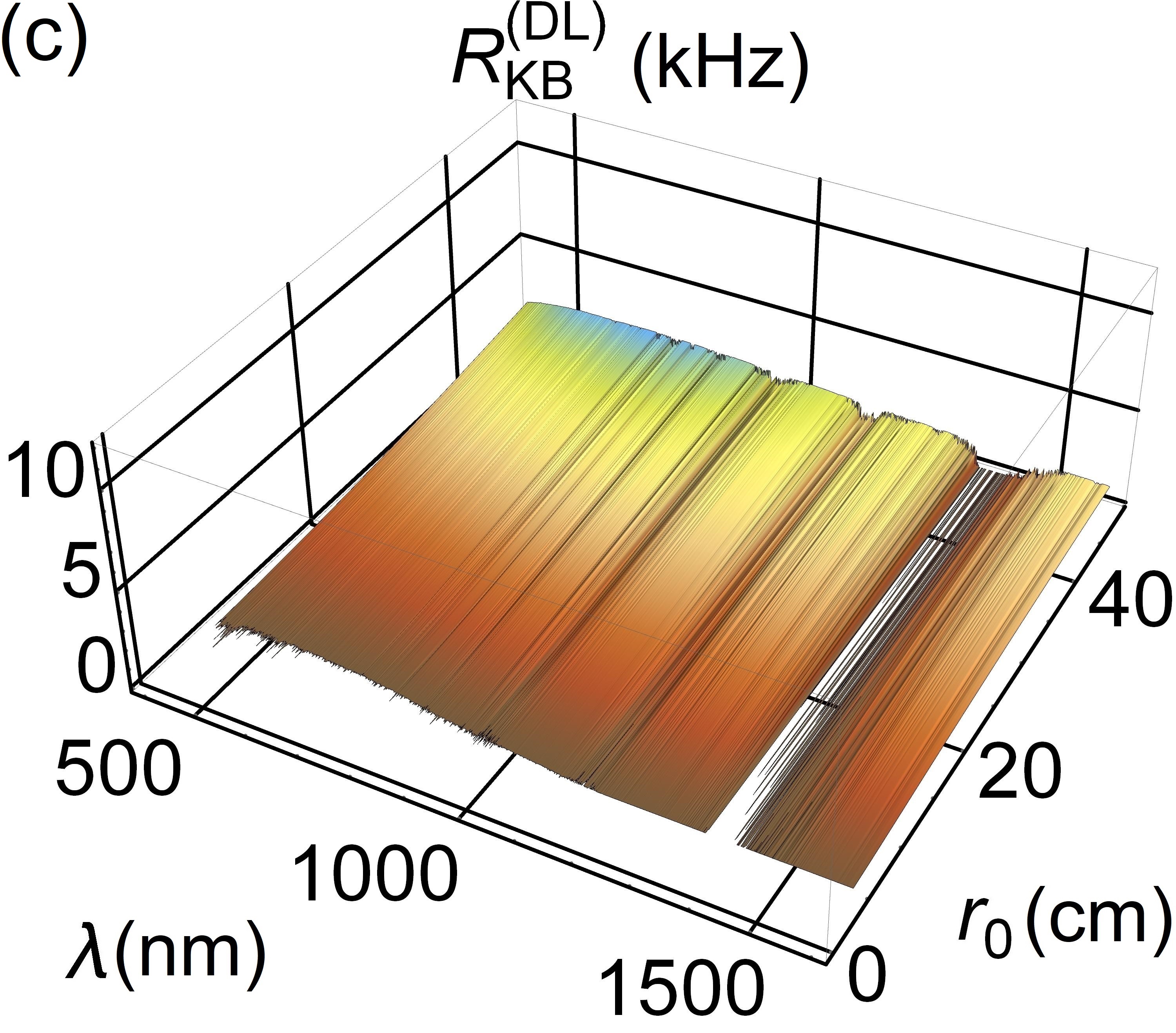} &
    \includegraphics[width=\linewidth]{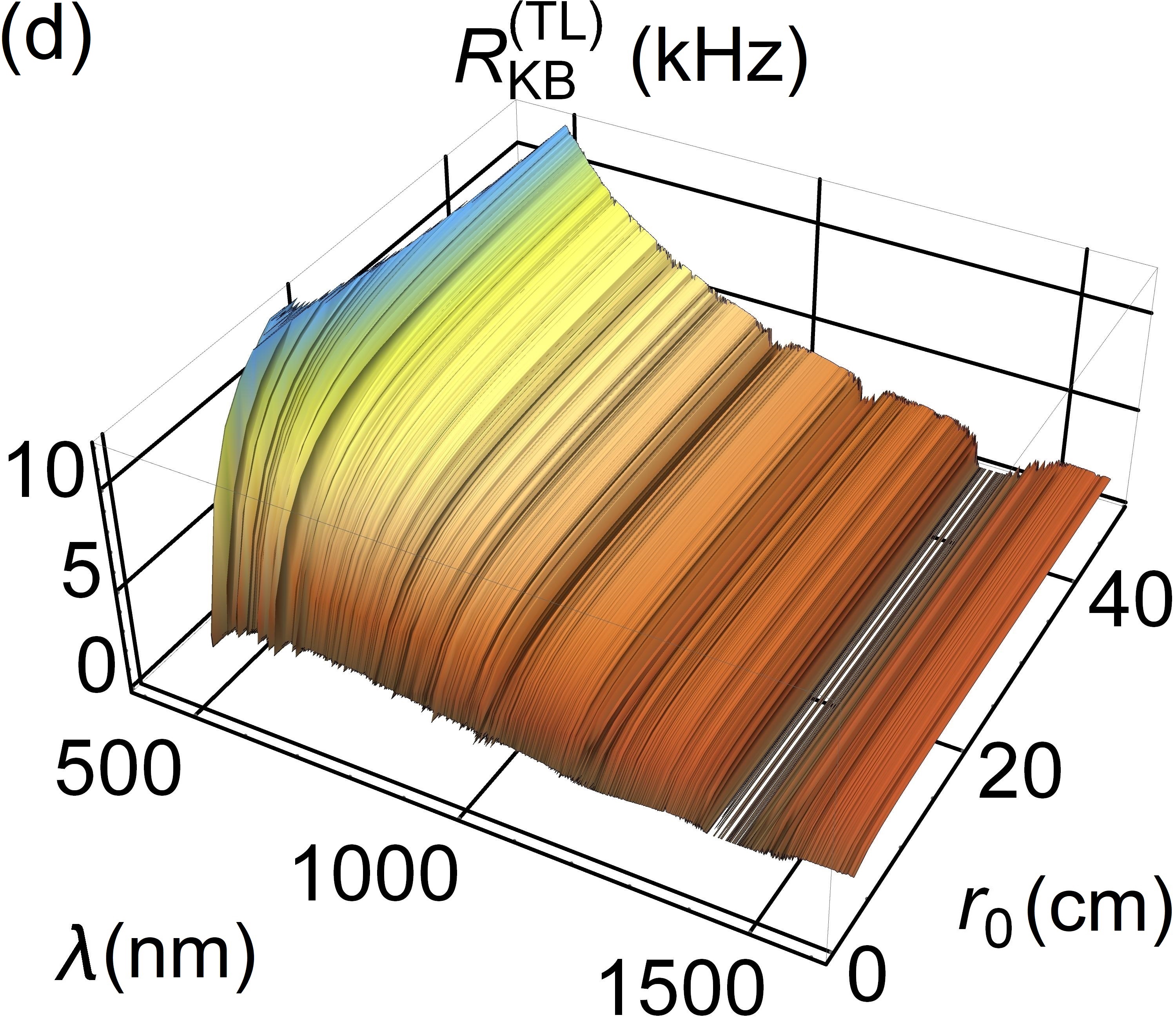} \\
  \end{tabular}
 \begin{tabular}{@{}p{0.95\linewidth}}
    \includegraphics[width=\linewidth]{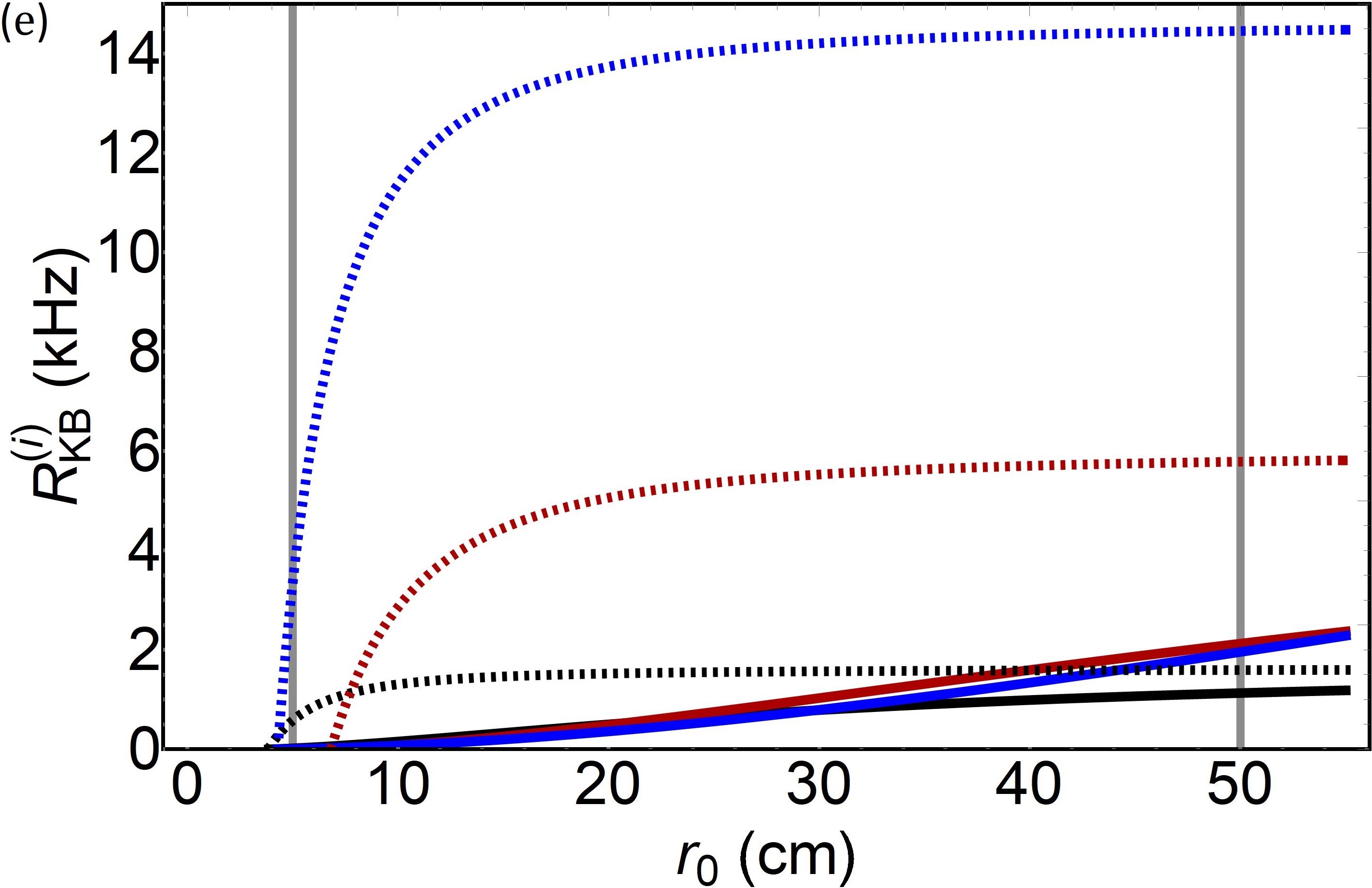} 
  \end{tabular}
 \begin{tabular}{@{}p{0.95\linewidth}}
    \includegraphics[width=\linewidth]{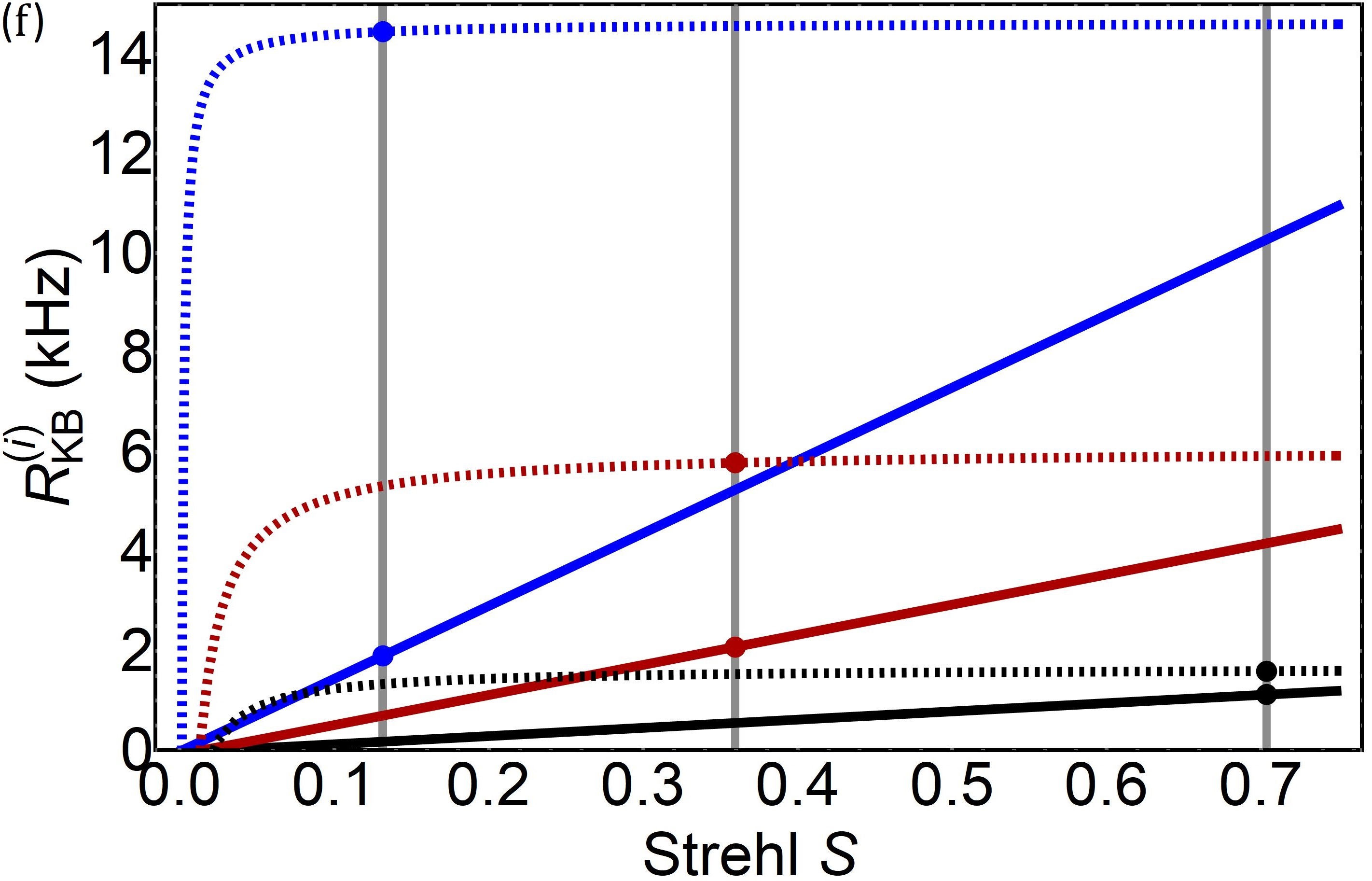} 
  \end{tabular}
  \caption{
The (a-b) atmospheric transmission $\eta_{\mathrm{trans}}$ and spectral radiance $H_{\mathrm{b}}$, and (c-d) the key-bit rate $R_{\mathrm{KB}}^{(i)}$ for summer solstice with high visibility (50-km), plotted as a function of $r_0$. In this case we have narrowed the spectral filter width to $\Delta \lambda = 0.05$-nm. In (e-f) we have chosen 1550, 781, and 405 nm (black, red, and blue respectively), and the solid and dashed curves indicate the DL and TL strategies, respectively. In (e) the vertical line at 5 cm corresponds to no AO correction and the line at 50 cm corresponds to the effective $r_0$ of a $f_c =200$-Hz AO system. In (f) the $\lambda$ dependence of $S$ produces three vertical lines indicating AO correction.
} \label{fig:SumSolZenHigh}
\end{figure}

\begin{figure}[t!]
  \centering
  \begin{tabular}{@{}p{0.475\linewidth}@{\quad}p{0.475\linewidth}@{}}
    \includegraphics[width=\linewidth]{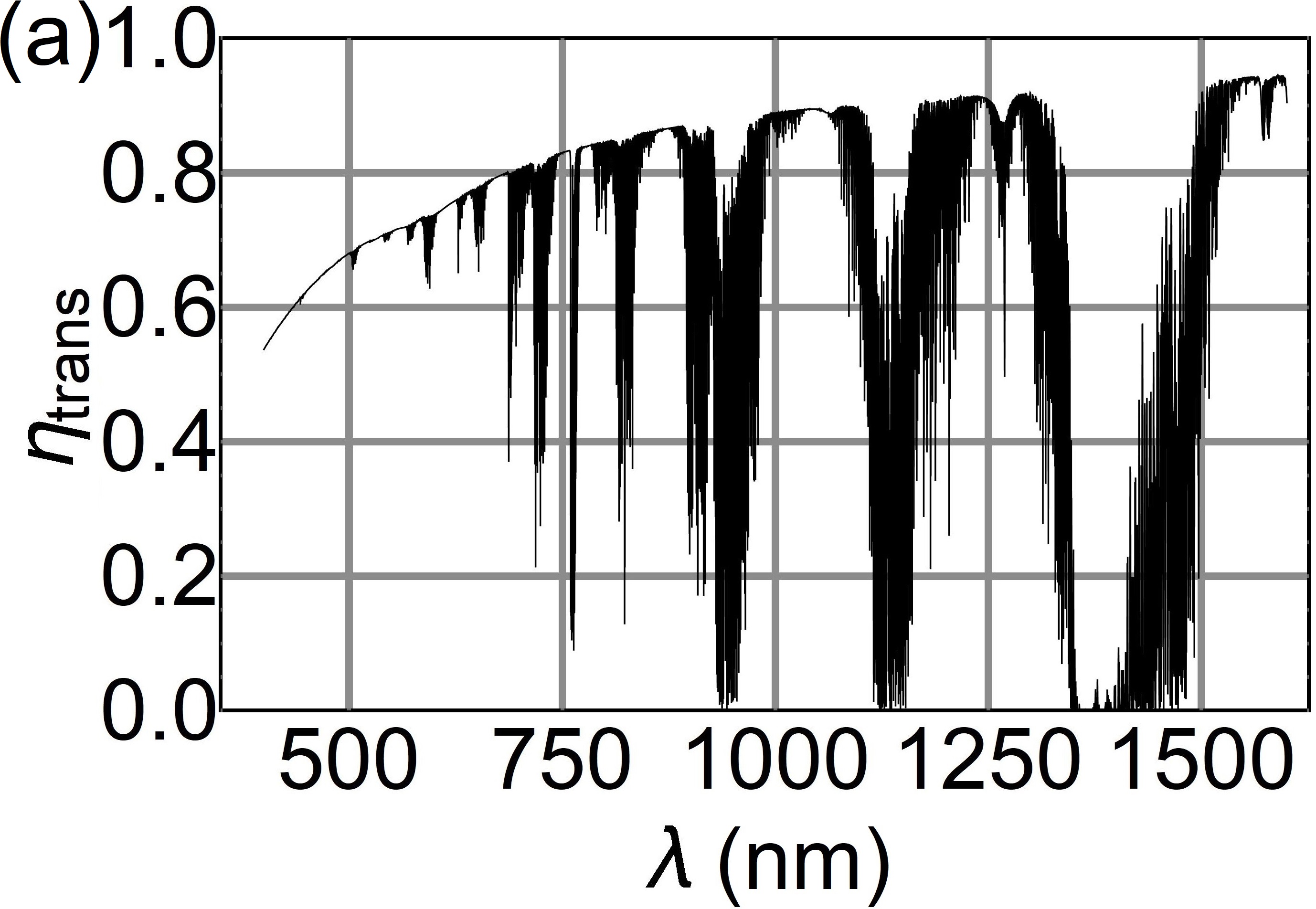} &
    \includegraphics[width=\linewidth]{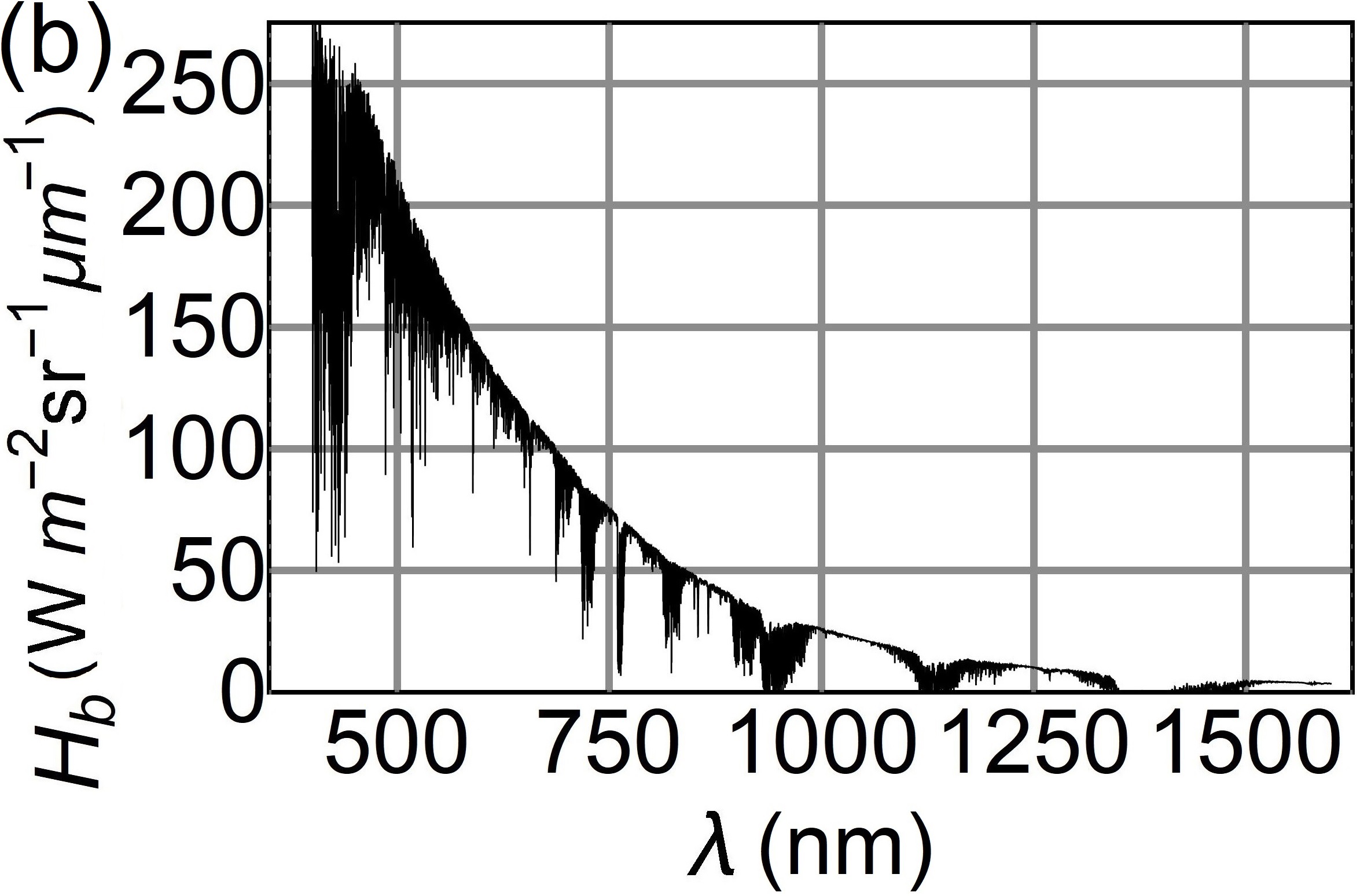} \\
    \includegraphics[width=\linewidth]{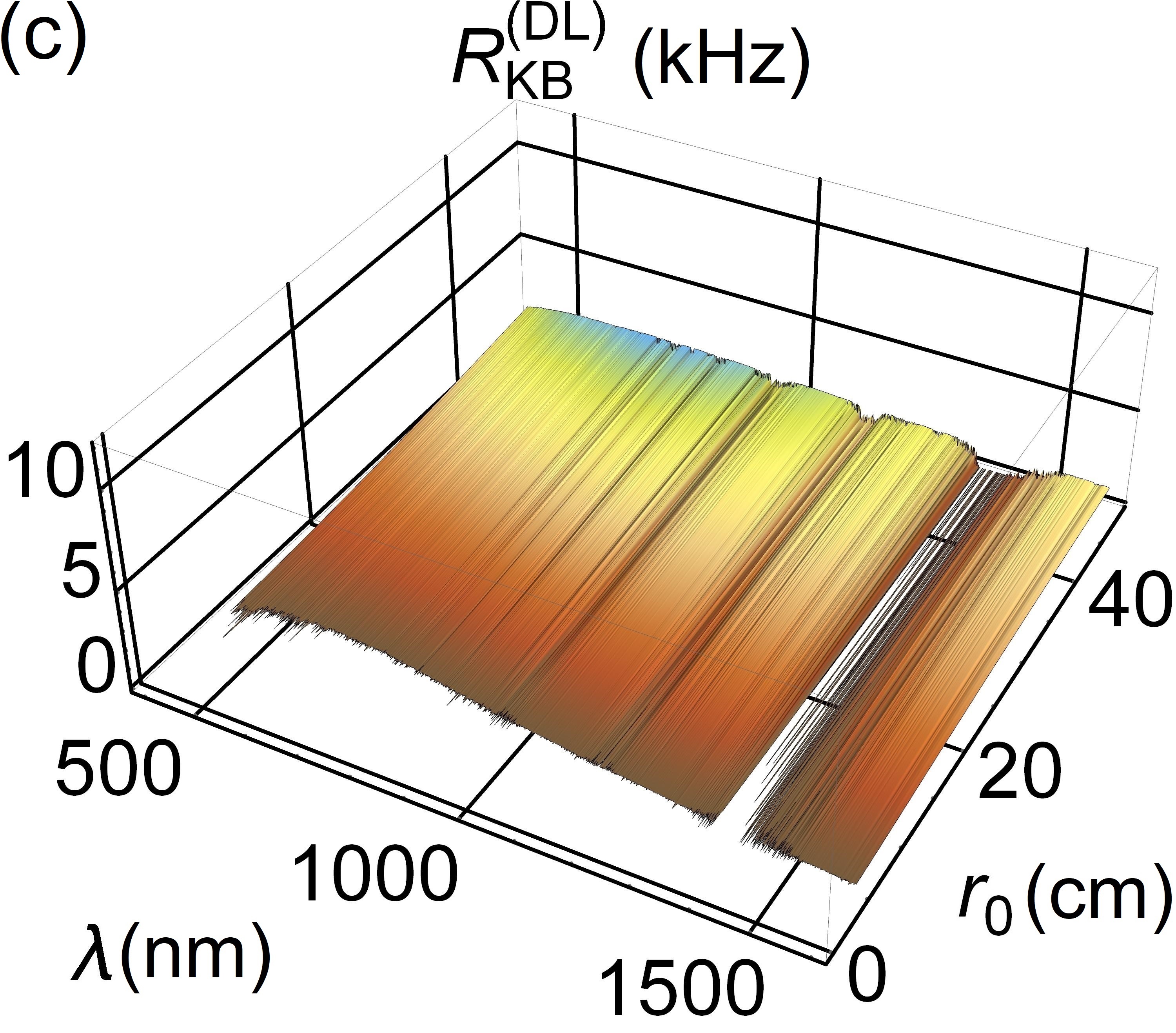} &
    \includegraphics[width=\linewidth]{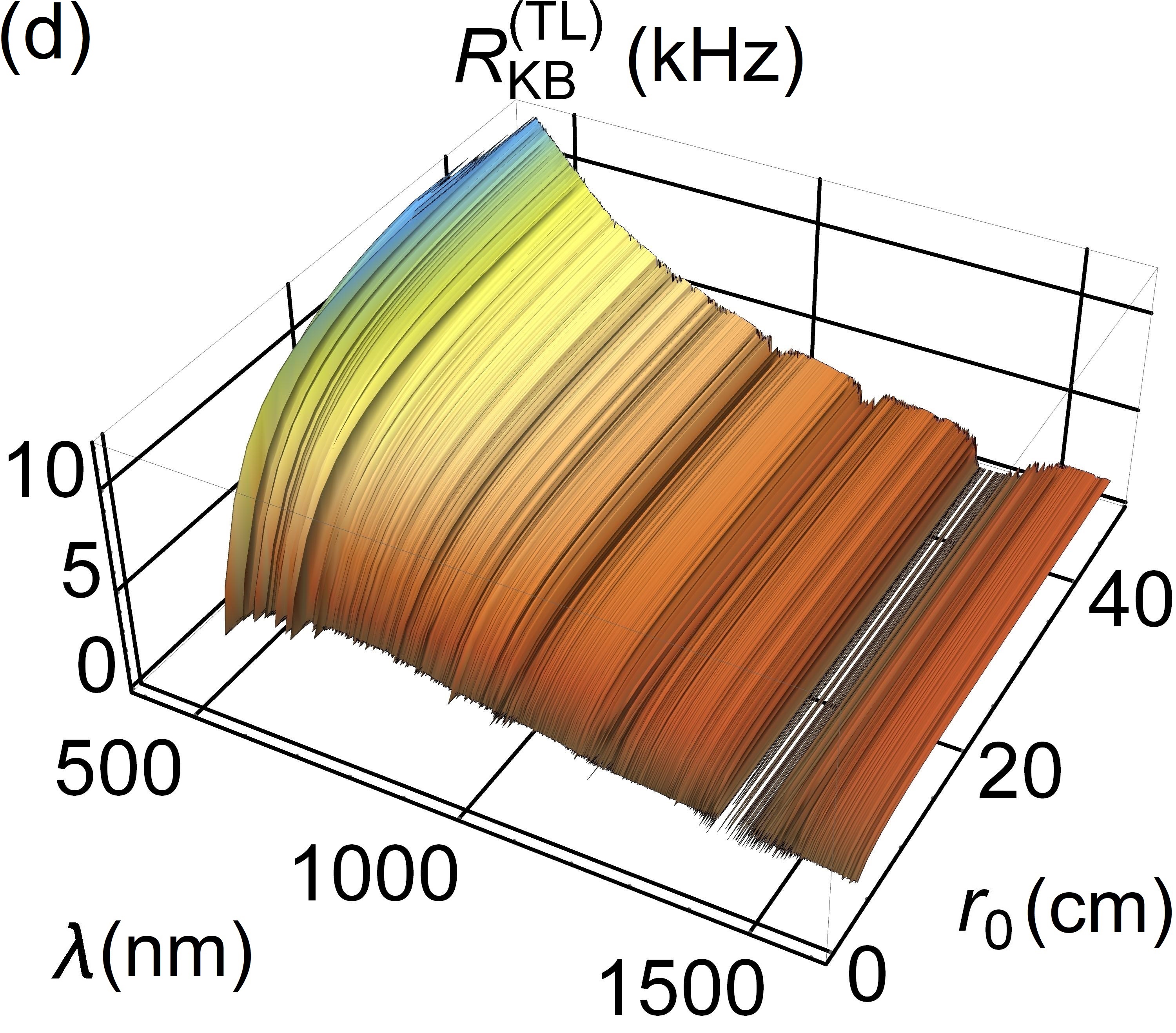} \\
  \end{tabular}
 \begin{tabular}{@{}p{0.95\linewidth}}
    \includegraphics[width=\linewidth]{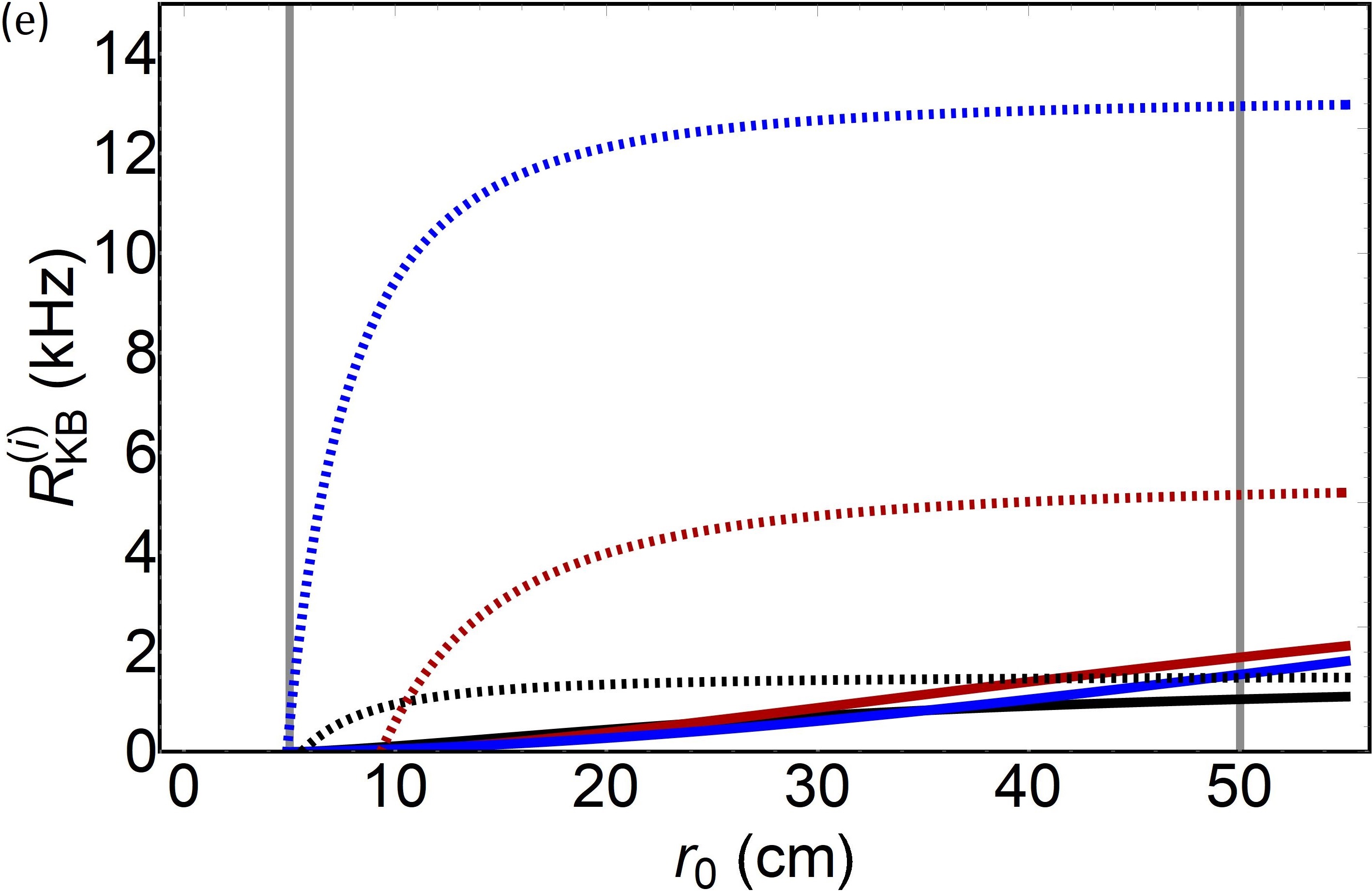} 
  \end{tabular}
 \begin{tabular}{@{}p{0.95\linewidth}}
    \includegraphics[width=\linewidth]{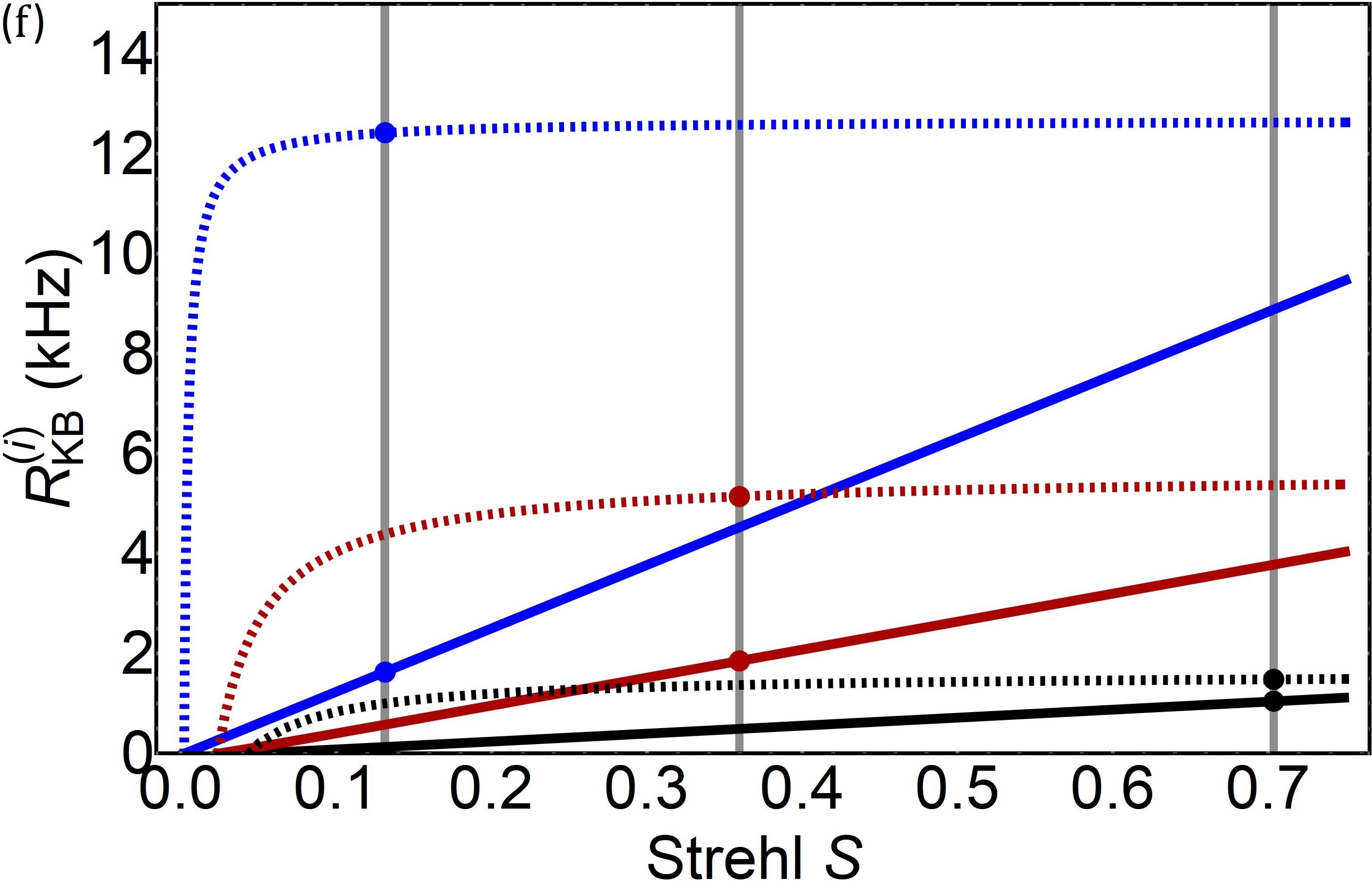} 
  \end{tabular}
  \caption{
The (a-b) atmospheric transmission $\eta_{\mathrm{trans}}$ and spectral radiance $H_{\mathrm{b}}$, and (c-d) the key-bit rate $R_{\mathrm{KB}}^{(i)}$ for summer solstice with medium visibility (23-km), plotted as a function of $r_0$. In this case we have narrowed the spectral filter to $\Delta \lambda = 0.05$-nm. In (e-f) we have chosen 1550, 781, and 405 nm (black, red, and blue respectively), and the solid and dashed curves indicate the DL and TL strategies, respectively. In (e) the vertical line at 5 cm corresponds to no AO correction and the line at 50 cm corresponds to the effective $r_0$ of a $f_c =200$-Hz AO system. In (f) the $\lambda$ dependence of $S$ produces three vertical lines indicating AO correction.   
} \label{fig:SumSolZenMed}
\end{figure}

\pagebreak
\clearpage

\newpage

\noindent would enable relatively high performance at shorter wavelengths.
Comparing the two conditions, we see that the grouping at low Strehl in Fig.~\ref{fig:WinSolZenMed}(f) is replaced with a more dispersed trend in Fig.~\ref{fig:WinSolZenLow}(f).
This is a result of the relative sky radiance at the two visibility conditions. 
For example, for medium visibility, 431 nm is $\sim$40$\times$ brighter than 1550 nm, but only $\sim$15$\times$ brighter for low visibility.
When considering the actual \textit{number} of photons, using the DL spatial filtering strategy and a 1-nm spectral filter, one finds that there are actually $\sim$1.2$\times$ \textit{more} 1550-nm photons at medium visibility and $\sim$3$\times$ more at low visibility.
In effect, the breakout in the curve is caused by the relatively larger QBER's at the longer wavelengths. 
This can be traced back to the relationship between the FOV's of the systems with different wavelengths.
The sky may be dimmer at 1550 nm, but a system at that wavelength is more susceptible to that noise as compared to shorter wavelengths.
Therefore, this reveals further robustness of the short wavelength strategy under lower visibility conditions where the relative sky brightness increases for longer wavelengths.

\subsection{Summer Solstice}
For the summer solstice, since the sun is approximately 15 degrees from zenith, we will only consider high (50-km) and medium (23-km) visibilities.
Thus far, we have assumed a relatively large 1-nm spectral filter for our down-link architecture.
This emphasizes the robustness of particular wavelength and strategy choices, and also suggests the level of performance possible when using entangled photon sources with comparable bandwidth.
However, in this case the spectral radiance is significantly larger due to the channel proximity to the sun angle and we would like to demonstrate the level of filtering necessary to generate key bits without the aid of AO.
Therefore, we narrow the spectral filter to $\Delta \lambda = 0.05$ nm for both visibility conditions.
In practice, one should use the most aggressive filtering possible and be careful to examine how the choice of filter affects the number of background photons within that spectral window $N_{\mathrm{b}}(\lambda_0,\, \Delta \lambda)$ (see Eq.~\ref{eq:Nb}).
In effect, the choice of filter and subsequent number of background photons can change the optimal wavelength for key generation.
For example, reducing the filter from 1 nm to 0.05 nm takes advantage of a narrow dip in spectral radiance and causes the optimal wavelength to shift from  431 nm to near 405 nm (see App.~\ref{sec:appendixB}).
Therefore, in this subsection the blue curves correspond to 405 nm.

Figures~\ref{fig:SumSolZenHigh}(a-b) and \ref{fig:SumSolZenMed}(a-b) show the atmospheric transmission and spectral radiance for high and medium visibility, respectively.
In Fig.~\ref{fig:SumSolZenHigh}(c-e) and \ref{fig:SumSolZenMed}(c-e) we plot the $r_0$ dependence of $R_{\mathrm{KB}}^{(i)}$ for the two summer solstice visibility conditions.
One will see that the spectral radiances are considerably higher as compared to the winter solstice condition.
One will also notice the relative high performance of shorter wavelengths and the TL strategy, which persists even for more challenging atmospheric conditions.
Interestingly, the tighter filtering permits key-bit yield even at $r_0\approx 5$ cm with both 1550  and 405 nm for the high-visibility condition.
In Fig.~\ref{fig:SumSolZenHigh}(f) and \ref{fig:SumSolZenMed}(f) we plot $R_{\mathrm{KB}}^{(i)}$ as a  function of system Strehl. 
Tighter spectral filtering reveals the trend in Strehl even more clearly, that is, short wavelengths with the TL strategy permit very high performance even with relatively low Strehl.  
Therefore, our simulation demonstrates how one can choose an optimal wavelength and filtering strategy for even the most challenging down-link conditions.

\section{Conclusion}

In this article we investigate the optimal wavelength for free-space quantum communication over space-Earth quantum channels, particularly in daytime conditions where filtering sky-noise photons is a formidable challenge. 
Ultimately, the performance of a free-space quantum-communication system depends on the amount of signal and noise passing through the optical receiver spatial filter. 
However, the performance is an effect of quantum phenomenon and a simple analysis can be quite misleading. 
Therefore, we integrate the physics of focusing in the presence of atmospheric turbulence with the decoy-state BB84-QKD protocol, thus establishing an actual quantum-performance metric.
We carefully examine the wavelength and Fried spatial-coherence length $r_0$ dependence of each component of the protocol.  
Namely, we investigate the dependence on the optical receiver field of view, the resulting number of sky-noise photons, the quantum bit error rate, and the signal-to-noise ratio.  

Ultimately, we derive the QKD bit yield as a function of wavelength and spatial coherence $r_0$, and investigate two different spatial-filtering strategies.
Although the quantum bit error rate and signal-to-noise ratio do not change considerably with the two strategies, one strategy has a clear advantage due to the boost in signal.
We show that, in general, shorter wavelengths outperform longer wavelengths for a wide range of channel conditions.
For our site condition, there is a relatively large dip in spectral radiance $H_{\mathrm{b}}$ near 431 nm that permits the highest QKD-system performance over the entire visible spectrum and into the telecom band when using a 1-nm spectral filter.
This persists for several channel conditions representative of a space-Earth down-link architecture, namely winter and summer solstices with visibilities ranging from 50 to 5 km.
We show how aggressive spectral filtering can permit high performance under challenging channel conditions, but a trade study is necessary in the case of attenuation by a spectral filter narrower than the bandwidth of the signal photons.

We also cast the optimization problem in terms of higher-order AO, which allows one to correct for atmospheric turbulence, in effect operating their optical receiver closer to the diffraction limit.
This allows one to use very tight spatial filtering and relax other filtering requirements if necessary.
For example, in order to accommodate an entanglement based protocol with broader-band photons. 
Adaptive optics systems for shorter wavelengths are in general more difficult to construct due to, for example, requiring more wavefront sensor sub-apertures due to the wavelength dependence of the Fried coherence length $r(\lambda)$.
However, our results show that even a relatively low performance AO system, for example an AO corrected Strehl of $\sim$0.1 at a short wavelength, can provide significant performance benefit relative to longer wavelengths operating near the diffraction limit.
We contend that in the context of the global-scale quantum internet mediated by free-space links, these engineering concerns pose relatively straightforward problems and warrant further investigation and investment.

For space-based networks where there are no atmospheric effects, the optical receivers would naturally operate near the diffraction limit, and shorter wavelengths have the clear advantage due to the increased geometric aperture-to-aperture coupling.
To complete a system design of a global-scale quantum network, an angle-dependent study including the full effects on the Fried coherence and the Greenwood frequencies should be conducted for cases with stronger turbulence and large Zenith angles.
Lastly, the corresponding analysis for the up-link scenario should include the spectral radiance due to Earth shine and turbulence-induced beam spreading.

\begin{acknowledgments}
This work was supported by the Office of the Secretary of Defense (OSD) ARAP Defense Optical Channel Program (DOC-P).
\end{acknowledgments}

\bibliography{bibliography/bibliography}

\begin{thebibliography}{34}
\expandafter\ifx\csname natexlab\endcsname\relax\def\natexlab#1{#1}\fi
\expandafter\ifx\csname bibnamefont\endcsname\relax
  \def\bibnamefont#1{#1}\fi
\expandafter\ifx\csname bibfnamefont\endcsname\relax
  \def\bibfnamefont#1{#1}\fi
\expandafter\ifx\csname citenamefont\endcsname\relax
  \def\citenamefont#1{#1}\fi
\expandafter\ifx\csname url\endcsname\relax
  \def\url#1{\texttt{#1}}\fi
\expandafter\ifx\csname urlprefix\endcsname\relax\def\urlprefix{URL }\fi
\providecommand{\bibinfo}[2]{#2}
\providecommand{\eprint}[2][]{\url{#2}}

\bibitem[{\citenamefont{Friis}(1971)}]{friis1971introduction}
\bibinfo{author}{\bibfnamefont{H.~T.} \bibnamefont{Friis}},
  \bibinfo{journal}{IEEE spectrum} \textbf{\bibinfo{volume}{8}},
  \bibinfo{pages}{55} (\bibinfo{year}{1971}).

\bibitem[{\citenamefont{Alexander}(1997)}]{alexander1997optical}
\bibinfo{author}{\bibfnamefont{S.~B.} \bibnamefont{Alexander}},
  \emph{\bibinfo{title}{Optical communication receiver design}},
  \bibinfo{number}{37} (\bibinfo{publisher}{IET}, \bibinfo{year}{1997}).

\bibitem[{\citenamefont{Van~Meter}(2014)}]{van2014quantum}
\bibinfo{author}{\bibfnamefont{R.}~\bibnamefont{Van~Meter}},
  \emph{\bibinfo{title}{Quantum networking}} (\bibinfo{publisher}{John Wiley \&
  Sons}, \bibinfo{year}{2014}).

\bibitem[{\citenamefont{Boone et~al.}(2015)\citenamefont{Boone, Bourgoin,
  Meyer-Scott, Heshami, Jennewein, and Simon}}]{boone2015entanglement}
\bibinfo{author}{\bibfnamefont{K.}~\bibnamefont{Boone}},
  \bibinfo{author}{\bibfnamefont{J.-P.} \bibnamefont{Bourgoin}},
  \bibinfo{author}{\bibfnamefont{E.}~\bibnamefont{Meyer-Scott}},
  \bibinfo{author}{\bibfnamefont{K.}~\bibnamefont{Heshami}},
  \bibinfo{author}{\bibfnamefont{T.}~\bibnamefont{Jennewein}},
  \bibnamefont{and} \bibinfo{author}{\bibfnamefont{C.}~\bibnamefont{Simon}},
  \bibinfo{journal}{Physical Review A} \textbf{\bibinfo{volume}{91}},
  \bibinfo{pages}{052325} (\bibinfo{year}{2015}).

\bibitem[{\citenamefont{Wehner et~al.}(2018)\citenamefont{Wehner, Elkouss, and
  Hanson}}]{wehner2018quantum}
\bibinfo{author}{\bibfnamefont{S.}~\bibnamefont{Wehner}},
  \bibinfo{author}{\bibfnamefont{D.}~\bibnamefont{Elkouss}}, \bibnamefont{and}
  \bibinfo{author}{\bibfnamefont{R.}~\bibnamefont{Hanson}},
  \bibinfo{journal}{Science} \textbf{\bibinfo{volume}{362}}
  (\bibinfo{year}{2018}).

\bibitem[{\citenamefont{Jacobs and Franson}(1996)}]{jacobs1996quantum}
\bibinfo{author}{\bibfnamefont{B.}~\bibnamefont{Jacobs}} \bibnamefont{and}
  \bibinfo{author}{\bibfnamefont{J.}~\bibnamefont{Franson}},
  \bibinfo{journal}{Optics Letters} \textbf{\bibinfo{volume}{21}},
  \bibinfo{pages}{1854} (\bibinfo{year}{1996}).

\bibitem[{\citenamefont{Buttler et~al.}(2000)\citenamefont{Buttler, Hughes,
  Lamoreaux, Morgan, Nordholt, and Peterson}}]{buttler2000daylight}
\bibinfo{author}{\bibfnamefont{W.~T.} \bibnamefont{Buttler}},
  \bibinfo{author}{\bibfnamefont{R.~J.} \bibnamefont{Hughes}},
  \bibinfo{author}{\bibfnamefont{S.~K.} \bibnamefont{Lamoreaux}},
  \bibinfo{author}{\bibfnamefont{G.~L.} \bibnamefont{Morgan}},
  \bibinfo{author}{\bibfnamefont{J.~E.} \bibnamefont{Nordholt}},
  \bibnamefont{and} \bibinfo{author}{\bibfnamefont{C.~G.}
  \bibnamefont{Peterson}}, \bibinfo{journal}{Physical Review Letters}
  \textbf{\bibinfo{volume}{84}}, \bibinfo{pages}{5652} (\bibinfo{year}{2000}).

\bibitem[{\citenamefont{Hughes et~al.}(2002)\citenamefont{Hughes, Nordholt,
  Derkacs, and Peterson}}]{hughes2002practical}
\bibinfo{author}{\bibfnamefont{R.~J.} \bibnamefont{Hughes}},
  \bibinfo{author}{\bibfnamefont{J.~E.} \bibnamefont{Nordholt}},
  \bibinfo{author}{\bibfnamefont{D.}~\bibnamefont{Derkacs}}, \bibnamefont{and}
  \bibinfo{author}{\bibfnamefont{C.~G.} \bibnamefont{Peterson}},
  \bibinfo{journal}{New journal of physics} \textbf{\bibinfo{volume}{4}},
  \bibinfo{pages}{43} (\bibinfo{year}{2002}).

\bibitem[{\citenamefont{Shan et~al.}(2006)\citenamefont{Shan, Sun, Luo, Tan,
  and Zhan}}]{shan2006free}
\bibinfo{author}{\bibfnamefont{X.}~\bibnamefont{Shan}},
  \bibinfo{author}{\bibfnamefont{X.}~\bibnamefont{Sun}},
  \bibinfo{author}{\bibfnamefont{J.}~\bibnamefont{Luo}},
  \bibinfo{author}{\bibfnamefont{Z.}~\bibnamefont{Tan}}, \bibnamefont{and}
  \bibinfo{author}{\bibfnamefont{M.}~\bibnamefont{Zhan}},
  \bibinfo{journal}{Applied physics letters} \textbf{\bibinfo{volume}{89}},
  \bibinfo{pages}{191121} (\bibinfo{year}{2006}).

\bibitem[{\citenamefont{Peloso et~al.}(2009)\citenamefont{Peloso, Gerhardt, Ho,
  Lamas-Linares, and Kurtsiefer}}]{peloso2009daylight}
\bibinfo{author}{\bibfnamefont{M.~P.} \bibnamefont{Peloso}},
  \bibinfo{author}{\bibfnamefont{I.}~\bibnamefont{Gerhardt}},
  \bibinfo{author}{\bibfnamefont{C.}~\bibnamefont{Ho}},
  \bibinfo{author}{\bibfnamefont{A.}~\bibnamefont{Lamas-Linares}},
  \bibnamefont{and}
  \bibinfo{author}{\bibfnamefont{C.}~\bibnamefont{Kurtsiefer}},
  \bibinfo{journal}{New Journal of Physics} \textbf{\bibinfo{volume}{11}},
  \bibinfo{pages}{045007} (\bibinfo{year}{2009}).

\bibitem[{\citenamefont{Heim et~al.}(2010)\citenamefont{Heim, Elser, Bartley,
  Sabuncu, Wittmann, Sych, Marquardt, and Leuchs}}]{heim2010atmospheric}
\bibinfo{author}{\bibfnamefont{B.}~\bibnamefont{Heim}},
  \bibinfo{author}{\bibfnamefont{D.}~\bibnamefont{Elser}},
  \bibinfo{author}{\bibfnamefont{T.}~\bibnamefont{Bartley}},
  \bibinfo{author}{\bibfnamefont{M.}~\bibnamefont{Sabuncu}},
  \bibinfo{author}{\bibfnamefont{C.}~\bibnamefont{Wittmann}},
  \bibinfo{author}{\bibfnamefont{D.}~\bibnamefont{Sych}},
  \bibinfo{author}{\bibfnamefont{C.}~\bibnamefont{Marquardt}},
  \bibnamefont{and} \bibinfo{author}{\bibfnamefont{G.}~\bibnamefont{Leuchs}},
  \bibinfo{journal}{Applied Physics B} \textbf{\bibinfo{volume}{98}},
  \bibinfo{pages}{635} (\bibinfo{year}{2010}).

\bibitem[{\citenamefont{Garc{\'\i}a-Mart{\'\i}nez
  et~al.}(2013)\citenamefont{Garc{\'\i}a-Mart{\'\i}nez, Denisenko, Soto,
  Arroyo, Orue, and Fernandez}}]{garcia2013high}
\bibinfo{author}{\bibfnamefont{M.}~\bibnamefont{Garc{\'\i}a-Mart{\'\i}nez}},
  \bibinfo{author}{\bibfnamefont{N.}~\bibnamefont{Denisenko}},
  \bibinfo{author}{\bibfnamefont{D.}~\bibnamefont{Soto}},
  \bibinfo{author}{\bibfnamefont{D.}~\bibnamefont{Arroyo}},
  \bibinfo{author}{\bibfnamefont{A.}~\bibnamefont{Orue}}, \bibnamefont{and}
  \bibinfo{author}{\bibfnamefont{V.}~\bibnamefont{Fernandez}},
  \bibinfo{journal}{Applied optics} \textbf{\bibinfo{volume}{52}},
  \bibinfo{pages}{3311} (\bibinfo{year}{2013}).

\bibitem[{\citenamefont{Carrasco-Casado
  et~al.}(2014)\citenamefont{Carrasco-Casado, Denisenko, and
  Fernandez}}]{carrasco2014correction}
\bibinfo{author}{\bibfnamefont{A.}~\bibnamefont{Carrasco-Casado}},
  \bibinfo{author}{\bibfnamefont{N.}~\bibnamefont{Denisenko}},
  \bibnamefont{and}
  \bibinfo{author}{\bibfnamefont{V.}~\bibnamefont{Fernandez}},
  \bibinfo{journal}{Optical Engineering} \textbf{\bibinfo{volume}{53}},
  \bibinfo{pages}{084112} (\bibinfo{year}{2014}).

\bibitem[{\citenamefont{Gruneisen et~al.}(2015)\citenamefont{Gruneisen,
  Flanagan, Sickmiller, Black, Stoltenberg, and
  Duchane}}]{gruneisen2015modeling}
\bibinfo{author}{\bibfnamefont{M.~T.} \bibnamefont{Gruneisen}},
  \bibinfo{author}{\bibfnamefont{M.~B.} \bibnamefont{Flanagan}},
  \bibinfo{author}{\bibfnamefont{B.~A.} \bibnamefont{Sickmiller}},
  \bibinfo{author}{\bibfnamefont{J.~P.} \bibnamefont{Black}},
  \bibinfo{author}{\bibfnamefont{K.~E.} \bibnamefont{Stoltenberg}},
  \bibnamefont{and} \bibinfo{author}{\bibfnamefont{A.~W.}
  \bibnamefont{Duchane}}, \bibinfo{journal}{Optics Express}
  \textbf{\bibinfo{volume}{23}}, \bibinfo{pages}{23924} (\bibinfo{year}{2015}).

\bibitem[{\citenamefont{Gruneisen et~al.}(2016)\citenamefont{Gruneisen,
  Sickmiller, Flanagan, Black, Stoltenberg, and
  Duchane}}]{gruneisen2016adaptive}
\bibinfo{author}{\bibfnamefont{M.~T.} \bibnamefont{Gruneisen}},
  \bibinfo{author}{\bibfnamefont{B.~A.} \bibnamefont{Sickmiller}},
  \bibinfo{author}{\bibfnamefont{M.~B.} \bibnamefont{Flanagan}},
  \bibinfo{author}{\bibfnamefont{J.~P.} \bibnamefont{Black}},
  \bibinfo{author}{\bibfnamefont{K.~E.} \bibnamefont{Stoltenberg}},
  \bibnamefont{and} \bibinfo{author}{\bibfnamefont{A.~W.}
  \bibnamefont{Duchane}}, \bibinfo{journal}{Optical Engineering}
  \textbf{\bibinfo{volume}{55}}, \bibinfo{pages}{026104}
  (\bibinfo{year}{2016}).

\bibitem[{\citenamefont{Gruneisen et~al.}(2017)\citenamefont{Gruneisen,
  Flanagan, and Sickmiller}}]{gruneisen2017modeling}
\bibinfo{author}{\bibfnamefont{M.~T.} \bibnamefont{Gruneisen}},
  \bibinfo{author}{\bibfnamefont{M.~B.} \bibnamefont{Flanagan}},
  \bibnamefont{and} \bibinfo{author}{\bibfnamefont{B.~A.}
  \bibnamefont{Sickmiller}}, \bibinfo{journal}{Optical Engineering}
  \textbf{\bibinfo{volume}{56}}, \bibinfo{pages}{126111}
  (\bibinfo{year}{2017}).

\bibitem[{\citenamefont{Liao et~al.}(2017)\citenamefont{Liao, Yong, Liu,
  Shentu, Li, Lin, Dai, Zhao, Li, Guan et~al.}}]{liao2017long}
\bibinfo{author}{\bibfnamefont{S.-K.} \bibnamefont{Liao}},
  \bibinfo{author}{\bibfnamefont{H.-L.} \bibnamefont{Yong}},
  \bibinfo{author}{\bibfnamefont{C.}~\bibnamefont{Liu}},
  \bibinfo{author}{\bibfnamefont{G.-L.} \bibnamefont{Shentu}},
  \bibinfo{author}{\bibfnamefont{D.-D.} \bibnamefont{Li}},
  \bibinfo{author}{\bibfnamefont{J.}~\bibnamefont{Lin}},
  \bibinfo{author}{\bibfnamefont{H.}~\bibnamefont{Dai}},
  \bibinfo{author}{\bibfnamefont{S.-Q.} \bibnamefont{Zhao}},
  \bibinfo{author}{\bibfnamefont{B.}~\bibnamefont{Li}},
  \bibinfo{author}{\bibfnamefont{J.-Y.} \bibnamefont{Guan}},
  \bibnamefont{et~al.}, \bibinfo{journal}{Nature Photonics}
  \textbf{\bibinfo{volume}{11}}, \bibinfo{pages}{509} (\bibinfo{year}{2017}).

\bibitem[{\citenamefont{Vasylyev et~al.}(2017)\citenamefont{Vasylyev, Semenov,
  Vogel, G{\"u}nthner, Thurn, Bayraktar, and Marquardt}}]{vasylyev2017free}
\bibinfo{author}{\bibfnamefont{D.}~\bibnamefont{Vasylyev}},
  \bibinfo{author}{\bibfnamefont{A.}~\bibnamefont{Semenov}},
  \bibinfo{author}{\bibfnamefont{W.}~\bibnamefont{Vogel}},
  \bibinfo{author}{\bibfnamefont{K.}~\bibnamefont{G{\"u}nthner}},
  \bibinfo{author}{\bibfnamefont{A.}~\bibnamefont{Thurn}},
  \bibinfo{author}{\bibfnamefont{{\"O}.}~\bibnamefont{Bayraktar}},
  \bibnamefont{and}
  \bibinfo{author}{\bibfnamefont{C.}~\bibnamefont{Marquardt}},
  \bibinfo{journal}{Physical Review A} \textbf{\bibinfo{volume}{96}},
  \bibinfo{pages}{043856} (\bibinfo{year}{2017}).

\bibitem[{\citenamefont{Arteaga-D{\'\i}az
  et~al.}(2019)\citenamefont{Arteaga-D{\'\i}az, Ocampos-Guill{\'e}n, and
  Fernandez}}]{arteaga2019enabling}
\bibinfo{author}{\bibfnamefont{P.}~\bibnamefont{Arteaga-D{\'\i}az}},
  \bibinfo{author}{\bibfnamefont{A.}~\bibnamefont{Ocampos-Guill{\'e}n}},
  \bibnamefont{and}
  \bibinfo{author}{\bibfnamefont{V.}~\bibnamefont{Fernandez}}, in
  \emph{\bibinfo{booktitle}{2019 21st International Conference on Transparent
  Optical Networks (ICTON)}} (\bibinfo{organization}{IEEE},
  \bibinfo{year}{2019}), pp. \bibinfo{pages}{1--4}.

\bibitem[{\citenamefont{Gruneisen et~al.}(2020)\citenamefont{Gruneisen,
  Eickhoff, Newey, Stoltenberg, Morris, Bareian, Harris, Oesch, Oliker,
  Flanagan et~al.}}]{gruneisen2020adaptive}
\bibinfo{author}{\bibfnamefont{M.~T.} \bibnamefont{Gruneisen}},
  \bibinfo{author}{\bibfnamefont{M.~L.} \bibnamefont{Eickhoff}},
  \bibinfo{author}{\bibfnamefont{S.~C.} \bibnamefont{Newey}},
  \bibinfo{author}{\bibfnamefont{K.~E.} \bibnamefont{Stoltenberg}},
  \bibinfo{author}{\bibfnamefont{J.~F.} \bibnamefont{Morris}},
  \bibinfo{author}{\bibfnamefont{M.}~\bibnamefont{Bareian}},
  \bibinfo{author}{\bibfnamefont{M.~A.} \bibnamefont{Harris}},
  \bibinfo{author}{\bibfnamefont{D.~W.} \bibnamefont{Oesch}},
  \bibinfo{author}{\bibfnamefont{M.~D.} \bibnamefont{Oliker}},
  \bibinfo{author}{\bibfnamefont{M.~B.} \bibnamefont{Flanagan}},
  \bibnamefont{et~al.}, \bibinfo{journal}{arXiv preprint arXiv:2006.07745}
  (\bibinfo{year}{2020}).

\bibitem[{\citenamefont{Nordholt et~al.}(2002)\citenamefont{Nordholt, Hughes,
  Morgan, Peterson, and Wipf}}]{nordholt2002present}
\bibinfo{author}{\bibfnamefont{J.~E.} \bibnamefont{Nordholt}},
  \bibinfo{author}{\bibfnamefont{R.~J.} \bibnamefont{Hughes}},
  \bibinfo{author}{\bibfnamefont{G.~L.} \bibnamefont{Morgan}},
  \bibinfo{author}{\bibfnamefont{C.~G.} \bibnamefont{Peterson}},
  \bibnamefont{and} \bibinfo{author}{\bibfnamefont{C.~C.} \bibnamefont{Wipf}},
  in \emph{\bibinfo{booktitle}{Free-Space Laser Communication Technologies
  XIV}} (\bibinfo{organization}{International Society for Optics and
  Photonics}, \bibinfo{year}{2002}), vol. \bibinfo{volume}{4635}, pp.
  \bibinfo{pages}{116--126}.

\bibitem[{\citenamefont{Bourgoin et~al.}(2013)\citenamefont{Bourgoin,
  Meyer-Scott, Higgins, Helou, Erven, Huebel, Kumar, Hudson, D’Souza, Girard
  et~al.}}]{bourgoin2013comprehensive}
\bibinfo{author}{\bibfnamefont{J.}~\bibnamefont{Bourgoin}},
  \bibinfo{author}{\bibfnamefont{E.}~\bibnamefont{Meyer-Scott}},
  \bibinfo{author}{\bibfnamefont{B.~L.} \bibnamefont{Higgins}},
  \bibinfo{author}{\bibfnamefont{B.}~\bibnamefont{Helou}},
  \bibinfo{author}{\bibfnamefont{C.}~\bibnamefont{Erven}},
  \bibinfo{author}{\bibfnamefont{H.}~\bibnamefont{Huebel}},
  \bibinfo{author}{\bibfnamefont{B.}~\bibnamefont{Kumar}},
  \bibinfo{author}{\bibfnamefont{D.}~\bibnamefont{Hudson}},
  \bibinfo{author}{\bibfnamefont{I.}~\bibnamefont{D’Souza}},
  \bibinfo{author}{\bibfnamefont{R.}~\bibnamefont{Girard}},
  \bibnamefont{et~al.}, \bibinfo{journal}{New J. Phys}
  \textbf{\bibinfo{volume}{15}}, \bibinfo{pages}{023006}
  (\bibinfo{year}{2013}).

\bibitem[{\citenamefont{Bennett and Brassard}(1984)}]{bennett1984proc}
\bibinfo{author}{\bibfnamefont{C.}~\bibnamefont{Bennett}} \bibnamefont{and}
  \bibinfo{author}{\bibfnamefont{G.~i.} \bibnamefont{Brassard}},
  \emph{\bibinfo{title}{Proc. ieee int. conf. on computers, systems and signal
  processing, bangalore, india}} (\bibinfo{year}{1984}).

\bibitem[{\citenamefont{Bennett and Brassard}(2020)}]{bennett2020quantum}
\bibinfo{author}{\bibfnamefont{C.~H.} \bibnamefont{Bennett}} \bibnamefont{and}
  \bibinfo{author}{\bibfnamefont{G.}~\bibnamefont{Brassard}},
  \bibinfo{journal}{arXiv preprint arXiv:2003.06557}  (\bibinfo{year}{2020}).

\bibitem[{\citenamefont{Ma et~al.}(2005)\citenamefont{Ma, Qi, Zhao, and
  Lo}}]{ma2005practical}
\bibinfo{author}{\bibfnamefont{X.}~\bibnamefont{Ma}},
  \bibinfo{author}{\bibfnamefont{B.}~\bibnamefont{Qi}},
  \bibinfo{author}{\bibfnamefont{Y.}~\bibnamefont{Zhao}}, \bibnamefont{and}
  \bibinfo{author}{\bibfnamefont{H.-K.} \bibnamefont{Lo}},
  \bibinfo{journal}{Physical Review A} \textbf{\bibinfo{volume}{72}},
  \bibinfo{pages}{012326} (\bibinfo{year}{2005}).

\bibitem[{\citenamefont{Fried}(1966)}]{fried1966optical}
\bibinfo{author}{\bibfnamefont{D.~L.} \bibnamefont{Fried}},
  \bibinfo{journal}{JOSA} \textbf{\bibinfo{volume}{56}}, \bibinfo{pages}{1372}
  (\bibinfo{year}{1966}).

\bibitem[{\citenamefont{Shapiro}(2011)}]{shapiro2011scintillation}
\bibinfo{author}{\bibfnamefont{J.~H.} \bibnamefont{Shapiro}},
  \bibinfo{journal}{Physical Review A} \textbf{\bibinfo{volume}{84}},
  \bibinfo{pages}{032340} (\bibinfo{year}{2011}).

\bibitem[{\citenamefont{Tyson}(2015)}]{tyson2015principles}
\bibinfo{author}{\bibfnamefont{R.~K.} \bibnamefont{Tyson}},
  \emph{\bibinfo{title}{Principles of adaptive optics}}
  (\bibinfo{publisher}{CRC press}, \bibinfo{year}{2015}).

\bibitem[{\citenamefont{Sasiela}(2012)}]{sasiela2012electromagnetic}
\bibinfo{author}{\bibfnamefont{R.~J.} \bibnamefont{Sasiela}},
  \emph{\bibinfo{title}{Electromagnetic wave propagation in turbulence:
  evaluation and application of Mellin transforms}}, vol.~\bibinfo{volume}{18}
  (\bibinfo{publisher}{Springer Science \& Business Media},
  \bibinfo{year}{2012}).

\bibitem[{\citenamefont{Andrews}(2004)}]{andrews2004field}
\bibinfo{author}{\bibfnamefont{L.~C.} \bibnamefont{Andrews}},
  \emph{\bibinfo{title}{Field guide to atmospheric optics}}
  (\bibinfo{publisher}{SPIE Press,}, \bibinfo{year}{2004}).

\bibitem[{\citenamefont{Tomaello et~al.}(2011)\citenamefont{Tomaello,
  Dall'Arche, Naletto, and Villoresi}}]{tomaello2011intersatellite}
\bibinfo{author}{\bibfnamefont{A.}~\bibnamefont{Tomaello}},
  \bibinfo{author}{\bibfnamefont{A.}~\bibnamefont{Dall'Arche}},
  \bibinfo{author}{\bibfnamefont{G.}~\bibnamefont{Naletto}}, \bibnamefont{and}
  \bibinfo{author}{\bibfnamefont{P.}~\bibnamefont{Villoresi}}, in
  \emph{\bibinfo{booktitle}{Quantum Communications and Quantum Imaging IX}}
  (\bibinfo{organization}{International Society for Optics and Photonics},
  \bibinfo{year}{2011}), vol. \bibinfo{volume}{8163}, p.
  \bibinfo{pages}{816309}.

\bibitem[{\citenamefont{Bonato et~al.}(2009)\citenamefont{Bonato, Tomaello,
  Da~Deppo, Naletto, and Villoresi}}]{bonato2009feasibility}
\bibinfo{author}{\bibfnamefont{C.}~\bibnamefont{Bonato}},
  \bibinfo{author}{\bibfnamefont{A.}~\bibnamefont{Tomaello}},
  \bibinfo{author}{\bibfnamefont{V.}~\bibnamefont{Da~Deppo}},
  \bibinfo{author}{\bibfnamefont{G.}~\bibnamefont{Naletto}}, \bibnamefont{and}
  \bibinfo{author}{\bibfnamefont{P.}~\bibnamefont{Villoresi}},
  \bibinfo{journal}{NJP} \textbf{\bibinfo{volume}{11}}, \bibinfo{pages}{045017}
  (\bibinfo{year}{2009}).

\bibitem[{\citenamefont{Hardy}(1998)}]{hardy1998adaptive}
\bibinfo{author}{\bibfnamefont{J.~W.} \bibnamefont{Hardy}},
  \emph{\bibinfo{title}{Adaptive optics for astronomical telescopes}},
  vol.~\bibinfo{volume}{16} (\bibinfo{publisher}{Oxford University Press on
  Demand}, \bibinfo{year}{1998}).

\bibitem[{\citenamefont{Noll}(1976)}]{noll1976zernike}
\bibinfo{author}{\bibfnamefont{R.~J.} \bibnamefont{Noll}},
  \bibinfo{journal}{JOsA} \textbf{\bibinfo{volume}{66}}, \bibinfo{pages}{207}
  (\bibinfo{year}{1976}).

\end{thebibliography}

\pagebreak
\clearpage

\newpage 

\appendix
\section{Higher-Order AO Correction}\label{sec:appendixA}

\subsection{Residual Error}\label{sec:appendixARPE}
In an AO system, the aberrations of the incoming wavefront are corrected via a fast-steering mirror (FSM) and a deformable mirror (DM) imprinted with the conjugate of the higher-order wavefront error \cite{hardy1998adaptive, tyson2015principles}. 
The system is run in a closed-loop configuration in which a small portion of the light output from the DM is split off into a wavefront sensor which measures the residual phase error (RPE) of the wavefront, and is used to update the FSM and DM.
This makes the RPE perhaps the most fundamental parameter related to AO system performance.
For the open-loop case, the RPE variance in radians $\sigma_{\phi}^2$ depends primarily on the spatial coherence at the receiver aperture. Considering the tip, tilt, and higher-order aberrations one can write \cite{tyson2015principles, noll1976zernike}
\begin{equation}\label{eq:RPE}
\sigma_{\phi,\mathrm{OL}}^2 = 1.03 \Big( \dfrac{D_{\mathrm{R}}}{r(\lambda)} \Big) ^{5/3},
\end{equation}
where the OL subscript indicates open-loop operation and $r(\lambda)$ can be calculated according to \cite{tyson2015principles}
\begin{equation}\label{eq:rlambda}
\begin{split}
r(\lambda) = \Big[ 0.423 \, k^2 \, \mathrm{sec}(\theta_\mathrm{z}) \int_0^a dh\, C^2_n (h) \Big]^{-3/5},
\end{split}
\end{equation}
where $C^2_n (h)$ is the atmospheric turbulence structure parameter, $\theta_\mathrm{z}$ is the zenith angle, $a$ is the altitude of the light source, and $k=2\pi / \lambda$.
One should note that the wavelength dependence of Eq.~\ref{eq:rlambda} is the origin of the wavelength dependence of Eq.~\ref{eq:FRIED}, that is, $(k^2)^{-3/5} \propto (\lambda^{-2})^{-3/5} = \lambda^{6/5}$.
Furthermore, as specified in the main text, we assume $r_0$ is the value calculated at 500 nm and therefore $r_0 \equiv r(500\mbox{ }\mathrm{nm})$ using Eq.~\ref{eq:rlambda}.

A closed-loop AO system has many contributions to the RPE \cite{hardy1998adaptive}. 
However, for our down-link scenario with a bright cooperative AO beacon, the RPE is dominated by the systems ability to keep pace with the temporal 
fluctuations characterized by the tracking-Greenwood frequency \cite{tyson2015principles}
\begin{equation}\label{eq:fTG}
\begin{split}
f_{\mathrm{TG}} &=  5.268\times10^{-2} \, D_{\mathrm{R}}^{-1/6} \, k \\
&\times \Big[ \sec (\theta_\mathrm{z}) \int_0^a dh\, C^2_n (h) \, v^{2}_\mathrm{w} (h) \Big]^{1/2}
\end{split}
\end{equation}
and the higher-order atmospheric fluctuations characterized by the Greenwood frequency \cite{tyson2015principles}
\begin{equation}\label{eq:fG}
\begin{split}
f_\mathrm{G} &= \Big[ 0.1022 \, k^2 \, \mathrm{sec}(\theta_\mathrm{z}) \int_0^a dh\, C^2_n (h) \, v^{5/3}_\mathrm{w} (h) \Big]^{3/5},
\end{split}
\end{equation}
where $v_\mathrm{w}(h)$ is the altitude-dependent wind-velocity profile.
The total closed-loop RPE can be written in terms of the tracking closed-loop bandwidth $f_{tc}$ and the higher-order closed-loop bandwidth $f_c$ according to \cite{tyson2015principles, hardy1998adaptive}
\begin{equation}\label{eq:RPECL}
\sigma_{\phi,\mathrm{CL}}^2 = \Big(\dfrac{\pi}{2} \dfrac{f_\mathrm{TG}}{f_{tc}} \Big) ^{2} + \Big( \dfrac{f_\mathrm{G}}{f_c} \Big) ^{5/3}.
\end{equation}
This expression is independent of $r_0$ as long as ones wavefront sensor is able to sufficiently spatially resolve the wavefront error.
Strong intensity variations in the optical field can degrade the performance of an AO system based on certain wavefront sensors.  
Such scintillation effects are typically attributed to deep turbulence, large zenith angle, or horizontal propagation conditions, whereas this article considers slant-path turbulence and zenith angles where the scintillation effects are typically much less severe.

Both Eq.~\ref{eq:RPE} and Eq.~\ref{eq:RPECL} can be used to find the optical-path-difference (OPD) variance 
\begin{equation}\label{eq:RPEnm}
\sigma_{\mathrm{OPD}}^2 = \sigma_{\phi}^2 \Big( \dfrac{\lambda}{2\pi} \Big)^2.
\end{equation}
The OPD variance is a useful quantity because it is independent of wavelength.
Using Eqs.~\ref{eq:RPE}, \ref{eq:RPEnm}, and \ref{eq:FRIED} one can show that
\begin{equation}\label{eq:RPEnm2}
\sigma_{\mathrm{OPD,OL}}^2 = 1.03 \Big( \dfrac{\lambda_0}{2 \pi} \Big)^2  \Big( \dfrac{D_{\mathrm{R}}}{r_0} \Big)^{5/3},
\end{equation}
where $r_0$ is the Fried coherence length measured at $\lambda_0$=500 nm.
The open-loop OPD variance $\sigma_{\mathrm{OPD,OL}}^{2}$ is a property of the atmosphere that is directly measured by the wavefront sensor when the AO system is in the open-loop configuration, that is, a flat DM and no tip/tilt correction.
From Eq.~\ref{eq:RPEnm2} one can see that it can be used to infer $r_0$.
For the closed-loop case, we combine Eqs.~\ref{eq:fTG}, \ref{eq:fG}, \ref{eq:RPECL}, and \ref{eq:RPEnm} to find
\begin{equation}\label{eq:RPEnmCL}
\begin{split}
\sigma_{\mathrm{OPD,CL}}^2 &= 0.1022 \, f_c^{-5/3} \\
&\times \mathrm{sec}(\theta_\mathrm{z}) \int_0^a dh\, C^2_n (h) \, v^{5/3}_\mathrm{w} (h)\\
& + 2.775 \times 10^{-3} \Big( \dfrac{\pi}{2} \Big)^2 f_{tc}^{-2} D_{\mathrm{R}}^{-1/3} \\
&\times \sec (\theta_\mathrm{z}) \int_0^a dh\, C^2_n (h) \, v^{2}_\mathrm{w} (h).
\end{split}
\end{equation}
One can see that the closed-loop OPD variance $\sigma_{\mathrm{OPD,CL}}^{2}$ depends on the relationship between the closed-loop bandwidths and the temporal component of the turbulence characterized by the integrals.

\subsection{Effective Fried Coherence Length
and Closed-Loop Bandwidth}\label{sec:appendixAEffective}
We wish to interpret the $r_0$ dependence of the key-bit yield $R_{\mathrm{KB}}$ and each of the contributing phenomenon in terms of AO.
To do so, we would like to establish an effective $r_0$ corresponding to residual-turbulence effects after AO correction.
Hence, we will assume for the moment that the RPE variance for open- and closed-loop operation are equal.
This permits one to equate Eqs.~\ref{eq:RPE} and \ref{eq:RPECL} and use Eq.~\ref{eq:FRIED} to solve for $r_0$:
\begin{equation}\label{eq:r0CL}
\begin{split}
r_0^{(\mathrm{CL})} &= 1.03^{3/5} \, \Big( \dfrac{\lambda_0}{\lambda} \Big)^{6/5} D_{\mathrm{R}}\\
& \times \Bigg[ \Big( \dfrac{f_{\mathrm{G}}}{f_c} \Big)^{5/3} + \Big(  \dfrac{\pi}{2} \dfrac{f_{\mathrm{TG}}}{f_{tc}} \Big)^{2} \Bigg]^{-3/5},
\end{split}
\end{equation}
where the superscript CL indicates that this $r_0$ is the effective spatial coherence during closed-loop operation.
In other words, with AO, a given combination of $f_{\mathrm{G}}$, $f_c$, $f_{\mathrm{TG}}$, and $f_{tc}$ yields the same optical receiver performance that would be achieved without AO in an atmosphere described by $r_0^{(\mathrm{CL})}$.
Similarly, one can solve for $f_c$ and find 
\begin{equation}\label{eq:fcCL}
\begin{split}
f_c^{(\mathrm{OL})} &= f_{\mathrm{G}} \Bigg[ 1.03 \Big( \dfrac{D_{\mathrm{R}}}{r_0} \Big)^{5/3} \Big( \dfrac{\lambda_0}{\lambda} \Big)^{2} \\
&- \Big( \dfrac{\pi}{2} \dfrac{f_{\mathrm{TG}}}{f_{tc}} \Big)^{2} \Bigg]^{-3/5},
\end{split}
\end{equation}
where in this case the superscript OL indicates that this $f_c$ is the effective closed-loop bandwidth of open-loop operation.
In other words, $f_c^{(\mathrm{OL})}$ is the closed-loop bandwidth that yields no improvement over open-loop optical receiver performance.
It also serves as the lower bound when investigating the $f_c$ dependence of Eq.~\ref{eq:RPECL}.
Either of these equations can be used to show the relative improvement AO can provide. 
In the main text we do this by including vertical lines indicating the effective $r_0$ of AO correction and by plotting the key-bit rate as a function of $f_c$ in Fig.~\ref{fig:WinSolZenHighVisSKBP2}. 

In Ref.~\cite{gruneisen2020adaptive} we carefully show how one can calculate the Fried coherence and slew dependent Greenwood frequency for a circular orbit passing through zenith. 
The emphasis therein was to indicate the equivalent horizontal-path propagation length corresponding to slant-path propagation.
Since in the present analysis we have limited our study to zenith, we set $\theta_\mathrm{z}=0$ and using the slant-path expressions with $D_{\mathrm{R}}$=1 m, the 1$\times$HV$_{5/7}$ turbulence profile, and the Bufton wind model we find
\begin{equation}\label{eq:ATMOSPHERE}
\begin{split}
r(500 \, \mathrm{nm})=r_0 \approx 5 \, \mathrm{cm},\\
f_{\mathrm{G}}(500 \, \mathrm{nm}) \approx 301 \, \mathrm{Hz},\\
f_{\mathrm{TG}}(500 \, \mathrm{nm}) \approx 43 \, \mathrm{Hz}.
\end{split}
\end{equation}
In our field experiment we built a $f_c=130$-Hz AO system for compensation of temporal characteristics corresponding to a 1.6-km horizontal channel where the maximum observed $f_{\mathrm{G}}$ was approximately 60 Hz.
For the LEO down-link case that we consider here, one would in practice build a faster AO system that can more effectively compensate for the temporal component of the turbulence which is enhanced due to slewing.
For example, we conducted detailed simulations considering both 200-Hz and 500-Hz AO systems in Refs.~\cite{gruneisen2016adaptive, gruneisen2017modeling}.
For this simulation, we will assume a tracking bandwidth $f_{tc} = 60$ Hz and consider these three higher-order AO bandwidths in order to demonstrate there is not a sharp cutoff in effectiveness and even a relatively slow system, that is, a system slower than the observed Greenwood frequency, can still provide a substantial boost in QKD performance if the proper spatial filtering strategy is implemented.  
Therefore, using $f_{\mathrm{TG}}=43$ Hz, $f_{\mathrm{G}}=301$ Hz, and $f_{tc} = 60$ Hz in Eq.~\ref{eq:r0CL} we find $r_0^{(\mathrm{CL})} \approx 37$, 50, and 74 cm for $f_c = 130$, 200, and 500 Hz respectively.
We will use these values in the main text to indicate examples of closed-loop AO operation.

 \begin{figure}[b]
	\includegraphics[width=1\columnwidth]{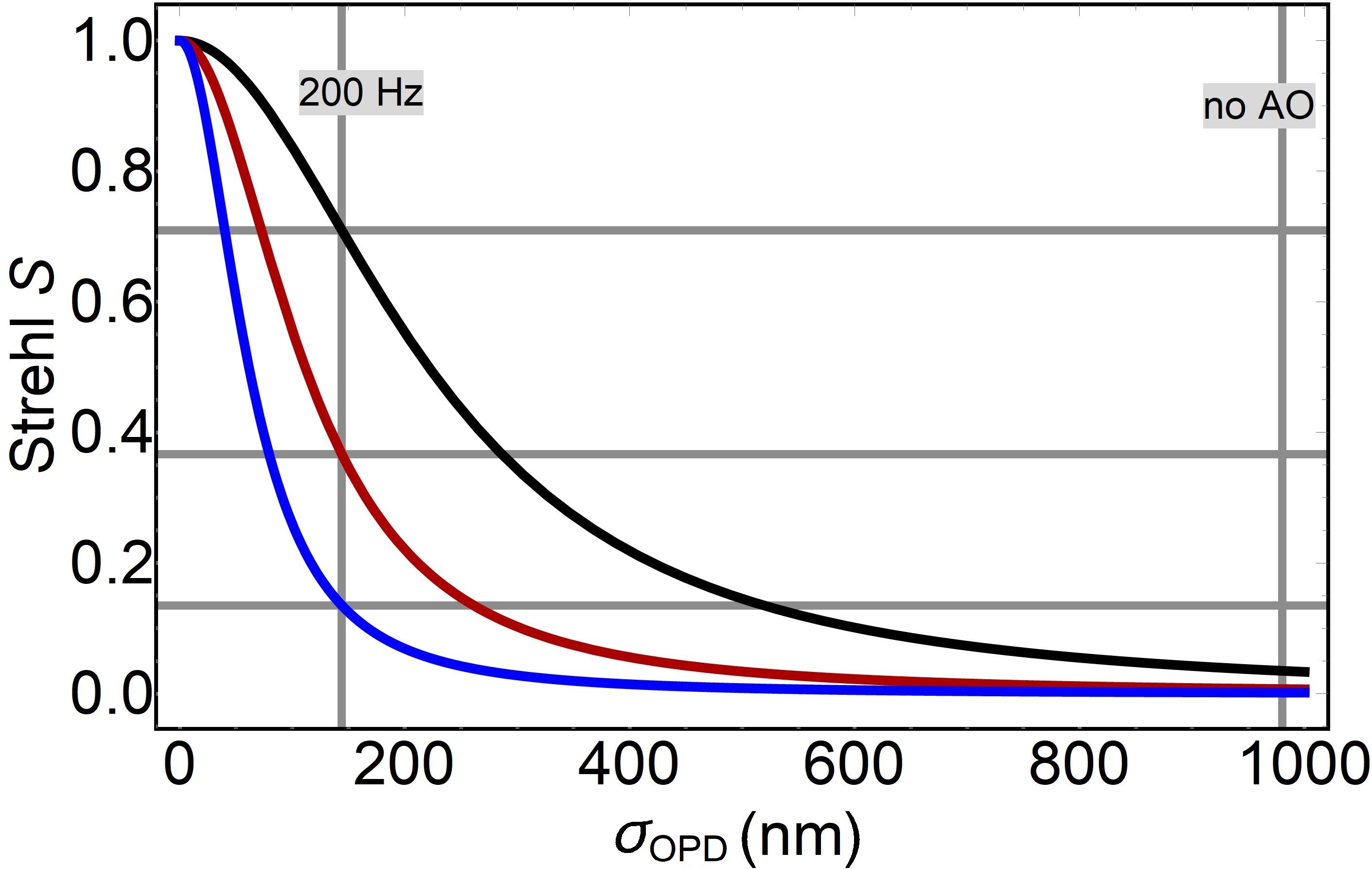}\\
	\caption{\label{fig:WinSolZenHighVisSnm}
The system Strehl for 1550, 781, and 431 nm (black, red, and blue respectively), plotted as a function of $\sigma_{\mathrm{OPD}}$. The vertical lines at 144 nm and 980 nm indicate the $\sigma_{\mathrm{OPD}}$ of a closed- and open-loop 200-Hz AO system correcting for an atmosphere characterized by $r_0=5$ cm, respectively. The horizontal lines indicate the achieved closed-loop Strehl at each wavelength. 
	}
\end{figure}
\subsection{Strehl}\label{sec:appendixAStrehl}
Another performance parameter closely related to AO systems is the system Strehl.
Using Eq.~\ref{eq:RPE} and \ref{eq:RPEnm} in Eq.~\ref{eq:SR}, one can rewrite the Strehl as
\begin{equation}\label{eq:SRPEnm}
S  = \Big[ 1 + \dfrac{1}{1.03} \, \sigma_{\mathrm{OPD}}^2 \Big( \dfrac{2\pi}{\lambda} \Big)^2 \Big]^{-6/5}.
\end{equation}
This is significant because it shows that a AO system has a wavelength-dependent system Strehl.
For the present analysis we use $r_0$ from Eq.~\ref{eq:ATMOSPHERE} and substitute Eq.~\ref{eq:RPEnm2} into Eq.~\ref{eq:SRPEnm} to find $\sigma_{\mathrm{OPD,OL}} \approx 980$ nm.
Using $\theta_z$=0, $D_{\mathrm{R}}$=1 m, the 1$\times$HV$_{5/7}$ turbulence profile, the Bufton wind model, and tracking bandwidth $f_{tc}=60$ Hz one can use Eq.~\ref{eq:RPEnmCL} and find $\sigma_{\mathrm{OPD,CL}} \approx 184$, 144, and 104 nm for $f_c$=130-, 200-, and 500-Hz higher-order-bandwidth AO systems, respectively.

 \begin{figure*}[t!]
	\includegraphics[width=1\textwidth]{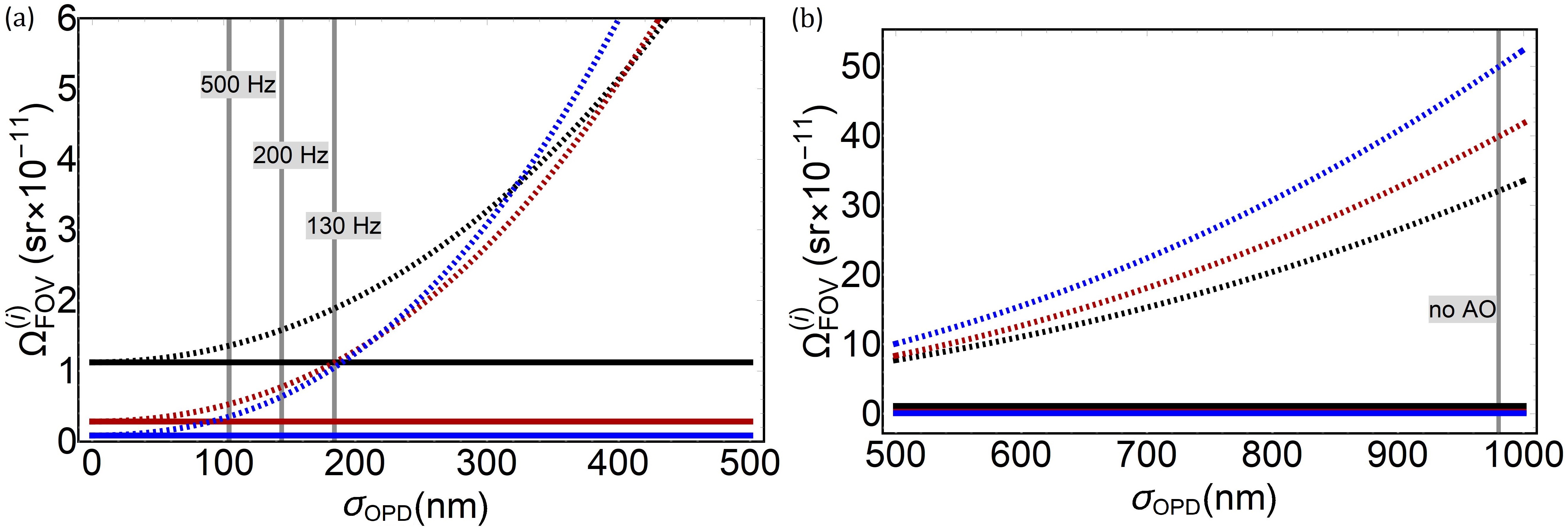}
	\caption{\label{fig:WinSolZenHighVisFOVsigmanm}
The linear angle FOV for 1550, 781, and 431 nm (black, red, and blue respectively), plotted as a function of $\sigma_{\mathrm{OPD}}$ for (a) 0 to 500 nm and (b) 500 to 1000 nm. The solid and dashed curves indicate the DL and TL FOV's, respectively. The line at $\sigma_{\mathrm{OPD}}=980$ nm corresponds to no AO correction whereas the lines at $\sigma_{\mathrm{OPD}}=184$, 144, and 104 nm correspond to full AO with 130-, 200-, and 500-Hz closed-loop bandwidths, respectively. 
	}
\end{figure*}
In Fig.~\ref{fig:WinSolZenHighVisSnm} we plot the system Strehl for 1550, 781, and 431 nm (black, red, and blue respectively) as a function of $\sigma_{\mathrm{OPD}}$. 
The vertical line on the left indicates the $\sigma_{\mathrm{OPD}}$ corresponding to closed-loop operation with $f_{tc}=60$ Hz and $f_c$=200 Hz.
The horizontal lines indicate the achieved closed-loop Strehl at each wavelength.
Therefore, one can see that a short-wavelength closed-loop AO system will operate at a much lower system Strehl.
In the main text we investigate how this affects the performance of the QKD system.

\subsection{Spot Size}\label{sec:appendixASpot}
The wavelength dependence of the sky background $H_{\mathrm{b}}$ and the geometric coupling are important components of the optimization problem, but the most fundamental physics pertains to the focused spot size and it's dependence on wavelength. 
Equation~\ref{eq:FRIED} shows that longer wavelengths are affected less by a given atmospheric condition, but Eq.~\ref{eq:DLSPOT} shows that, in the absence of turbulence, the focused spot is larger, resulting in either more loss at the spatial filter \textit{or} more noise, depending on the spatial filter strategy.
The shorter wavelength has a smaller focused spot but is impacted more by the atmosphere.
We will now investigate these competing trends in terms of the residual error.
Using Eq.~\ref{eq:SRPEnm} in Eq.~\ref{eq:TLSPOT} one can write 
\begin{equation}\label{eq:TLSPOT_RPEnm}
d_{\mathrm{spot}}^{(\mathrm{TL})}  = d_{\mathrm{spot}}^{(\mathrm{DL})} \Big[ 1 + \dfrac{1}{1.03} \, \sigma_{\mathrm{OPD}}^2 \Big( \dfrac{2\pi}{\lambda} \Big)^2 \Big]^{3/5},
\end{equation}
which can be used to find the TL FOV
\begin{equation}\label{eq:TLFOV_RPEnm}
\Omega_{\mathrm{FOV}}^{(\mathrm{TL})}=\pi \bigg( 1.22 \dfrac{ \lambda }{D_\mathrm{R}} \Big[
1+ \dfrac{1}{1.03} \, \sigma_{\mathrm{OPD}}^2 \Big( \dfrac{2\pi}{\lambda} \Big)^2
\Big]^{ 3/5} \bigg)^2 .
\end{equation}
In Fig.~\ref{fig:WinSolZenHighVisFOVsigmanm} we use Eq.~\ref{eq:TLFOV_RPEnm} to plot the solid-angle FOV as a function of OPD standard deviation $\sigma_{\mathrm{nm}}$ (this figure is an analog to Fig.~\ref{fig:WinSolZenHighVisLAFOV1550} which was a function of $r_0$).
This open-loop scenario can be seen in Fig.~\ref{fig:WinSolZenHighVisFOVsigmanm}(b) where one will notice that the strong wavelength dependent atmospheric effects are dominant here and the FOV's are larger for the short wavelengths.
In Fig.~\ref{fig:WinSolZenHighVisFOVsigmanm}(a) we plot the range for a well-functioning AO system.
One will see that in the range $300\,\mathrm{nm}< \sigma_{\mathrm{OPD}} <400$ nm there is a transition where the wavelength dependence of spot size begins to dominate and shorter wavelengths permit smaller FOV's.

\section{Optimal Wavelength for Decoy-State BB84-QKD}\label{sec:appendixB}
\begin{figure*}[]
	\includegraphics[width=1.0\textwidth]{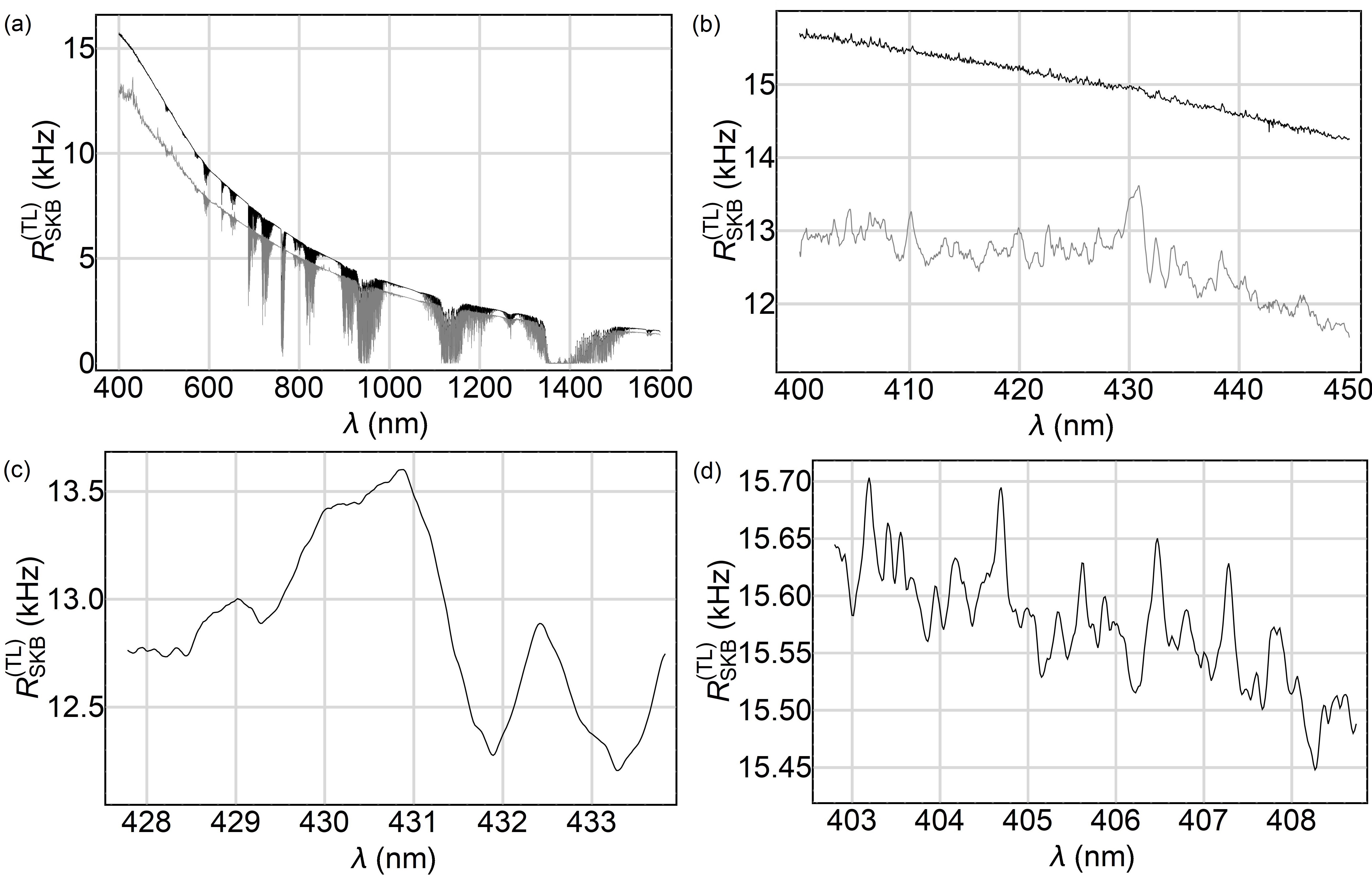}
	\caption{\label{fig:WinSolZenHighVisFineWavelength}
The key-bit rate $R_{\mathrm{KB}}^{(\mathrm{TL})}$ for winter solstice with high visibility and $r_0 = 50$ cm. In (a) and (b) we give key-bit rate for a 1-nm (0.05-nm) filter in gray (black). In (c) and (d) we zoom in to reveal a range of optimal wavelengths for a 1-nm and 0.05-nm filter, respectively.
	}
\end{figure*}
We chose wavelengths near 1550 and 780 nm because these are common wavelengths considered for space-Earth quantum communications.
To find a true optimal wavelength we investigate the wavelength dependence of the key-bit rate $R_{\mathrm{SKB}}^{(\mathrm{TL})}$ for 1:00 PM on the winter solstice with high visibility over a large wavelength range.
In Figs.~\ref{fig:WinSolZenHighVisFineWavelength}(a) and \ref{fig:WinSolZenHighVisFineWavelength}(b) we plot the key-bit rate for two different wavelength ranges. 
We assume an AO system with $f_{tc}=60$ Hz and $f_c=200$ Hz yielding $r_0^{(\mathrm{CL})} = 50$ cm.
Furthermore, the black and gray curves represent the key-bit rate for 0.05-nm and 1-nm spectral filters, respectively.
One can see that although the sky is generally brighter and transmission is poorer at shorter wavelengths, the key-bit rates are generally higher.
In Figs.~\ref{fig:WinSolZenHighVisFineWavelength}(b-d) we zoom in to the shorter wavelength range and reveal the optimal wavelength for the site condition using the two different filters.
For the 1-nm filter we find the optimal wavelength $\lambda_{\mathrm{opt}}^{(1\,\mathrm{nm})}=430.886$ nm and for the 0.05-nm filter there is a convenient peak at $\lambda_{\mathrm{opt}}^{(0.05\,\mathrm{nm})}=404.694$ nm (see Figs.~\ref{fig:WinSolZenHighVisFineWavelength}(c) and \ref{fig:WinSolZenHighVisFineWavelength}(d), respectively).


\end{document}